

\input amssym.tex 

\def\unredoffs{}
\tolerance=1000\hfuzz=2pt
\catcode`\@=11 
\ifx\hyperdef\UNd@FiNeD\def\hyperdef#1#2#3#4{#4}\def\hyperref#1#2#3#4{#4}\def\href#1#2{#2}\fi
\magnification=1200\unredoffs\baselineskip=16pt plus 2pt minus 1pt
\def\Date#1{\vfill\leftline{#1}\tenpoint\supereject%
\footline={\hss\tenrm\hyperdef\hypernoname{page}\folio\folio\hss}}%

{\count255=\time\divide\count255 by 60 \xdef\hourmin{\number\count255}
 \multiply\count255 by-60\advance\count255 by\time
 \xdef\hourmin{\hourmin:\ifnum\count255<10 0\fi\the\count255}
}
\def\date{\number\day.\number\month.\number\year\ at \hourmin}


\def\nolabels{\def\wrlabeL##1{}\def\eqlabeL##1{}\def\reflabeL##1{}}
\def\writelabels{\def\wrlabeL##1{\leavevmode\vadjust{\rlap{\smash%
{\line{{\escapechar=` \hfill\rlap{\sevenrm\hskip.03in\string##1}}}}}}}%
\def\eqlabeL##1{{\escapechar-1\rlap{\sevenrm\hskip.05in\string##1}}}%
\def\reflabeL##1{\noexpand\llap{\noexpand\sevenrm\string\string\string##1}}}
\nolabels

\global\newcount\secno \global\secno=0
\global\newcount\meqno \global\meqno=1
\def\s@csym{}

\def\newsec#1\par{\global\advance\secno by1%
{\toks0{#1}\message{(\the\secno. \the\toks0)}}%
\global\subsecno=0\eqnres@t\let\s@csym\secsym\xdef\secn@m{\the\secno}\noindent
{\bf\hyperdef\hypernoname{section}{\the\secno}{\the\secno.} #1}%
\writetoca{{\string\hyperref{}{section}{\the\secno}{\bf \the\secno\quad}} {\bf #1}}\par%
\nobreak\medskip\nobreak\noindent\ignorespaces}
\def\eqnres@t{\xdef\secsym{\the\secno.}\global\meqno=1\bigbreak\bigskip}
\def\sequentialequations{\def\eqnres@t{\bigbreak}}\xdef\secsym{}

\global\newcount\subsecno \global\subsecno=0
\def\subsec#1\par{\global\advance\subsecno by1%
{\toks0{#1}\message{(\s@csym\the\subsecno. \the\toks0)}}%
\global\subsubsecno=0%
\ifnum\lastpenalty>9000\else\bigbreak\fi
\noindent{\it\hyperdef\hypernoname{subsection}{\secn@m.\the\subsecno}%
{\secn@m.\the\subsecno.} #1}\writetoca{\string\hskip1.45cm
{\string\hyperref{}{subsection}{\secn@m.\the\subsecno}{\secn@m.\the\subsecno.}}
{#1}}\par\nobreak\medskip\nobreak\noindent\ignorespaces}

\global\newcount\subsubsecno \global\subsubsecno=0
\def\subsubsec#1\par{\global\advance\subsubsecno by1%
{\toks0{#1}\message{(\secn@m.\the\subsecno.\the\subsubsecno. \the\toks0)}}%
\global\subsubsubsecno=0%
\ifnum\lastpenalty>9000\else\bigbreak\fi
\noindent{\it\hyperdef\hypernoname{subsubsection}{\secn@m.\the\subsecno\the\subsubsecno}%
{\secn@m.\the\subsecno.\the\subsubsecno.} #1}
\par\nobreak\medskip\nobreak\noindent\ignorespaces}

\global\newcount\subsubsubsecno \global\subsubsubsecno=0
\def\subsubsubsec#1\par{\global\advance\subsubsubsecno by1%
{\toks0{#1}\message{(\secn@m.\the\subsecno.\the\subsubsecno.\the\subsubsubsecno \the\toks0)}}%
\ifnum\lastpenalty>9000\else\bigbreak\fi
\noindent{\it\hyperdef\hypernoname{subsubsection}{\secn@m.\the\subsecno\the\subsubsecno\the\subsubsubsecno}%
{\secn@m.\the\subsecno.\the\subsubsecno.\the\subsubsubsecno.} #1}%
\par\nobreak\medskip\nobreak\noindent\ignorespaces}


\def\newnewsec#1#2\par{\global\advance\secno by1%
{\toks0{#2}\message{(\secn@m. \the\toks0)}}%
\global\subsecno=0\eqnres@t\let\s@csym\secsym\xdef\secn@m{\the\secno}\noindent
\ifnum\lastpenalty>9000\else\bigbreak\fi
\noindent{\bf\hyperdef\hypernoname{section}{\secn@m}{\secn@m.} #2}%
\writetoca{{\string\hyperref{}{section}{\the\secno}{\bf \the\secno\quad}} {\bf #2}}
\DefWarn#1%
\xdef#1{\noexpand\hyperref{}{section}{\the\secno}%
{\the\secno}}\writedef{#1\leftbracket#1}\wrlabeL{#1=#1}%
\par\nobreak\medskip\nobreak\noindent\ignorespaces}

\def\newsubsec#1#2\par{\global\advance\subsecno by1%
{\toks0{#2}\message{(\secn@m.\the\subsecno. \the\toks0)}}%
\global\subsubsecno=0%
\ifnum\lastpenalty>9000\else\bigbreak\fi
\noindent{\it\hyperdef\hypernoname{subsection}{\secn@m.\the\subsecno}%
{\secn@m.\the\subsecno.} #2}
\DefWarn#1%
\xdef#1{\noexpand\hyperref{}{subsection}{\secn@m.\the\subsecno}%
{\secn@m.\the\subsecno}}\writedef{#1\leftbracket#1}\wrlabeL{#1=#1}%
\writetoca{\string\hskip1.45cm
{\string\hyperref{}{subsection}{\secn@m.\the\subsecno}{\secn@m.\the\subsecno.}}
{#2}}%
\par\nobreak\medskip\nobreak\noindent\ignorespaces}

\def\newsubsubsec#1#2\par{\global\advance\subsubsecno by1%
{\toks0{#2}\message{(\secn@m.\the\subsecno.\the\subsubsecno. \the\toks0)}}%
\global\subsubsubsecno=0%
\ifnum\lastpenalty>9000\else\bigbreak\fi
\noindent{\it\hyperdef\hypernoname{subsubsection}{\secn@m.\the\subsecno\the\subsubsecno}%
{\secn@m.\the\subsecno.\the\subsubsecno.} #2}
\DefWarn#1%
\xdef#1{\noexpand\hyperref{}{subsubsection}{\secn@m.\the\subsecno.\the\subsubsecno}%
{\secn@m.\the\subsecno.\the\subsubsecno}}\writedef{#1\leftbracket#1}\wrlabeL{#1=#1}%
\par\nobreak\medskip\nobreak\noindent\ignorespaces}

\def\newsubsubsubsec#1#2\par{\global\advance\subsubsubsecno by1%
{\toks0{#2}\message{(\secn@m.\the\subsecno.\the\subsubsecno.\the\subsubsubsecno \the\toks0)}}%
\ifnum\lastpenalty>9000\else\bigbreak\fi
\noindent{\it\hyperdef\hypernoname{subsubsection}{\secn@m.\the\subsecno\the\subsubsecno\the\subsubsubsecno}%
{\secn@m.\the\subsecno.\the\subsubsecno.\the\subsubsubsecno.} #2}
\DefWarn#1%
\xdef#1{\noexpand\hyperref{}{subsubsubsection}{\secn@m.\the\subsecno.\the\subsubsecno.\the\subsubsubsecno}%
{\secn@m.\the\subsecno.\the\subsubsecno.\the\subsubsubsecno}}\writedef{#1\leftbracket#1}\wrlabeL{#1=#1}%
\par\nobreak\medskip\nobreak\noindent\ignorespaces}

\def\appendix#1#2{\global\meqno=1\global\subsecno=0\global\subsubsecno=0\xdef\secsym{\hbox{#1.}}%
\bigbreak\bigskip\noindent{\bf Appendix \hyperdef\hypernoname{appendix}{#1}%
{#1.} #2}{\toks0{(#1. #2)}\message{\the\toks0}}%
\xdef\s@csym{#1.}\xdef\secn@m{#1}%
\writetoca{{\string\hyperref{}{appendix}{#1}{\bf {#1}\quad}} {\bf #2}}%
\par\nobreak\medskip\nobreak}

%
\def\checkm@de#1#2{\ifmmode{\def\f@rst##1{##1}\hyperdef\hypernoname{equation}%
{#1}{#2}}\else\hyperref{}{equation}{#1}{#2}\fi}
\def\eqnn#1{\DefWarn#1\xdef #1{(\noexpand\relax\noexpand\checkm@de%
{\s@csym\the\meqno}{\secsym\the\meqno})}%
\wrlabeL#1\writedef{#1\leftbracket#1}\global\advance\meqno by1}
\def\f@rst#1{\c@t#1a\em@ark}\def\c@t#1#2\em@ark{#1}
\def\eqna#1{\DefWarn#1\wrlabeL{#1$\{\}$}%
\xdef #1##1{(\noexpand\relax\noexpand\checkm@de%
{\s@csym\the\meqno\noexpand\f@rst{##1}1}{\hbox{$\secsym\the\meqno##1$}})}
\writedef{#1\numbersign1\leftbracket#1{\numbersign1}}\global\advance\meqno by1}
\def\eqn#1#2{\DefWarn#1%
\xdef #1{(\noexpand\hyperref{}{equation}{\s@csym\the\meqno}%
{\secsym\the\meqno})}$$#2\eqno(\hyperdef\hypernoname{equation}%
{\s@csym\the\meqno}{\secsym\the\meqno})\eqlabeL#1$$%
\writedef{#1\leftbracket#1}\global\advance\meqno by1}
\def\xeqn{\expandafter\xe@n}\def\xe@n(#1){#1}
\def\xeqna#1{\expandafter\xe@n#1}
\def\eqns#1{(\e@ns #1{\hbox{}})}
\def\e@ns#1{\ifx\UNd@FiNeD#1\message{eqnlabel \string#1 is undefined.}%
\xdef#1{(?.?)}\fi{\let\hyperref=\relax\xdef\next{#1}}%
\ifx\next\em@rk\def\next{}\else%
\ifx\next#1\xeqn#1\else\def\n@xt{#1}\ifx\n@xt\next#1\else\xeqna#1\fi
\fi\let\next=\e@ns\fi\next}
\def\DefWarn#1{}
%
\newskip\footskip\footskip14pt plus 1pt minus 1pt 
\def\footnotefont{\ninepoint}\def\f@t#1{\footnotefont #1\@foot}
\def\f@@t{\baselineskip\footskip\bgroup\footnotefont\aftergroup\@foot\let\next}
\setbox\strutbox=\hbox{\vrule height9.5pt depth4.5pt width0pt}
\global\newcount\ftno \global\ftno=0
\def\foot{\global\advance\ftno by1\def\foot@rg{\hyperref{}{footnote}%
{\the\ftno}{\the\ftno}\xdef\foot@rg{\noexpand\hyperdef\noexpand\hypernoname%
{footnote}{\the\ftno}{\the\ftno}}}\footnote{$^{\foot@rg}$}}
%
%
%
\global\newcount\refno \global\refno=1
\newwrite\rfile
\def\ref{[\hyperref{}{reference}{\the\refno}{\the\refno}]\nref}
\def\nref#1{\DefWarn#1%
\xdef#1{[\noexpand\hyperref{}{reference}{\the\refno}{\the\refno}]}%
\writedef{#1\leftbracket#1}%
\ifnum\refno=1\immediate\openout\rfile=\jobname.refs\fi
\chardef\wfile=\rfile\immediate\write\rfile{\noexpand\item{[\noexpand\hyperdef%
\noexpand\hypernoname{reference}{\the\refno}{\the\refno}]\ }%
\reflabeL{#1\hskip.31in}\pctsign}\global\advance\refno by1\findarg}
\def\findarg#1#{\begingroup\obeylines\newlinechar=`\^^M\pass@rg}
{\obeylines\gdef\pass@rg#1{\writ@line\relax #1^^M\hbox{}^^M}%
\gdef\writ@line#1^^M{\expandafter\toks0\expandafter{\striprel@x #1}%
\edef\next{\the\toks0}\ifx\next\em@rk\let\next=\endgroup\else\ifx\next\empty%
\else\immediate\write\wfile{\the\toks0}\fi\let\next=\writ@line\fi\next\relax}}
\def\striprel@x#1{} \def\em@rk{\hbox{}}
\def\lref{\begingroup\obeylines\lr@f}
\def\lr@f#1#2{\DefWarn#1\gdef#1{\let#1=\UNd@FiNeD\ref#1{#2}}\endgroup\unskip}
\def\semi{;\hfil\break}
\def\addref#1{\immediate\write\rfile{\noexpand\item{}#1}} 
\def\listrefs{\vfill\supereject\immediate\closeout\rfile\writestoppt
\baselineskip=\footskip\centerline{{\bf References}}\bigskip{\parindent=20pt%
\frenchspacing\escapechar=` \input \jobname.refs\vfill\eject}\nonfrenchspacing}
\def\startrefs#1{\immediate\openout\rfile=\jobname.refs\refno=#1}
\def\xref{\expandafter\xr@f}\def\xr@f[#1]{#1}
\def\refs#1{\count255=1[\r@fs #1{\hbox{}}]}
\def\r@fs#1{\ifx\UNd@FiNeD#1\message{reflabel \string#1 is undefined.}%
\nref#1{need to supply reference \string#1.}\fi%
\vphantom{\hphantom{#1}}{\let\hyperref=\relax\xdef\next{#1}}%
\ifx\next\em@rk\def\next{}%
\else\ifx\next#1\ifodd\count255\relax\xref#1\count255=0\fi%
\else#1\count255=1\fi\let\next=\r@fs\fi\next}
%

%
\newwrite\ffile\global\newcount\figno \global\figno=1
\def\fig{fig.~\hyperref{}{figure}{\the\figno}{\the\figno}\nfig}
\def\nfig#1{\DefWarn#1%
\xdef#1{fig.~\noexpand\hyperref{}{figure}{\the\figno}{\the\figno}}%
\writedef{#1\leftbracket fig.\noexpand~\xfig#1}%
\ifnum\figno=1\immediate\openout\ffile=\jobname.figs\fi\chardef\wfile=\ffile%
{\let\hyperref=\relax
\immediate\write\ffile{\noexpand\medskip\noexpand\item{Fig.\ %
\noexpand\hyperdef\noexpand\hypernoname{figure}{\the\figno}{\the\figno}. }
\reflabeL{#1\hskip.55in}\pctsign}}\global\advance\figno by1\findarg}
\def\xfig{\expandafter\xf@g}\def\xf@g fig.\penalty\@M\ {}
\def\figs#1{figs.~\f@gs #1{\hbox{}}}
\def\f@gs#1{{\let\hyperref=\relax\xdef\next{#1}}\ifx\next\em@rk\def\next{}\else
\ifx\next#1\xfig #1\else#1\fi\let\next=\f@gs\fi\next}
%
\def\figin{\epsfcheck\figin}\def\figins{\epsfcheck\figins}
\def\epsfcheck{\ifx\epsfbox\UnDeFiNeD
\message{(NO epsf.tex, FIGURES WILL BE IGNORED)}
\gdef\figin##1{\vskip2in}\gdef\figins##1{\hskip.5in}
\else\message{(FIGURES WILL BE INCLUDED)}%
\gdef\figin##1{##1}\gdef\figins##1{##1}\fi}
\def\figinsert{\goodbreak\topinsert}
\def\ifig#1#2#3{\DefWarn#1\xdef#1{fig.~\the\figno}
\writedef{#1\leftbracket fig.\noexpand~\the\figno}%
\figinsert\figin{\centerline{#3}}
\smallskip
\leftskip=0pt \rightskip=0pt
\baselineskip12pt\noindent
{{\bf Fig.~\the\figno}\ \ninepoint #2}
\medskip
\global\advance\figno by1\par\endinsert}
\newwrite\lfile
{\escapechar-1\xdef\pctsign{\string\%}\xdef\leftbracket{\string\{}
\xdef\rightbracket{\string\}}\xdef\numbersign{\string\#}}
\def\writedefs{\immediate\openout\lfile=label.defs \def\writedef##1{%
{\let\hyperref=\relax\let\hyperdef=\relax\let\hypernoname=\relax
 \immediate\write\lfile{\string\checkdef\string##1\rightbracket}}}}%
\def\writestop{\def\writestoppt{\immediate\write\lfile{\string\pageno
 \the\pageno\string\startrefs\leftbracket\the\refno\rightbracket
 \string\def\string\secsym\leftbracket\secsym\rightbracket
 \string\secno\the\secno\string\meqno\the\meqno}\immediate\closeout\lfile}}
\def\writestoppt{}\def\writedef#1{}

\def\seclab#1\par{\DefWarn#1%
\xdef #1{\noexpand\hyperref{}{section}{\the\secno}{\the\secno}}%
\writedef{#1\leftbracket#1}\wrlabeL{#1=#1}\par%
\nobreak\medskip\nobreak\noindent\ignorespaces}
\def\subseclab#1\par{\DefWarn#1%
\xdef #1{\noexpand\hyperref{}{subsection}{\the\secno.\the\subsecno}%
{\the\secno.\the\subsecno}}\writedef{#1\leftbracket#1}\wrlabeL{#1=#1}\par%
\nobreak\medskip\nobreak\noindent\ignorespaces}
\def\subsubseclab#1\par{\DefWarn#1%
\xdef#1{\noexpand\hyperref{}{subsubsection}{\the\secno.\the\subsecno.\the\subsubsecno}%
{\the\secno.\the\subsecno.\the\subsubsecno}}\writedef{#1\leftbracket#1}\wrlabeL{#1=#1}\par%
\nobreak\medskip\nobreak\noindent\ignorespaces}
\def\applab#1\par{\DefWarn#1%
\xdef#1{\noexpand\hyperref{}{appendix}{\secn@m}{\secn@m}}%
\writedef{#1\leftbracket#1}\wrlabeL{#1=#1}%
\par\nobreak\medskip\nobreak\noindent\ignorespaces}
\def\appsublab#1{\DefWarn#1%
\xdef #1{\noexpand\hyperref{}{appendix}{\secn@m.\the\subsecno}{\secn@m.\the\subsecno}}%
\writedef{#1\leftbracket#1}\wrlabeL{#1=#1}}
\newwrite\tfile \def\writetoca#1{}
\def\leaderfill{\leaders\hbox to 1em{\hss.\hss}\hfill}
\def\writetoc{\immediate\openout\tfile=\jobname.toc
   \def\writetoca##1{{\edef\next{\write\tfile{\noindent ##1
   \string\leaderfill{
   \string\hyperref{}{page}{\noexpand\number\pageno}%
   {\noexpand\number\pageno}} \par}}\next}}
}
\newread\ch@ckfile
\def\listtoc{\immediate\closeout\tfile\immediate\openin\ch@ckfile=\jobname.toc
\ifeof\ch@ckfile\message{no file \jobname.toc, no table of contents this pass}%
\else\closein\ch@ckfile\centerline{\bf Contents}\nobreak\medskip%
{\baselineskip=14.5pt\footnotefont\parskip=0pt\catcode`\@=11\input\jobname.toc
\catcode`\@=12\bigbreak\bigskip}\fi}
\catcode`\@=12 
\def\tenpoint{\def\rm{\fam0\tenrm}
\textfont0=\tenrm \scriptfont0=\sevenrm \scriptscriptfont0=\fiverm
\textfont1=\teni  \scriptfont1=\seveni  \scriptscriptfont1=\fivei
\textfont2=\tensy \scriptfont2=\sevensy \scriptscriptfont2=\fivesy
\textfont\itfam=\tenit \def\it{\fam\itfam\tenit}\def\footnotefont{\ninepoint}%
\textfont\bffam=\tenbf \def\bf{\fam\bffam\tenbf}\def\sl{\fam\slfam\tensl}\rm}
\font\ninerm=cmr9 \font\sixrm=cmr6 \font\ninei=cmmi9 \font\sixi=cmmi6
\font\ninesy=cmsy9 \font\sixsy=cmsy6 \font\ninebf=cmbx9
\font\nineit=cmti9 \font\ninesl=cmsl9 \skewchar\ninei='177
\skewchar\sixi='177 \skewchar\ninesy='60 \skewchar\sixsy='60
\def\ninepoint{\def\rm{\fam0\ninerm}
\textfont0=\ninerm \scriptfont0=\sixrm \scriptscriptfont0=\fiverm
\textfont1=\ninei \scriptfont1=\sixi \scriptscriptfont1=\fivei
\textfont2=\ninesy \scriptfont2=\sixsy \scriptscriptfont2=\fivesy
\textfont\itfam=\ninei \def\it{\fam\itfam\nineit}\def\sl{\fam\slfam\ninesl}%
\textfont\bffam=\ninebf \def\bf{\fam\bffam\ninebf}\rm}
%
\hyphenation{anom-aly anom-alies coun-ter-term coun-ter-terms}

\def\tikzcaption#1#2{\DefWarn#1\xdef#1{Fig.~\the\figno}
\writedef{#1\leftbracket Fig.\noexpand~\the\figno}%
{
\smallskip
\leftskip=20pt \rightskip=20pt \baselineskip12pt\noindent
{{\bf Fig.~\the\figno}\ \ninepoint #2}
\bigskip
\global\advance\figno by1 \par}}

\def\ntoalpha#1{%
\ifcase#1%
@%
\or A\or B\or C\or D\or E\or F\or G\or H\or I\or J\or K\or L\or M%
\fi
}

\global\newcount\appno \global\appno=1
\def\applab#1{\xdef #1{\ntoalpha{\appno}}\writedef{#1\leftbracket#1}\wrlabeL{#1=#1}
\global\advance\appno by1}

\def\preprint#1 #2\par{\rightline{\vbox{\baselineskip12pt\hbox{#1}\hbox{#2}}}\vskip2cm}
%
\def\title#1\par{\centerline{\bf #1}\nopagenumbers\pageno=0}
\def\author#1\par{\bigskip\bigskip\centerline{#1}}

\newcount\addressno

\def\email#1#2{
\footnote{\null}{\kern-\parindent \llap{$^#1$\hskip1pt}email: #2}}

\def\startcenter{%
  \par
  \begingroup
  \leftskip=0pt plus 1fil
  \rightskip=\leftskip
  \parindent=0pt
  \parfillskip=0pt
}
\def\stopcenter{\endgroup}

\def\address{\bigskip%
  \ifnum\the\addressno=0\else\stopcenter\endgroup\fi
  \advance\addressno by 1%
  \begingroup
  \startcenter
  \it
  \obeylines
  \addressAux
}
\def\addressAux#1{#1}

\def\abstract{\stopcenter\endgroup\bigskip\bigskip\noindent}

\def\Dsl{\,\raise.15ex\hbox{/}\mkern-13.5mu D} 
\def\dsl{\raise.15ex\hbox{/}\kern-.57em\partial}
 
\def\boxeqn#1{\vcenter{\vbox{\hrule\hbox{\vrule\kern3pt\vbox{\kern3pt
	\hbox{${\displaystyle #1}$}\kern3pt}\kern3pt\vrule}\hrule}}}


\def\ap{{\alpha^{\prime}}}

\def\a{\alpha}

\def\d{{\delta}}

\def\t{{\theta}}

\def\half{{1\over 2}}
\def\p{{\partial}}

\def\bar{\overline}
\def\({\left(}
\def\){\right)}

\def\cJ{{\cal J}}
\def\cK{{\cal K}}
\def\cI{{\cal I}}

\def\cY{{\cal Y}}
\def\cZ{{\cal Z}}




\def\Im{\mathop{{\rm Im}}} 
\def\sfrac#1/#2{\kern.1em\raise.5ex\hbox{\the\scriptfont0 #1}%
\kern-.1em/\kern-.15em\lower.25ex\hbox{\the\scriptfont0 #2}}

\font\tenshuffle=shuffle10 \font\sevenshuffle=shuffle7 \font\fiveshuffle=shuffle7 at 5pt
\def\shuffle{{%
\def\Dshuffle{\mathbin{\hbox{\tenshuffle\char'001}}}%
\def\Sshuffle{\mathbin{\hbox{\sevenshuffle\char'001}}}%
\def\SSshuffle{\mathbin{\hbox{\fiveshuffle\char'001}}}%
\mathchoice{\Dshuffle}{\Dshuffle}{\Sshuffle}{\SSshuffle}}}


\def\qed{\hbox{\hskip 3pt
\vbox{\hrule\hbox to 7pt{\vrule height 7pt\hfill\vrule}
\hrule}}\hskip3pt}

\overfullrule=0pt\relax

\frenchspacing

\def\checkdef#1#2{
\ifx\UndeFined#1%
	\def#1{#2}
\else
	\immediate\write16{*** BUG ***: the label \string#1 is already defined ***}
\fi
}
\newread\instream

\openin\instream= label.defs
\ifeof\instream\message{No labels in advance yet. Wait till next pass.}
\else\closein\instream \input label.defs
\fi
\writedefs

\def\arXiv:#1].{\hepthStrip#1 \nil}
\def\hepthStrip#1 #2\nil{\href{http://arxiv.org/abs/#1}{arXiv:#1 #2\unskip}].}


\input amssym
\input epsf
\input localpaper.defs


\title Towards the n-point one-loop superstring amplitude II:

\title Worldsheet functions and their duality to kinematics

\author
Carlos R. Mafra\email{\dagger}{c.r.mafra@soton.ac.uk}$^{\dagger}$ and
Oliver Schlotterer\email{\ddagger}{olivers@aei.mpg.de}$^{\ddagger,\ast}$

\address
$^\dagger$Mathematical Sciences and STAG Research Centre, University of Southampton,
Highfield, Southampton, SO17 1BJ, UK

\address
$^\ddagger$Max--Planck--Institut f\"ur Gravitationsphysik
Albert--Einstein--Institut, 14476 Potsdam, Germany

\address
$^\ast$Perimeter Institute for Theoretical Physics, Waterloo, ON N2L 2Y5, Canada

\abstract
This is the second installment of a series of three papers in which we describe
a method to determine higher-point correlation functions in one-loop
open-superstring amplitudes from first principles. In this second part, we study
worldsheet functions defined on a genus-one surface built from the coefficient
functions of the Kronecker--Einsenstein series. We construct two classes of
worldsheet functions whose properties lead to several simplifying features 
within our description of one-loop correlators with the
pure-spinor formalism. The first class is described by functions with prescribed
monodromies, whose characteristic shuffle-symmetry property leads to a
Lie-polynomial structure when multiplied by the local superfields from part I of
this series. The second class is given by so-called generalized elliptic
integrands (GEIs) that are constructed using the same combinatorial patterns 
of the BRST pseudo-invariant superfields from part~I. Both of them lead to 
compact and combinatorially rich expressions for the correlators in part III.
The identities obeyed by the two classes of worldsheet functions exhibit 
striking parallels with those of the superfield kinematics. We will refer to 
this phenomenon as a duality between worldsheet functions and kinematics.

\Date{December 2018}


\lref\wipI{
C.~R.~Mafra and O.~Schlotterer,
``Towards the n-point one-loop superstring amplitude I:
Pure spinors and superfield kinematics'', [arXiv:1812.10969 [hep-th]]\semi
C.~R.~Mafra and O.~Schlotterer,
``Towards the n-point one-loop superstring amplitude II:
Worldsheet functions and their duality to kinematics'', [arXiv:1812.10970 [hep-th]]\semi
C.~R.~Mafra and O.~Schlotterer,
``Towards the n-point one-loop superstring amplitude III:
 One-loop correlators and their double-copy structure'', [arXiv:1812.10971 [hep-th]]
}

\lref\FORM{
	J.A.M.~Vermaseren,
	``New features of FORM,''
	arXiv:math-ph/0010025.
\semi
	M.~Tentyukov and J.A.M.~Vermaseren,
	``The multithreaded version of FORM,''
	arXiv:hep-ph/0702279.
}

\lref\cresson{
	J. Cresson,
	``Calcul Moulien'', [math/0509548].
}

\lref\Zagier{
	D. Zagier, ``Periods of modular forms and Jacobi theta functions'',
	Invent.Math., 104, 449 (1991).
}

\lref\weil{
	A. Weil, ``Elliptic Functions according to Eisenstein and Kronecker'',
	Ergebnisse der Mathematik und ihrer Grenzgebiete 88,
	Springer-Verlag, 1976
}

\lref\oldMomKer{
	Z.~Bern, L.~J.~Dixon, M.~Perelstein and J.~S.~Rozowsky,
	``Multileg one loop gravity amplitudes from gauge theory,''
	Nucl.\ Phys.\ B {\bf 546}, 423 (1999).
	[hep-th/9811140].
}
\lref\MomKer{
	N.~E.~J.~Bjerrum-Bohr, P.~H.~Damgaard, T.~Sondergaard and P.~Vanhove,
	``The Momentum Kernel of Gauge and Gravity Theories,''
	JHEP {\bf 1101}, 001 (2011).
	[arXiv:1010.3933 [hep-th]].
}

\lref\GreenSW{
  M.~B.~Green, J.~H.~Schwarz and L.~Brink,
  ``N=4 Yang-Mills and N=8 Supergravity as Limits of String Theories,''
Nucl.\ Phys.\ B {\bf 198}, 474 (1982).
}

\lref\mumfordref{
D.~Mumford, M.~Nori and P.~Norman, ``Tata Lectures on Theta I, II,'' Birkh\"auser (1983, 1984).
}

\lref\TsuchiyaVA{
	A.~Tsuchiya,
  	``More on One Loop Massless Amplitudes of Superstring Theories,''
	Phys.\ Rev.\ D {\bf 39}, 1626 (1989).
}

\lref\TsuchiyaNF{
  A.~G.~Tsuchiya,
  ``On the pole structures of the disconnected part of hyper elliptic g loop M point super string amplitudes,''
[arXiv:1209.6117 [hep-th]].
}

\lref\TsuchiyaJOO{
  A.~G.~Tsuchiya,
  ``On new theta identities of fermion correlation functions on genus g Riemann surfaces,''
[arXiv:1710.00206 [hep-th]].
}

\lref\verlindes{
	E.P.~Verlinde and H.L.~Verlinde,
	``Chiral Bosonization, Determinants and the String Partition Function,''
	Nucl.\ Phys.\ B {\bf 288}, 357 (1987).
}

\lref\Ree{
	R. Ree, ``Lie elements and an algebra associated with shuffles'',
	Ann. Math. {\bf 62}, No. 2 (1958), 210--220.
}

\lref\lothaire{
	Lothaire, M., ``Combinatorics on Words'',
	(Cambridge Mathematical Library), Cambridge University Press (1997).
}

\lref\Richards{
	D.~M.~Richards,
	``The One-Loop Five-Graviton Amplitude and the Effective Action,''
	JHEP {\bf 0810}, 042 (2008).
	[arXiv:0807.2421 [hep-th]].
}

\lref\reutenauer{
	C.~Reutenauer,
	``Free Lie Algebras'', London Mathematical Society Monographs, 1993.
}

\lref\DolanEH{
	L.~Dolan and P.~Goddard,
  	``Current Algebra on the Torus,''
	Commun.\ Math.\ Phys.\  {\bf 285}, 219 (2009).
	[arXiv:0710.3743 [hep-th]].
}

\lref\BrownLevin{
	F.~Brown, A.~Levin,
	``Multiple elliptic polylogarithms,'' [arXiv:1110.6917 [math.NT]].
}
\lref\Kronecker{
	L.~Kronecker, ``Zur Theorie der elliptischen Funktionen,''
	Mathematische Werke IV, 313 (1881).
}

\lref\kiritsis{
	E.~Kiritsis,
  	``Introduction to superstring theory,''
	[hep-th/9709062].
}

\lref\MafraNWR{
	C.R.~Mafra and O.~Schlotterer,
	``One-loop superstring six-point amplitudes and anomalies in pure spinor superspace,''
	JHEP {\bf 1604}, 148 (2016).
	[arXiv:1603.04790 [hep-th]].
}

\lref\partI{
  C.R.~Mafra and O.~Schlotterer,
  ``Cohomology foundations of one-loop amplitudes in pure spinor superspace,''
[arXiv:1408.3605 [hep-th]].
}

\lref\nptString{
	C.~R.~Mafra, O.~Schlotterer and S.~Stieberger,
  ``Complete N-Point Superstring Disk Amplitude I. Pure Spinor Computation,''
Nucl.\ Phys.\ B {\bf 873}, 419 (2013).
[arXiv:1106.2645 [hep-th]].
}

\lref\BroedelVLA{
  J.~Broedel, C.R.~Mafra, N.~Matthes and O.~Schlotterer,
  ``Elliptic multiple zeta values and one-loop superstring amplitudes,''
JHEP {\bf 1507}, 112 (2015).
[arXiv:1412.5535 [hep-th]].
}

\lref\psf{
 	N.~Berkovits,
	``Super-Poincare covariant quantization of the superstring,''
	JHEP {\bf 0004}, 018 (2000)
	[arXiv:hep-th/0001035].
}
\lref\MPS{
  N.~Berkovits,
  ``Multiloop amplitudes and vanishing theorems using the pure spinor formalism for the superstring,''
JHEP {\bf 0409}, 047 (2004).
[hep-th/0406055].
}

\lref\MafraIOJ{
C.~R.~Mafra and O.~Schlotterer,
  ``Double-Copy Structure of One-Loop Open-String Amplitudes,''
Phys.\ Rev.\ Lett.\  {\bf 121}, no. 1, 011601 (2018).
[arXiv:1711.09104 [hep-th]].
}
\lref\oneloopMichael{
 M.~B.~Green, C.~R.~Mafra and O.~Schlotterer,
  ``Multiparticle one-loop amplitudes and S-duality in closed superstring theory,''
JHEP {\bf 1310}, 188 (2013).
[arXiv:1307.3534 [hep-th]].
}

\lref\DHokerPDL{
	E.~D'Hoker and D.~H.~Phong,
  	``The Geometry of String Perturbation Theory,''
	Rev.\ Mod.\ Phys.\  {\bf 60}, 917 (1988).
}
\lref\xerox{
	E.~D'Hoker and D.~H.~Phong,
  	``Conformal Scalar Fields and Chiral Splitting on Superriemann Surfaces,''
	Commun.\ Math.\ Phys.\  {\bf 125}, 469 (1989).
}

\lref\GreenMN{
	M.~B.~Green, J.~H.~Schwarz and E.~Witten,
	``Superstring Theory. Vol. 2: Loop Amplitudes, Anomalies And Phenomenology,''
	Cambridge  University Press (1987).
}

\lref\AnomalyGreen{
	M.B.~Green and J.H.~Schwarz,
  	``The Hexagon Gauge Anomaly in Type I Superstring Theory,''
  	Nucl.\ Phys.\ B {\bf 255} (1985) 93.
\semi
	M.B.~Green and J.H.~Schwarz,
  	``Anomaly Cancellation in Supersymmetric D=10 Gauge Theory and Superstring Theory,''
  	Phys.\ Lett.\ B {\bf 149} (1984) 117.
}

\lref\PolchinskiRQ{
  J.~Polchinski,
  ``String theory. Vol. 1: An introduction to the bosonic string,'' Cambridge University Press (2007).
}

\lref\FMS{
	D.~Friedan, E.J.~Martinec and S.H.~Shenker,
  	``Conformal Invariance, Supersymmetry and String Theory,''
  	Nucl.\ Phys.\ B {\bf 271} (1986) 93.
}

\lref\expPSS{
	N.~Berkovits,
  	``Explaining Pure Spinor Superspace,''
  	[hep-th/0612021].
}

\lref\NMPS{
  	N.~Berkovits,
  	``Pure spinor formalism as an N=2 topological string,''
	JHEP {\bf 0510}, 089 (2005).
	[hep-th/0509120].
}

\lref\Faytheta{
	J.D. Fay, ``Theta functions on Riemann surfaces'', Springer Notes in Mathematics 352 (Springer, 1973)
}

\lref\mathMZV{
T.~Terasoma, ``Selberg integrals and multiple zeta values,'' Compositio Mathematica {\bf 133}, 1 (2002)
\semi
F.~Brown, ``Multiple zeta values and periods of moduli spaces ${\cal M}_{0,n}$,''
Ann. Sci. Ec. Norm. Super. {\bf 42} (4), 371 (2009),
[math/0606419]
\semi
  J.~Broedel, O.~Schlotterer, S.~Stieberger and T.~Terasoma,
  ``All order $\alpha^{\prime}$-expansion of superstring trees from the Drinfeld associator,''
Phys.\ Rev.\ D {\bf 89}, no. 6, 066014 (2014).
[arXiv:1304.7304 [hep-th]].
}

\lref\AntoniadisVW{
I.~Antoniadis, C.~Bachas, C.~Fabre, H.~Partouche and T.~R.~Taylor,
  ``Aspects of type I - type II - heterotic triality in four-dimensions,''
Nucl.\ Phys.\ B {\bf 489}, 160 (1997).
[hep-th/9608012].
}

\lref\Zfunctions{
{\tt http://repo.or.cz/Zfunctions.git}
}

\input labelI.defs
\input labelIII.defs

\listtoc
\writetoc
\filbreak

\newsec Introduction

This is the second part of a series of papers \wipI\ (henceforth referred to as part I, II and III) 
in the quest of deriving the one-loop correlators of massless open- and closed-superstring
states using the pure-spinor formalism \refs{\psf,\MPS}. As detailed in the introduction 
of part I, the goal of these papers is to determine the correlators from first principles
including gauge invariance, supersymmetry, locality and single-valuedness. The present
work is dedicated to the implication of single-valuedness on how the correlators may depend
meromorphically on the punctures on a genus-one worldsheet. The key results are the following
\medskip
\item{i)} We present a bootstrap program to construct worldsheet functions for the correlators that 
share the differential structure and relations of their superspace kinematics. These parallels will 
be referred to as a duality between kinematics and worldsheet functions, and they endow 
one-loop amplitudes of the open superstring with a double-copy structure \MafraIOJ.
\item{ii)} We establish the notion of generalized elliptic integrands (GEIs) which mirror the combinatorics
of BRST invariant kinematic factors in the spirit of the duality between kinematics 
and worldsheet functions.
\medskip
\noindent 
These results will come to fruition in the assembly of one-loop correlators in
part III, also see appendix \convencorr\ for their representation that manifests
their double-copy structure. Since we will often refer to section and equation 
numbers from the papers I and III, these numbers will be prefixed by the 
roman numerals I and III accordingly.

\newnewsec\ZEsec Worldsheet functions at one loop

This section introduces the elementary worldsheet functions used in part III as
building blocks of multiparticle genus-one amplitudes.  These functions are
meromorphic and defined as the coefficients of a recent expansion \BrownLevin\
of the classical Kronecker--Eisenstein series \refs{\Kronecker, \Zagier}.  They
are quasi-periodic under $z \rightarrow z+\tau$ and therefore live on the
universal cover of an elliptic curve.

However, our goal is to study string scattering amplitudes that require
functions on an elliptic curve. For this purpose, we will later on consider
meromorphic functions defined on an enlarged space parameterized by the standard
vertex-insertion coordinates $z_i$ and the {\it loop momentum} $\ell^m$ (with
vector indices $m,n,p,\ldots=0,1,\ldots,9$ of the ten-dimensional Lorentz
group). Following the chiral-splitting formalism
\refs{\verlindes,\DHokerPDL,\xerox}, $\ell^m$ represents certain zero modes
associated with the worldsheet field $x^m(z,\bar z)$, cf.\ \Pizero.  The
interplay between $z_j$ and $\ell^m$ will then lead to the definition of {\it
generalized elliptic integrands} (GEIs) \MafraIOJ, which become doubly-periodic
under $z \rightarrow z+1$ and $z \rightarrow z+\tau$ upon integration of loop
momenta. The properties and explicit construction of GEIs will be the subject of
the subsequent discussions.

As reviewed in more detail in section \chiralsplitsec, chiral splitting allows to derive
open- and closed-string amplitudes from the same function ${\cal K}_n(\ell)$ of the
kinematic data. Open string $n$-point amplitudes at one loop descend from worldsheets 
of cylinder- and Moebius-strip topologies with punctures $z_j$ on the boundary,
\eqn\theamphere{
{\cal A}_n  =
\sum_{\rm top} C_{\rm top} \int_{D_{\rm top}}\!\!\!\! 
d\tau \, dz_2 \, d z_3 \, \ldots \, d z_{n} \, \int d^{D} \ell \ |{\cal I}_n(\ell)| \,  \langle {\cal K}_n(\ell)  \rangle \, ,
}
see \GreenMN\ for the integration domains $D_{\rm top}$ and the associated
color factors $C_{\rm top}$. Closed-string one-loop amplitudes in turn are given by
\eqn\theclosedamphere{
{\cal M}_n  =
 \int_{{\cal F}} 
d^2\tau \, d^2z_2 \, d^2 z_3 \, \ldots \, d^2 z_{n} \,
\int d^{D} \ell \ |{\cal I}_n(\ell)|^2 \,
\langle {\cal K}_n(\ell)\rangle \, \langle\tilde{\cal K}_n(-\ell)\rangle\,,
}
where ${\cal F}$ denotes the fundamental domain for the modular parameters
$\tau$ of the torus worldsheet. As a universal part of the underlying correlation
functions, both \theamphere\ and \theclosedamphere\ involve the Koba--Nielsen factor
(with $s_{ij}\equiv k_i\cdot k_j$ and conventions where $2\ap = 1$ for open and
$\ap = 2$ for closed strings)
\eqn\KNfactor{
{\cal I}_n(\ell) \equiv  \exp\Big( \sum^n_{i<j} s_{ij} \log \theta_1(z_{ij},\tau)
+  \sum_{j=1}^n  z_j(\ell\cdot k_j) + {\tau \over 4\pi i} \ell^2 \Big)\,.
}
The leftover factors of ${\cal K}_n(\ell)$ in the loop integrands carry the dependence
on the superspace polarizations and are referred to as {\it correlators}, see
part III
for their construction. The brackets $\langle \ldots \rangle$ in the above integrands denote the
zero-mode integration of the spinor variables $\lambda^\alpha$ and $\theta^\alpha$
of the pure-spinor formalism \psf, and the odd Jacobi theta function in \KNfactor\
is defined by ($q\equiv e^{2\pi i \tau}$)
\eqn\basicsD{
\theta_{1}(z,\tau) \equiv  2q^{1/8}\sin(\pi z)  \! \prod_{n=1}^{\infty}
(1-q^n)
\bigl(1 - q^ne^{2\pi i z}\bigr)\bigl(1 - q^{n}e^{-2\pi i z}\bigr)\,.
}
Note that the open-string worldsheets relevant to \theamphere\ can be obtained
from a torus via suitable involutions \refs{\AntoniadisVW, \PolchinskiRQ}, that is why
the subsequent periodicity requirements will be tailored to the torus topology.

\newsubsec\EKsec The Kronecker--Eisenstein series

Our starting point to describe the dependence of the correlators ${\cal K}_n(\ell)$
on the worldsheet punctures is the Kronecker--Eisenstein series $F(z,\alpha,\tau)$
\refs{\Kronecker,\Zagier}. Its Laurent series in the second variable
defines meromorphic functions $g^{(n)}(z,\tau)$ \BrownLevin,
\eqn\EisKron{
F(z,\a,\tau) \equiv { \t_1'(0,\tau) \t_1(z+\a,\tau) \over
\t_1(\a,\tau)  \t_1(z,\tau)} \equiv \sum_{n=0}^{\infty} \a^{n-1} g^{(n)}(z,\tau) \ .
}
The simplest instances of these functions are $g^{(0)}(z,\tau) = 1 $ and ($\partial \equiv {\partial \over \partial z}$)
\eqn\lopEE{
g^{(1)}(z,\tau)= \partial \log \theta_1(z,\tau) \ , \ \ \ \  
g^{(2)}(z,\tau)= {1\over 2} \Big[
(\partial \log \theta_1(z,\tau))^2 - \wp(z,\tau) \Big]\,,
}
where
$\wp(z,\tau) = -\p^2\log\t_1(z,\tau) - {\rm G}_2(\tau)$
is the Weierstrass function and ${\rm G}_{2k}(\tau)$ denotes the holomorphic Eisenstein series\foot{Note that
the lattice-sum representation \holeis\ of ${\rm G}_2$ is not absolutely convergent and requires the
specification of a summation prescription
$
{\rm G}_2(\tau) = \sum_{n\in \Bbb Z \setminus \{0\}}   {1\over n^{2}}+\sum_{m\in\Bbb Z \setminus \{0\}}\sum_{n\in\Bbb Z} {1 \over(m\tau+n)^{2}}$.}
\eqn\holeis{
{\rm G}_{2k}(\tau) =
\sum_{(m,n)\in\Bbb Z\times\Bbb Z \setminus \{(0,0)\}}{1\over(m\tau+n)^{2k}}  = - g^{(2k)}(0,\tau) \,.
}
See the appendix~\appgs\ for the explicit expansions of $g^{(n)}(z,\tau)$ for $n\le5$
in terms of Jacobi theta functions.

It is important to note that the function $g^{(1)}(z,\tau)$ has a simple pole
$\sim {1\over z}$ at the origin while all $g^{(n)}(z,\tau)$ for $n\ge 2$ are
non-singular\foot{Note, however, that $g^{(k)}(z,\tau)$ for $k\ge 2$ have a
simple pole at $z=\tau$ and in fact at all lattice points $z=m\tau + n$ with
$m,n\in\Bbb Z$ and $m\neq 0$.} as $z\to0$. Furthermore, the heat equation
$4\pi i \p_\tau \theta_1(z,\tau)= \p^2 \theta_1(z,\tau)$ implies that
\eqn\dtau{
{\p\over\p\tau}\log\theta_1(z,\tau) = {1 \over 2\pi i} \Big\{ g^{(2)}(z,\tau)
- \half {\rm G}_2(\tau) \Big\}\,.
}
Similarly, one can obtain the $\tau$-derivatives of the
above $g^{(n)}$ from the mixed heat equation
\eqn\mixedheat{
{\p \over \p \tau} F(z,\alpha,\tau) = {1\over 2\pi i }
{\p^2 F(z,\alpha,\tau) \over \p z \, \p \alpha}\,, \qquad
{\p \over \p \tau}  g^{(n)}(z,\tau) = {n \over 2\pi i} \, \p g^{(n+1)}(z,\tau)\,,
}
and these relations will be instrumental when analyzing boundary terms
with respect to $\tau$ in one-loop correlators later on.

\newsubsubsec\doublysec Monodromies of the $g^{(n)}$-functions

\ifig\figuA{Parameterization
of the torus through the lattice $\Bbb C/(\Bbb Z {+} \tau \Bbb Z)$
with an identification
of points $z$ with their translates $z{+}1$ and $z{+}\tau$ along the $A$- and $B$-cycle.}
{\epsfxsize=0.70\hsize\epsfbox{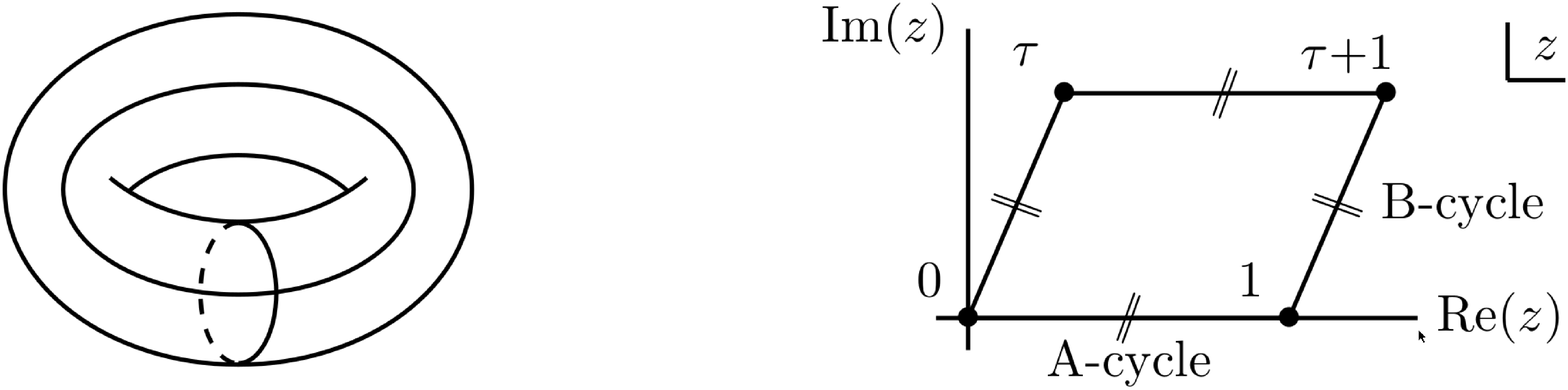}}

In the parameterization of the torus depicted in \figuA, translations around the
$A$- and $B$-cycle amount to shifts by $1$ and $\tau$, respectively.
The quasi-periodicity of the Jacobi theta function \basicsD,
\eqn\basicsF{
\theta(z{+}1,\tau)=- \theta(z,\tau) \ , \ \ \ \ \ \
\theta(z{+}\tau,\tau)=- e^{-i\pi \tau -2\pi i z} \theta(z,\tau)\,,
}
results in the following monodromies of the Kronecker--Eisenstein series \EisKron\ \refs{\Kronecker,\Zagier}
\eqnn\shift
$$\eqalignno{
F(z+1,\alpha,\tau)  &= F(z,\alpha,\tau)\,,&\shift\cr
F(z+\tau,\alpha,\tau)  &= e^{-2\pi i \alpha} F(z,\alpha,\tau)\,.
}$$
It then follows from the expansion \EisKron\ that
the functions $g^{(n)}(z,\tau)$ are single-valued around the $A$-cycle
but have non-trivial $B$-cycle monodromy,
\eqnn\gnmonodromy
$$\eqalignno{
g^{(n)}(z+1,\tau) &= g^{(n)}(z,\tau)\,,&\gnmonodromy\cr
g^{(n)}(z+\tau,\tau) &= \sum_{k=0}^n {(-2\pi i )^k \over k!} g^{(n-k)}(z,\tau)\,.
}$$
For instance,
\eqn\shiftonetwo{
g^{(1)}(z+\tau,\tau) = - 2\pi i \ , \ \ \ \ \ \ 
g^{(2)}(z+\tau,\tau) = - 2\pi i g^{(1)}(z,\tau)  + {1\over 2}(2\pi i)^2 \ .
}
From now on, in order to compactly represent the dependence on the external punctures
$z_1,z_2,\ldots,z_n$ in string correlators, we will use the shorthand
\eqn\gij{
g^{(n)}_{ij} \equiv g^{(n)}(z_i-z_j,\tau)\,.
}

\subsubsec{Weight counting}
\par\subsubseclab\seconethreefour

The integrand of $n$-point one-loop open-string amplitudes \theamphere\ can be
written in terms of loop momenta, holomorphic Eisenstein series \holeis\
excluding ${\rm G}_2$ and the above $g^{(m)}_{ij}$ (possibly including their
$z$-derivatives) \refs{\TsuchiyaVA, \BroedelVLA}. As a necessary condition for
modular invariance of the closed-string amplitude \theclosedamphere, the overall
powers of $\ell, g^{(m)}_{ij}$ and ${\rm G}_k$ have to obey the following
selection rule: Once we assign the following weights to these constituents,
\medskip\smallskip
\centerline{
\vbox{\offinterlineskip
\halign{
\strut#\hfil\vrule width0.6pt&
\hfil\quad#\quad\hfil\vrule width0.6pt&
\hfil\quad#\quad\hfil\vrule width0.6pt&
\hfil\quad#\quad\hfil\vrule width0.6pt&
\hfil\quad#\quad\hfil\vrule width0.6pt&
\quad#\hfil\cr
$\ $term & $2\pi i \! \!$ & $\ell \! \!$ & $\p_{z_j} \! \! \!$ & ${\rm G}_k \! \!$ & $g^{(m)}_{ij}$ \cr
\noalign{\hrule height 0.6pt}
weight $\ $& $1\! \! $ & $1\! \! $ & $1\! \! \!$ & $k\! \!$ & $\,m$\cr
}}}
\smallskip\smallskip
\noindent each term in the $n$-point open-string correlator ${\cal K}_n(\ell)$ must have
weight $n{-}4$.  The notion of weight in the table is conserved in each term of
the monodromies \gnmonodromy, and the same will hold in the subsequent Fay
relations and total derivatives.

\subsec Fay identities

In the subsequent discussions of one-loop open-string correlators,
the Fay identity \mumfordref
\eqn\FayKron{
F(z_1,\alpha_1,\tau)F(z_2,\alpha_2,\tau) =
F(z_1,\alpha_1+\alpha_2,\tau) F(z_2-z_1,\alpha_2,\tau)
+(1\leftrightarrow 2)
}
plays a crucial role when expanded in terms of its coefficient
functions from \EisKron\ \BroedelVLA,
\eqnn\Faygn
$$\eqalignno{
g^{(n)}_{12} g^{(m)}_{23} &= - g_{13}^{(m+n)}
+ \sum_{j=0}^{n}(-1)^j {m-1+j \choose j}
g_{13}^{(n-j)} g_{23}^{(m+j)} \cr
& \ \ \ \ \   \ \ \ \ \   \ \ \ \ \  \,
+ \sum_{j=0}^{m}(-1)^j {n-1+j \choose j} g_{13}^{(m-j)} g_{12}^{(n+j)}\,. &\Faygn
}$$
Its simplest instance can be viewed as the one-loop
counterpart of the tree-level partial fraction identity
$(z_{12} z_{23})^{-1} + {\rm cyc}(1,2,3)=0$,
\eqn\lopEa{
g^{(1)}_{12}g^{(1)}_{23} + g^{(2)}_{12} + {\rm cyc}(1,2,3) = 0\,.
}
Additional instances relevant to the worldsheet functions
that appear in one-loop correlators for up to and including nine points
are given by
\eqnn\weightthreeFay
$$\eqalignno{
\gg1(1,2)\gg2(2,3) &=\gg1(1,3)\gg2(2,3) + \gg1(1,2)\gg2(1,3)-\gg1(1,3)\gg2(1,2)
+ \gg3(1,2) -\gg3(1,3) - 2\gg3(2,3)\,,&\weightthreeFay\cr
g^{(2)}_{12} g^{(2)}_{23} &= g^{(2)}_{12} g^{(2)}_{13}+g^{(2)}_{13} g^{(2)}_{23}
-2g^{(1)}_{13} g^{(3)}_{12} - 2 g^{(1)}_{13} g^{(3)}_{23}
+ 3 g^{(4)}_{12}  - g^{(4)}_{13} + 3 g^{(4)}_{23}\,,\cr
g^{(1)}_{12} g^{(3)}_{23} &=
- g^{(2)}_{12} g^{(2)}_{13} + g^{(1)}_{13} g^{(3)}_{12}
+g^{(1)}_{12} g^{(3)}_{13} + g^{(1)}_{13} g^{(3)}_{23}
- g^{(4)}_{12} - g^{(4)}_{13} - 3 g^{(4)}_{23}\,,\cr
\gg2(1,2)\gg3(2,3) &=-\gg5(1,3) + 6\gg5(2,3) - 4\gg5(1,2) + \gg2(1,3)\gg3(2,3)
- 3\gg1(1,3)\gg4(2,3)+\gg3(1,3)\gg2(1,2) -
2\gg2(1,3)\gg3(1,2)+3\gg1(1,3)\gg4(1,2)\,,\cr
\gg1(1,2)\gg4(2,3) &=-\gg5(1,3) -4\gg5(2,3) +\gg5(1,2) + \gg1(1,3)\gg4(2,3)
+ \gg4(1,3)\gg1(1,2) - \gg3(1,3)\gg2(1,2)+
\gg2(1,3)\gg3(1,2)-\gg1(1,3)\gg4(1,2)\,.
}$$
Note that the
label $2$ (corresponding to $z_2$) appears twice in the monomials of the left-hand side
in the above identities while appearing at most once in the monomials of the right-hand side.
This property can be exploited to rewrite arbitrary products of $g^{(n)}_{ij}$-functions 
in a canonical way. Since any repeated label can be eliminated this
way, for convenience in a product $g^{(n)}_{ij}g^{(m)}_{jk}$ one can use the Fay 
identities if the repeated label $j$ is the smallest among $i,j$ and $k$
(which can be obtained from a relabeling of \weightthreeFay). In addition,
Fay identities involving $z$-derivatives of $g^{(n)}(z,\tau)$ are
easy to obtain from \lopEa\ and
\weightthreeFay, and can be similarly written in a canonical way.

Linear combinations of the above Fay identities can be used to derive
identities involving Eisenstein series ${\rm G}_n$.
For instance, from $g^{(3)}_{ii}=0$ and $g^{(4)}_{ii}=-{\rm G}_4$,
the limit $z_3\to z_1$ of the expressions \weightthreeFay\ for
$\gg2(1,2)\gg2(2,3) + 2\gg1(1,2)\gg3(2,3)$ and $\gg2(1,2)\gg3(2,3) +
3\gg1(1,2)\gg4(2,3)$ implies
\eqn\Gfourspoil{
2\gg4(1,2) + \gg2(1,2)\gg2(1,2) - 2\gg1(1,2)\gg3(1,2) - 3 {\rm G}_4 =0\,,\quad
5\gg5(1,2) + \gg2(1,2)\gg3(1,2) - 3\gg1(1,2)\gg4(1,2) - 3{\rm G}_4\gg1(1,2) =0\,,
}
and similar relations can be obtained at higher weights.
The weight-four identity in \Gfourspoil\ will often be used in 
proposing an expression for the
eight-point correlator, see section \EightPointsec.

\newsubsec\totalderivsec Total derivatives

Correlators ${\cal K}_n(\ell)$ are always accompanied by the Koba--Nielsen factor
${\cal I}_n(\ell)$ given by \KNfactor, when they enter open- and closed-string amplitudes, see \theamphere\
and \theclosedamphere. One can show that its derivatives with respect to worldsheet positions $z_i$ and
modulus $\tau$ are given by
\eqnn\zderiv
\eqnn\tauderiv
$$\eqalignno{
{\p\over\p z_i}\cI_n(\ell) &=\big(
\ell \cdot k_i + \sum_{j\neq i}^n s_{ij} g^{(1)}_{ij}\big)\cI_n(\ell)\,, &\zderiv\cr
{\p\over \p\tau}\cI_n(\ell) &= { 1 \over 2\pi i}
\Big( {1\over 2} \ell^2 + \sum^n_{i<j} s_{ij} g^{(2)}_{ij} \Big)\cI_n(\ell)\,, &\tauderiv
}$$
where \dtau\ and $\sum_{i<j}^n s_{ij}=0$ have been used in \tauderiv.
Given the integrations over $z_j$ and $\tau$ in the amplitudes \theamphere\ and \theclosedamphere, 
one can therefore set the following total derivatives to zero within one-loop correlators,
\eqnn\zderivzero
\eqnn\tauderivzero
$$\eqalignno{
\Big( \ell \cdot k_i + \sum_{j\neq i}^n s_{ij} g^{(1)}_{ij} \Big) f(z,\tau,\ldots)
+   {\p f(z,\tau,\ldots)\over\p z_i} &\cong 0\,,
\qquad\forall \,f(z,\tau, \ldots)\,,&\zderivzero\cr
\Big({1\over 2} \ell^2 + \sum_{1\leq i<j}^n s_{ij} g^{(2)}_{ij} \Big) f(z,\tau,\ldots)
+  2\pi i {\p f(z,\tau,\ldots)\over\p\tau} &\cong 0,\qquad \forall \,f(z,\tau, \ldots)\,,&\tauderivzero
}$$
where $f(z,\tau,\ldots)$ is an arbitrary function on the worldsheet.

The absence of boundary terms w.r.t.\ $z_j$ follows from the short-distance behavior\foot{The
cancellation of $ |z_{ij}|^{s_{ij}}$ as $z_i\rightarrow z_j$ is
obvious in the kinematic region where ${\rm Re} (s_{ij}) >0$ and otherwise
follows from analytic continuation.} $|{\cal I}_n(\ell) |
\rightarrow |z_{ij}|^{s_{ij}}$ of the Koba--Nielsen factor \KNfactor\ as $z_i\rightarrow z_j$.
It is well known from discussions of the anomaly cancellation in the open superstring that
the boundaries of moduli space can give non-vanishing contributions 
from individual worldsheet topologies \refs{\AnomalyGreen,\FMS}. Hence,
blindly discarding total derivatives w.r.t.\ the modulus $\tau$ would generically
lead to inconsistencies. However, when summing over the different
worldsheet topologies these inconsistencies are canceled for the gauge group
$SO(32)$; since this will always be the case for the open superstring
we may freely discard total derivatives in $\tau$.

\newnewsec\GEIsec Generalized elliptic integrands

When using the chiral-splitting method \refs{\verlindes,\DHokerPDL,\xerox} to handle the joint zero 
mode $\ell^m$ of\foot{In the pure-spinor formalism, the worldsheet fields $\partial x^m(z)$ and 
$\bar \partial x^m(\bar z)$ enter the vertex operators in their spacetime-supersymmetric combinations $\Pi^m(z)$
and $\bar \Pi^m(z)$ \psf.} $\partial x^m(z)$ and $\bar \partial x^m(\bar z)$,
superstring scattering integrands of \theamphere\ and \theclosedamphere\ involve a loop-momentum
dependent Koba--Nielsen factor \KNfactor.
As explained in \xerox, the integrands of superstring amplitudes containing the
loop momentum $\ell^m$ do not need to be single-valued as functions of $z_i$. Instead,
it is sufficient to attain single-valuedness after the loop momentum is integrated
out. Here ``single-valued'' is used in its conventional sense; it refers to
functions $f(z_i)$ left invariant as the coordinates $z_i$ are transported
around the $A$ and $B$ homology cycles of \figuA. In this work, the chiral-splitting
method will be used but the concept of single-valuedness will be extended to
invariant functions of $(z_i, \ell^m)$ under a simultaneous variation of both
$z_i$ and $\ell^m$ along the cycles. Let us now present the reasoning that motivated this idea.

\newsubsec\antiKfive Motivating and defining generalized elliptic integrands

As we will see in section~\secninetwo, the evaluation of the five-point one-loop
amplitude of the open superstring using the standard rules of the pure-spinor
formalism (and some mild assumptions) gives rise to the following integrand:
\eqn\explain{
\cK_5(\ell)=\ell_m V_1 T^m_{2,3,4,5} + \big[V_{12}T_{3,4,5}\gg1(1,2) +
(2\leftrightarrow3,4,5)\big] + \big[V_1T_{23,4,5}\gg1(2,3) +
(2,3|2,3,4,5)\big]\,.
}
The kinematic factors $V_1,V_{12}, T^m_{2,3,4,5} ,T_{3,4,5}$ in pure-spinor
superspace \expPSS\
are reviewed in section \LocalBBsec.
Throughout this work, the notation $+(a_1,\ldots,a_p|a_1,\ldots,a_{p+q})$ instructs to sum over all
ordered combinations of $p$ the labels $a_i$ taken from the set $\{a_1,a_2,\ldots,a_{p+q}\}$, 
leading for instance to a total of six permutations of $V_1T_{23,4,5}\gg1(2,3)$ in \explain.

Having obtained \explain, it was natural to ask about its $B$-cycle monodromies using
the relations \shiftonetwo.
Ignoring the term with
the loop momentum for a moment, it is easy to see that
the correlator \explain\ changes by
$-2\pi i\big[V_{12}T_{3,4,5} + (2\leftrightarrow3,4,5)]$ as $z_1$ goes around
the $B$-cycle. Recalling the vanishing of $k_1^m V_1 T^m_{2,3,4,5}
+\bigl[ V_{12} T_{3,4,5} + (2 \leftrightarrow 3,4,5) \bigr]$ in the BRST cohomology,
see \QJex, suggests the following speculation:
if the loop momentum changed as $\ell^m\to\ell^m - 2\pi i k_1^m$ at the
same time as $z_1$ goes around the $B$-cycle, then the integrand \explain\ would
be single valued as a function of both $z_1$ and $\ell^m$.

As it stands the above speculation is not compelling enough as we did not
consider how the Koba--Nielsen factor \KNfactor\ behaves under these changes.
Luckily, the quasi-periodicity $\theta_1(z{+}\tau,\tau) =-
e^{-i\pi \tau - 2\pi i z} \theta_1(z,\tau)$ of the odd Jacobi theta
function \basicsD\ implies that the absolute value of the Koba--Nielsen factor
is {\it invariant} under the simultaneous transformation of $z_1\rightarrow z_1 {+} \tau$ 
and $\ell^m \to \ell^m - 2\pi i k_1^m$,
\eqn\KNKNH{
 \big|{\cal I}_n(\ell - 2\pi i k_1) \big|_{z_1 \rightarrow z_1 + \tau} =
 \big|{\cal I}_n(\ell )\big|\,.
}
Hence, the loop-integrated open- and closed-string expressions $\int d^D \ell \, |{\cal I}_n(\ell)
| \langle \cK_5(\ell) \rangle$ and  $\int d^D \ell \, |{\cal I}_n(\ell)|^2
 \langle \cK_5(\ell) \rangle \langle \tilde \cK_5(-\ell) \rangle$ 
will still lead to single-valued functions of the punctures in the
conventional sense of \xerox. But the above reasoning suggests that one can even talk about
single-valued chirally-split superstring integrands by also letting the loop momentum change along
the $B$-cycle. Furthermore, the same analysis can be performed for shifts
along the $A$-cycle (without any modification of the loop momentum as $z_1 \rightarrow z_1+1$), 
motivating the following definition:
\medskip
\proclaim Definition 1 (GEI).
A generalized elliptic integrand (GEI) is a single-valued
function $f(z_i,\ell,\tau,k_j)$ of the
lattice coordinates $z_j$, $j=1, \ldots,n$, the loop momentum $\ell^m$,
the modular parameter $\tau$ and the external momenta $k_j^m$ such that
\eqn\MonFinv{
f(z'_j,\ell',\tau,k_j) = f(z_j,\ell,\tau,k_j)
}
as $z_j$ and $\ell^m$ go around the $A$ and $B$ cycles
\eqnn\shifts
$$\eqalignno{
\hbox{$A$-cycle}:\quad(z'_j,\ell') &= (z_j + 1,\ell)\,, &\shifts\cr
\hbox{$B$-cycle}:\quad(z'_j,\ell') &= (z_j + \tau,\ell - 2\pi i k_j)\,.
}$$
By their dependence on $\ell^m$ and $k_j^m$, GEIs may have free vector indices
$f^{m_1 m_2\ldots}(z_j,\ell,\tau,k_j)$.

\par
As the absolute value of the Koba--Nielsen factor is by itself a GEI, the five-point
example \explain\ suggests that superstring correlators are given
by GEIs in the above sense,
\eqn\KNKNcorr{
\cK_n(\ell - 2\pi i k_j) \big|_{z_j \rightarrow z_j + \tau} = \cK_n(\ell )\,.
}
We will see that this observation harbors valuable constructive
input to the derivation of correlators from first principles.
Furthermore, the argument above suggests a deeper connection between
BRST invariance of pure-spinor superspace expressions
and GEIs. As we will see in the following sections, this synergy
is quite powerful and leads to many interesting results.

Integrands depending on $\ell,k,z$ and $\tau$ satisfying the key property \MonFinv\
were used for the first time in~\MafraIOJ, where the acronym GEI was coined.
As detailed in section \secthree, integrating the GEIs in $n$-point closed-string integrands
over $\int d^D \ell \, |{\cal I}_n(\ell) |^2$ yields {\it modular forms}
of weight $(n{-}4,n{-}4)$ and leads to modular invariant closed-string
amplitudes \theclosedamphere.

\newsubsec\monodromysec The linearized-monodromy operator

Given a monomial in $g_{ij}^{(n)}$, the monodromies as $z_j \rightarrow
z_j+\tau$ are polynomials in $2\pi i$ by \gnmonodromy. We will be interested
in combinations of $g^{(n)}_{ij}$ and the loop momentum such that the
monodromies are compensated by shifts $\ell \rightarrow \ell - 2\pi i k_j$ and
the defining property \KNKNcorr\ of GEIs is attained. In order to efficiently
identify GEIs, we formally truncate the combined transformations of
$g_{ij}^{(n)}$ and $\ell$ to the linear order in $2\pi i$ and study the
operator
\eqn\covB{
\delta_j \ell = - 2 \pi i k_j\,,\quad
\delta_j g^{(n)}_{jm} = - 2 \pi i g^{(n-1)}_{jm}\,, \quad
 n\ge 1 \, ,
}
where $\delta_{j}g^{(0)}_{jm}=0$ and $\delta_{j}g^{(n)}_{im}=0$
for all $i,m\neq j$. This operator probes the {\it linearized monodromy}
w.r.t. a given puncture $\delta_j : \ z_j \rightarrow z_j+\tau$ with the
accompanying shift $\ell \rightarrow \ell - 2\pi i k_j$. Accordingly, it is
understood to obey a Leibniz property
\eqn\Leibdelta{
\delta_j \big(f_1(\ell,z_j)  f_2(\ell,z_j) \big) = f_1(\ell,z_j) \big( \delta_j f_2(\ell,z_j) \big)
+ f_2(\ell,z_j) \big( \delta_j f_1(\ell,z_j) \big) 
}
for arbitrary functions $f_i$ of the loop momentum and the punctures.
It is convenient to assemble the linearized monodromies w.r.t.\ all of
$z_1,z_2,\ldots,z_n$ into a single operator as
\eqn\covC{
D = -{1\over 2\pi i} \sum_{j=1}^n \Omega_j \delta_j \, ,
}
where we have introduced formal variables $\Omega_j$ to track the contribution of
the $j^{\rm th}$ puncture. Then, \covB\ and the shorthand notation 
$\Omega_{ij} \equiv\Omega_i - \Omega_j$ give rise to
\eqn\covD{
D g^{(n)}_{ij} = \Omega_{ij} g^{(n-1)}_{ij}\,, \qquad
D \ell^m =  \sum_{j=1}^n \Omega_j k_j^m = \sum_{j=2}^n \Omega_{j1} k_j^m \, ,
}
where momentum conservation $k^m_1 = -k^m_2 - \cdots - k^m_n$ has been
used in the last relation. For example,
\eqn\Dexs{
D\gg1(1,2) = \Omega_{12}\,,\qquad D\gg2(1,2) = \Omega_{12}\gg1(1,2)
\,,
\qquad
D\big(\gg1(1,2)\ell^m \big)
=\Omega_{12}\ell^m +  \gg1(1,2) \sum_{j=2}^n \Omega_{j1} k_j^m\,.
}
Note that $D$ will be later on argued to play a role similar to the BRST operator
$Q$ of the pure-spinor formalism. One can enforce that $D$ shares the nilpotency $Q^2=0$ by defining the
formal variables $\Omega_j$ to be fermionic\foot{The conditional nilpotency of
$D$ for fermionic formal variables $\Omega_j$ follows from the fact that
linearized monodromies \covB\ w.r.t.\ different punctures commute, $\d_i\d_j =
\d_j\d_i$. This commutativity property follows from \covB\ and \Leibdelta.}.
However, the choice of statistics for the $\Omega_j$ won't affect any
calculation done in this work, so we defer this decision to follow-up
research.

Since the linearized monodromy operator $D$ only picks the terms linear in
$2\pi i$ that arise from the transformation $z_j \rightarrow z_j+\tau$ and
$\ell \rightarrow \ell -2\pi ik_j$, invariance $DE=0$ is only a {\it
necessary} condition for $E$ to be a GEI.  It remains to check if the higher
orders in $2\pi i$ also drop out from the image of $E$ under the above shift
of $z_j$ and $\ell$. For all solutions to $DE=0$ studied in this work, we have
checked that they constitute a GEI on a case-by-case basis, and it would be
interesting to find a general argument. In many cases, single-valuedness can 
be seen from the generating-function techniques in later sections.

\newnewsec\dualitysec Bootstrapping shuffle-symmetric worldsheet functions

In this section we will construct a system of worldsheet functions ${\cal Z}$ for
superstring correlators on a genus-one Riemann surface by analogies with kinematic 
factors. When the latter are organized in terms of Berends--Giele superfields as 
detailed in part I, their variation under the pure-spinor BRST operator will be used as a
prototype to prescribe monodromy variations for the ${\cal Z}$-functions.
As a consequence, the combinatorics of BRST-invariant kinematic factors can be
borrowed to anticipate $D$-invariant combinations of ${\cal Z}$-functions, i.e.\ GEIs.

The correspondence between the pure-spinor BRST charge $Q$ acting on superfields
and the monodromy operator $D$ acting on functions is the first facet of a
duality between kinematics and worldsheet functions. Further aspects of the
duality will be presented in section \newdualitysec\ that lead to a variety of
applications. In particular, the duality between kinematics and worldsheet
functions implies a double-copy structure of open-superstring one-loop
amplitudes discussed in \MafraIOJ\ and expanded in part III.

\newsubsec\shufflesec Shuffle-symmetric worldsheet functions

In the computation of tree-level correlators for $n$-point open-string
amplitudes \nptString, the nested OPE singularities were captured by
worldsheet functions of the following form\foot{Note that the worldsheet
functions $( z_{12} z_{23}\ldots z_{p-1,p})^{-1}$ at tree level arise from cyclic
Parke--Taylor factors $( z_{12} z_{23}\ldots z_{p-1,p}z_{p,n} z_{n,1})^{-1}$
in an ${\rm SL}_2$-frame where $z_n \rightarrow \infty$.}
\eqn\shufA{
\cZ^{\rm tree}_{123\ldots p} \equiv {1\over z_{12} z_{23}\ldots z_{p-1,p}}\,.
}
It follows from partial-fraction relations such as $(z_{12} z_{23})^{-1} + {\rm cyc}(1,2,3)=0$
that the tree-level functions satisfy {\it shuffle symmetries}\foot{The 
{\it shuffle product} of words $A$ and $B$ of length $n$ and $m$ generates 
all ${(n{+}m)!\over n!m!}$ possible ways to interleave the
letters of $A$ and $B$ without changing their orderings within $A$ and $B$,
see \Shrecurs\ for a recursive definition. A more elaborate account on the 
combinatorics on words can be found in section \wordssec,
based on the mathematics literature \refs{\lothaire,\reutenauer}.} (e.g.\
${\cal Z}^{\rm tree}_{1\shuffle 23}={\cal Z}^{\rm tree}_{123}{+}{\cal Z}^{\rm tree}_{213}
{+}{\cal Z}^{\rm tree}_{231}=0$) \cresson
\eqn\shufB{
{\cal Z}^{\rm tree}_{A\shuffle B}=0, \quad \forall \ A,B\neq \emptyset\,.
}
Since the appearance of shuffle-symmetric worldsheet functions \shufA\ at
tree level can be traced back to the short-distance behavior of vertex
operators, the same structure must persist at higher genus. Therefore we
assume that the short-distance singularities at one loop arise from analogous
chains built from functions $g^{(1)}(z,\tau)= {1\over z}+{\cal O}(z)$
\eqn\assumeZ{
g^{(1)}_{12}g^{(1)}_{23} \ldots g^{(1)}_{p-1,p}\,.
}
As a fundamental starting point in obtaining one-loop $n$-point correlators of
the open superstring, the worldsheet functions associated with nested OPE
singularities will be required to obey shuffle symmetries like their
tree-level counterparts, i.e.,
\eqn\shufOne{
\cZ^{\rm 1-loop}_{ \ldots, A\shuffle B, \ldots } = 0\,,\qquad\forall \ A,B\neq\emptyset\,.
}
At multiplicity $p=2$, antisymmetry of $g^{(1)}_{12} = - g^{(1)}_{21}$ 
suffices to make it shuffle-symmetric. However, for the tentative one-loop counterpart
$g^{(1)}_{12} g^{(1)}_{23}$ of $\cZ^{\rm tree}_{123}$ it is easy to
see that the Fay identity \lopEa\ prevents the shuffle relation $\cZ^{\rm
tree}_{1\shuffle 23} =0$ from generalizing. Luckily, the same Fay identity
also suggests how to restore the shuffle symmetry without altering the
pole at $z_i\to z_j$ by adding non-singular $g^{(2)}_{ij}$-functions. One
can check via \lopEa\ that both~of
\eqn\twocands{
{\cal Z}^{\rm (i)}_{123}  \equiv g^{(1)}_{12} g^{(1)}_{23} + {1\over 2}(g^{(2)}_{12} +g^{(2)}_{23})\,, \qquad
{\cal Z}^{\rm (ii)}_{123}  \equiv g^{(1)}_{12} g^{(1)}_{23} + g^{(2)}_{12} +g^{(2)}_{23} - g^{(2)}_{13}
}
share the desired shuffle symmetry of $\cZ^{\rm tree}_{123} $. Also at higher
multiplicity, the non-singular functions $g^{(n)}_{ij}$ with $n\geq 2$ admit
various shuffle symmetric completions of $g^{(1)}_{12} g^{(1)}_{23}\ldots
g^{(1)}_{p-1,p}$ which reproduce the singularity structure of \shufA\ and
qualify as one-loop counterparts of ${\cal Z}^{\rm tree}_{123\ldots p}$. From
the availability of two shuffle-symmetric multiplicity-three candidates in
\twocands, one can anticipate that many more options arise at higher
multiplicities. In the next subsection we will identify a guiding principle to
prefer ${\cal Z}^{\rm (ii)}_{123} $ over ${\cal Z}^{\rm (i)}_{123}$
in our representations of one-loop correlators and to select
higher-multiplicity generalizations.

\newsubsec\DQdualsec Duality between monodromy and BRST variations

We will now prescribe the monodromy variation $D\cZ_{12\ldots p}$ of shuffle-symmetric
worldsheet functions by analogies with Berends--Giele superfields $M_{12\ldots p}$ that share
the shuffle symmetry and are reviewed in section \BRSTsec. The idea is to impose the combinatorics 
of the BRST variation $QM_{12\ldots p}$ to carry over to the worldsheet functions,
 ${\cal Z}_{12\ldots p} \leftrightarrow M_{12\ldots p}$. This relationship is at
the heart of an emerging proposal for a {\it duality between worldsheet
functions kinematics} -- monodromy variations are taken to be dual to BRST
variations.

\subsubsec Scalar monodromy variations

The Berends--Giele superfields at one loop have multiple slots, starting with
the scalar kinematics $M_A M_{B,C,D}$ of section \BGmapsec. Accordingly, the
simplest one-loop worldsheet functions should inherit the slot structure
${\cal Z}_{A,B,C,D}$ with shuffle symmetries in all of $A,B,C,D$. Throughout
this work, whenever multiparticle labels $A,B,\ldots$ in a subscript are separated 
by a comma rather than a vertical bar, then they are understood to be freely interchangeable,
${\cal Z}_{A,B,\ldots}={\cal Z}_{B,A,\ldots}$. 

The BRST variation \QMs\ of $M_{B,C,D}$ can be written as linear combinations of the BRST invariants
$C_{i|P,Q,R}$ \partI\ reviewed in section \BRSTpseudosec. Accordingly, the corresponding $D$-variations 
of ${\cal Z}_{A,B,C,D}$ should be written in terms of GEIs $E_{i|A,B,C,D}$, i.e.\ $D$-invariant 
combinations of simpler ${\cal Z}$-functions. More explicitly, the parallel is taken to be
\eqnn\protoA
\eqnn\protoC
$$\eqalignno{
Q M_{A,B,C} &= C_{a_1| a_2\ldots a_{|A|} ,B , C }
- C_{a_{|A|}| a_1\ldots a_{|A|-1} , B , C } + (A \leftrightarrow B,C)\,,&\protoA\cr
D \cZ_{A,B,C,D} &= \Omega_{a_1}E_{a_1| a_2\ldots a_{|A|} ,B , C , D}
- \Omega_{a_{|A|}}E_{a_{|A|}| a_1\ldots a_{|A|-1} , B , C ,D} + (A
\leftrightarrow B,C,D)\,,\quad\qquad{}&\protoC
}$$
where the length of the word $A=a_1 a_2\ldots a_{|A|}$ is denoted by $|A|$, and
the bookkeeping variables $\Omega_j$ of \covC\ always follow the special
label of $E_{j|\ldots}$, e.g.
\eqn\protoB{
\eqalign{
Q M_{1,2,3} &= 0\,,\cr
D\cZ_{1,2,3,4}&=0\,,
}\qquad
\eqalign{
Q M_{12,3,4} &= C_{1|2,3,4} - C_{2|1,3,4}\,,\cr
D\cZ_{12,3,4,5}&=\Omega_1E_{1|2,3,4,5} - \Omega_2 E_{2|1,3,4,5}\,.
}}
The $E_{i|\ldots}$ on the right-hand sides will be defined in analogy\foot{We
note the mismatch between the slots of $E_{i|A,B,C}$ defined in analogy with
the BRST invariants and the slots of $E_{i|A,B,C,D}$ appearing in the
right-hand side of \protoC. This difference is inconsequential for functions
up to multiplicity nine and can be bypassed by defining the {\it extension} of
scalar GEIs by $E_{i|A,B,C,D}\equiv E_{i|A,B,C}$ and by adding extra
permutations to the tensorial GEIs.} with $C_{i|\ldots}$, and this analogy will
be reflected by the notation: The duality between superfields and ${\cal
Z}_{A,B,C,D}$ as well as the resulting correspondence between $Q$ and $D$
imply that BRST invariants $C_{i|A,B,C}$ should be dualized to GEIs. By the
vertical-bar notation, the symmetries $C_{i|A,B,\ldots}=C_{i|B,A,\ldots}$
$E_{i|A,B,\ldots}=E_{i|B,A,\ldots}$ do not extend to the external-state label $i$
in the first entry.

At this point, we can identify a preferred choice among the two
multiplicity-three candidates \twocands.
Based on \Dexs, we have
\eqnn\protoD
$$\eqalignno{
D{\cal Z}^{\rm (i)}_{123} &= \Omega_1(g_{23}^{(1)} + {1\over 2} g^{(1)}_{12})
+ {1\over 2}\Omega_2 g^{(1)}_{12} + (1{\leftrightarrow} 3) \, , \ \ \ \
D{\cal Z}^{\rm (ii)}_{123} =\Omega_{13}( g^{(1)}_{12}{+}g^{(1)}_{23}{+}g^{(1)}_{31})\,, \ \ &\protoD
}
$$
where the second variation will later be shown to be equal to $\Omega_{13}E_{1|23,4,5}$. Since it is easy to see that
$D{\cal Z}^{\rm (i)}_{123}$ is not single-valued, only the second option has
the required structure \protoC\ on the right-hand side. In order to reconcile
the expression for ${\cal Z}^{\rm (ii)}_{123}$ with the slot structure of
scalar worldsheet functions ${\cal Z}_{A,B,C,D}$ in \protoC, from now on we use the
notation ${\cal Z}_{123,4,5,6} ={\cal Z}^{\rm (ii)}_{123}$, also see section \sixwssec.

\subsubsec Tensorial monodromy variations

The same ideas can be reused at higher tensor ranks $r$ to infer tensorial
worldsheet functions $\cZ^{m_1 \ldots m_r}_{A,B,C,..}$ involving loop momenta, external
momenta and $g^{(n)}_{ij}$, with shuffle symmetries in multiple slots $A,B,\ldots$. 
These tensorial functions will be constructed by imposing their linearized
monodromies to follow the BRST variation of tensorial kinematic building
blocks $M^{m_1 \ldots m_r}_{A,B,C,\ldots}$ in pure-spinor superspace. Explicitly, the map is
\eqn\MTocZ{
QM^{m_1 \ldots m_r}_{A,B,C,\ldots} \longleftrightarrow
D\cZ^{m_1 \ldots m_r}_{A,B,C,\ldots} \, ,
}
and we will use the following results for the left-hand side \partI,
\eqnn\Massumption
$$\eqalignno{
QM^{m_1 m_2 \ldots m_r}_{A,B,C,\ldots} &=
\delta^{(m_1 m_2} {\cal Y}^{m_3 \ldots m_r)}_{A,B,C,\ldots} &\Massumption\cr
&\quad{}+C^{m_1 m_2 \ldots m_r}_{a_1|a_2\ldots a_{|A|},B,C,\ldots}
 - C^{m_1 m_2 \ldots m_r}_{a_{|A|}|a_1\ldots a_{|A|-1},B,C,\ldots} +
 (A\leftrightarrow B,C,\ldots)\cr
 &\quad{}+\delta_{|A|,1}  rk^{(m_1}_{a_1}
C^{m_2\ldots m_r)}_{a_1|B,C,\ldots}
+ (A\leftrightarrow B,C,\ldots)\,,
}$$
with tensorial anomaly superfields ${\cal Y}^{m_1m_2\ldots}_{A,B,\ldots}$ and (pseudo-)invariants\foot{The defining
property of pseudo-invariants is that their BRST variation is entirely expressible in terms of
anomaly superfields \partI.} 
$C^{m_1 m_2\ldots}_{1|A,B,\ldots}$ \partI, e.g.
\eqnn\forinst
$$\eqalignno{
Q M^m_{1,2,3,4} &= k_1^m C_{1|2,3,4} + (1\leftrightarrow 2,3,4) 
&\forinst \cr
Q M^m_{12,3,4,5} &= C^m_{1|2,3,4,5} - C^m_{2|1,3,4,5} + \big[ k_3^m C_{3|12,4,5}
+ (3\leftrightarrow 4,5)\big]\,.
}$$
Here and in the following, Lorentz indices are (anti)symmetrized such that each
inequivalent term has unit coefficient, e.g.\ $k_{1}^{(m_1} k_{2}^{m_2} \ldots k_{r}^{m_r)} \equiv
k_{1}^{m_1} k_{2}^{m_2} \ldots k_{r}^{m_r}  + {\rm perm}(m_1,\ldots,m_r)$, for a total of $r!$ terms.
In case of symmetric tensors, imposing unit coefficients leads to fewer terms such 
as $\d^{(mn}k^{p)}\equiv \d^{mn}k^p + \d^{mp}k^n + \d^{np}k^m$, and expanding the 
symmetrization of $\delta^{(m_1 m_2} {\cal Y}^{m_3 \ldots m_r)}_{A,B,C,\ldots} $ in \Massumption\ yields ${r \choose 2}$ terms.

The vectorial BRST invariants $C^m_{i|A,\ldots}$ on the right-hand sides of \Massumption\ and \forinst\ are 
composed of $M_{A,B,C}$, $M^m_{A,B,C,D}$ and external momenta. Similarly, we will later on obtain 
vectorial GEIs $E^m_{i|A,B,\ldots}$
involving $\ell^m$, $k_j^m$ and $g^{(n)}_{ij}$ by following the same composition rules. According 
to the duality \MTocZ, the defining property of $\cZ^{m_1 \ldots m_r}_{A,B,C,..} $ is
\eqnn\assumption
$$\eqalignno{
D\cZ^{m_1 m_2 \ldots m_r}_{A,B,C,\ldots} &=\Big[
\Omega_{a_1} E^{m_1 m_2 \ldots m_r}_{a_1|a_2\ldots a_{|A|},B,C,\ldots}
 - \Omega_{a_{|A|}}
 E^{m_1 m_2 \ldots m_r}_{a_{|A|}|a_1\ldots a_{|A|-1},B,C,\ldots} \cr
&\quad{} +\delta_{|A|,1} \Omega_{a_1} k^{(m_1}_{a_1}
E^{m_2\ldots m_r)}_{a_1|B,C,\ldots}
+ (A\leftrightarrow B,C,\ldots)\Big]\,, &\assumption
}$$
where the anomalous superfield in the first line of \Massumption\ does not
have any worldsheet counterpart. The examples in \forinst\ then translate into
\eqnn\DZmEx
$$\eqalignno{
D\cZ^m_{1,2,3,4,5} &= \Omega_1 k_1^m E_{1|2,3,4,5} + (1\leftrightarrow2,3,4,5)
&\DZmEx \cr
D\cZ^m_{12,3,4,5,6} &= \Omega_1 E^m_{1|2,3,4,5,6}
- \Omega_2 E^m_{2|1,3,4,5,6} 
+ \big[ \Omega_3 k_3^m E_{3|12,4,5,6}
+(3\leftrightarrow 4,5,6) \big]
\,.
}$$

\subsubsec Refined bootstrap equations

The duality between superspace kinematics and worldsheet functions suggests
to introduce a notion of refined ${\cal Z}$-functions defined via monodromies
\eqn\MTocZref{
Q{\cal J}^{m_1 \ldots m_r}_{A_1,\ldots,A_d|B_1,B_2,\ldots} \longleftrightarrow
D\cZ^{m_1 \ldots m_r}_{A_1,\ldots,A_d|B_1,B_2,\ldots}\,,
}
where the Berends--Giele superfields ${\cal J}$ are derived from the refined
building blocks of section \RefBBsec. The number $d\geq 1$ of slots on 
the left of the vertical bar is referred to as the degree of refinement. The 
left-hand side of \MTocZref\ is given in terms of refined anomaly superfields 
${\cal Y}^{m_1m_2\ldots}_{A_1,\ldots,A_d|B,\ldots}$ and (pseudo-)invariants
$P^{m_1\ldots m_r}_{1|A_1,\ldots,A_d|B,\ldots}$ \partI,
\eqnn\JCgen
$$\eqalignno{
&\quad Q{\cal J}^{m_1\ldots m_r}_{A_1,\ldots,A_d | B_1,B_2,\ldots} =
\d^{(m_1 m_2} {\cal Y}^{m_3\ldots m_r)}_{A_1,\ldots,A_d
|B_1,\ldots,}&\JCgen\cr
& \ \ + \big[\cY^{m_1\ldots m_r}_{A_2,\ldots,A_d |A_1,B_1,\ldots, }
+(A_1{\leftrightarrow} A_2,\ldots, A_d)\big]\cr
& \ \ + \big[\d_{|A_1|,1} k_{a_1}^p
 P^{pm_1\ldots m_r}_{a_1|A_2,\ldots,A_d |B_1,\ldots }
+ (A_1 {\leftrightarrow} A_2,\ldots, A_d) \big]\cr
& \ \ +\bigl[\d_{|B_1|,1} k_{b_1}^{(m_1}
P^{m_2\ldots m_r)}_{b_1|A_1,\ldots,A_d|B_2,\ldots}
+ (B_1{\leftrightarrow} B_2,\ldots ) \bigr]\cr
&\ \ + \big[P^{m_1\ldots m_r}_{a_1|a_2\ldots a_{|A_1|},A_2,\ldots,A_d |B_1,\ldots}
-P^{m_1\ldots m_r}_{a_{|A_1|}|a_1\ldots a_{|A_1|-1},A_2,\ldots,A_d |B_1,\ldots}
+ (A_1 {\leftrightarrow} A_2,\ldots, A_d) \big] \cr
& \ \ +\bigl[P^{m_1\ldots m_r}_{b_1|A_1,\ldots,A_d|b_2\ldots b_{|B_1|},B_2,\ldots}
- P^{m_1\ldots m_r}_{b_{|B_1|}|A_1,\ldots, A_d| b_1\ldots b_{|B_1|-1}, B_2,\ldots}
+ (B_1 \leftrightarrow B_2,\ldots )\bigr]\,,
}$$
for instance,
\eqnn\forinstref
$$\eqalignno{
Q {\cal J}^m_{1|23,4,5,6,7} &= {\cal Y}^m_{1,23,4,5,6,7}
+ k_1^p C^{mp}_{1|23,4,5,6,7} &\forinstref  \cr
&\quad{} + \big[ k_4^m P_{4|1|23,5,6,7}+ (4\leftrightarrow 5,6,7)\big]
+ P^m_{2|1|3,4,5,6,7}  -  P^m_{3|1|2,4,5,6,7}\,.
}$$
Accordingly, the refined versions of the worldsheet functions comprising 
$\ell,k_j^m$ and $g_{ij}^{(n)}$ are characterized by the following monodromies
\eqnn\JCmono
$$\eqalignno{
&\quad D {\cal Z}^{m_1\ldots m_r}_{A_1,\ldots,A_d | B_1,B_2,\ldots} =
\big[\d_{|A_1|,1} \Omega_{a_1} k_{a_1}^p E^{pm_1\ldots m_r}_{a_1|A_2,\ldots,A_d |B_1,\ldots }
+ (A_1 {\leftrightarrow} A_2,\ldots, A_d)\big]  &\JCmono\cr
& \ \ + \bigl[\d_{|B_1|,1} \Omega_{b_1}k_{b_1}^{(m_1}  E^{m_2\ldots m_r)}_{b_1|A_1,\ldots,A_d|B_2,\ldots}
+ (B_1\leftrightarrow B_2,\ldots)\bigr]\cr
&\ \ + \big[\Omega_{a_1}  E^{m_1\ldots m_r}_{a_1|a_2\ldots a_{|A_1|},A_2,\ldots,A_d |B_1,\ldots}
- \Omega_{a_{|A_1|}}E^{m_1\ldots m_r}_{a_{|A_1|}|a_1\ldots a_{|A_1|-1},A_2,\ldots,A_d |B_1,\ldots }
+ (A_1 {\leftrightarrow} A_2,\ldots, A_d) \big] \cr
& \ \ +\bigl[\Omega_{b_1}   E^{m_1\ldots m_r}_{b_1|A_1,\ldots, A_d| b_2\ldots b_{|B_1|}, B_2,\ldots}
-\Omega_{b_{|B_1|}} E^{m_1\ldots m_r}_{b_{|B_1|}|A_1,\ldots, A_d| b_1\ldots b_{|B_1|-1}, B_2,\ldots}
+ (B_1 \leftrightarrow B_2,\ldots)\bigr]\,,
}$$
where the anomalous superfields in the first line of \JCgen\ do
not have any worldsheet counterpart. The right-hand side of \JCmono\
features refined GEIs $E^{m_1\ldots }_{i|A_1,\ldots, A_d| B_1,\ldots}$
which will enter the correlators discussed in part III as the coefficients of refined
superfields. For example, the monodromy variation dual to \forinstref\ reads
\eqnn\DrefEx
$$\eqalignno{
D\cZ^m_{1|23,4,\ldots,8} &= \Omega_1 k_1^p E^{mp}_{1|23,4,\ldots,8} &\DrefEx\cr
&\quad{}+ \big[ k_4^m \Omega_4 E_{4|1|23,5,\ldots,8}+ (4\leftrightarrow 5,\ldots,8)\big] +E^m_{2|1|3,4,\ldots,8}  -  E^m_{3|1|2,4,\ldots,8}
\,.
}$$
The above patterns were discovered upon studying correlators
previously obtained by various other considerations at multiplicities four,
five and six. At higher multiplicities, the existence of worldsheet functions
subject to \assumption\ and \JCmono\ is a working hypothesis---so far confirmed by
explicit construction up to and including eight points.

\newsubsubsec\Gnambiguitysec An ambiguity caused by Eisenstein series

Given a solution $\cZ_{A,B, \ldots}$ to monodromy-variation equations, it is always possible
to deform it by an arbitrary GEI. A partial resolution to this ambiguity is quite natural in view of
the defining properties of one-loop correlators: We require the words $A,B,\ldots$ of $\cZ_{A,B, \ldots}$ 
to reflect tree-level-like singularities $(z_{a_1a_2} z_{a_2a_3} \ldots z_{a_{|A|-1}a_{|A|}})^{-1}
(z_{b_1b_2} z_{b_2b_3} \ldots z_{b_{|B|-1}b_{|B|}})^{-1}$, cf.\ \shufA. This requirement
fixes the most singular term to be $(g^{(1)}_{a_1a_2} g^{(1)}_{a_2a_3} \ldots
g^{(1)}_{a_{|A|-1}a_{|A|}})$ and should prevent the addition of non-constant
functions with vanishing monodromies, as they would necessarily modify this
singularity structure.
At the level
of unrefined scalar GEIs, this follows from the fact that non-constant elliptic functions always involve
singularities as $z_i \rightarrow z_j$, and we expect this property to carry over to
tensorial and refined GEIs.

However, this requirement cannot determine the presence
(or absence) of terms proportional to a holomorphic Eisenstein series ${\rm G}_n$,
for they are monodromy invariant ($D{\rm G}_n = 0$) as well as
constant functions on the worldsheet (${\p {\rm G}_n\over \p z_j}=0$). The construction of $\cZ_{A,B, \ldots}$
and GEIs from $g^{(n)}_{ij}$ automatically qualifies holomorphic Eisenstein series ${\rm G}_n = - g^{(n)}_{ii}$
as possible constituents. Moreover, ${\rm G}_n $ are known to arise in ($n\geq 8$)-point
one-loop correlators from the spin sums in the RNS formalism \refs{\TsuchiyaVA, \BroedelVLA}.

By the weight counting of section \seconethreefour, the first instance where the above ambiguity may 
affect the expressions for shuffle-symmetric functions happens at eight points. And
indeed, we will see in section~\EightPointsec\ that the eight-point correlator
is plagued by unwanted appearances of ${\rm G}_4$ whose kinematic coefficient 
remains undetermined in this work.

\newsubsubsec\Liesymsec Lie-symmetric worldsheet functions

From the discussion in section \BGmapsec, Berends--Giele superfields $M_{A,B,C}$
subject to shuffle symmetries can be translated to local building blocks
$T_{A,B,C}$ that satisfy Lie symmetries (cf.\ section \multisec). The dictionary
in \TABCtoMABC\ boils down to the KLT-matrix $S(\cdot|\cdot)_i$ \oldMomKer\
(also known as the {\it momentum kernel} \MomKer) that cancels the kinematic
poles of the Berends--Giele currents and is recursively defined by
\eqn\KLTagainA{
S(P,j|Q,j,R)_i = (k_{iQ}\cdot k_j)S(P|Q,R)_i,\qquad
S(\emptyset|\emptyset)_i = 1\,,
}
for instance
\eqn\KLTagainB{
S(2|2)_1 = (k_1\cdot k_2),\quad S(23|23)_1 = (k_{12}\cdot
k_3)(k_1\cdot k_2),\quad S(23|32)_1 = (k_1\cdot k_3)(k_1\cdot k_2)\,.
}
In analogous fashion, one can also define
worldsheet functions that satisfy Lie symmetries. To this effect we define, in
analogy with \TABCtoMABC,
\eqnn\cZToZ
\eqnn\KLTE
$$\eqalignno{
Z^{(s)m_1 \ldots}_{aA,bB, \ldots} &\equiv \sum_{A',B', \ldots} S(A|A')_a
S(B|B')_b \cdots \cZ^{m_1 \ldots}_{aA',bB', \ldots}\,,&\cZToZ\cr
E^{(s)m_1 \ldots}_{1|aA,bB,\ldots}&\equiv
\sum_{A',B',\ldots}S(A|A')_a S(B|B')_b \cdots E^{m_1 \ldots}_{1|aA',bB',
\ldots}\,,&\KLTE
}$$
where the matrix $S(A|A')_a$ defined in \KLTagainA\ contributes $|A|$ 
powers of $s_{ij}=k_i\cdot k_j$. For example, 
\eqnn\LieZex
$$\eqalignno{
Z^{(s)}_{1,2,3,4} &=  \cZ_{1,2,3,4} ,\quad
Z^{(s)}_{12,3,4,5} = s_{12}\cZ_{12,3,4,5} &\LieZex
\cr
Z^{(s)}_{12,34,5,6} &= s_{12}s_{34}\cZ_{12,34,5,6} ,\quad
Z^{(s)}_{123,4,5,6} = (s_{13}+s_{23})s_{12}\cZ_{123,4,5,6}
+ s_{13}s_{12}\cZ_{132,4,5,6}\,.
}$$
One can explicitly check that $Z^{(s)}_{123,4,5,6}$
indeed obeys the Lie symmetries in $A=123$;
$Z^{(s)}_{123,4,5,6} + Z^{(s)}_{213,4,5,6} = 0$, and
$Z^{(s)}_{123,4,5,6} + Z^{(s)}_{231,4,5,6} + Z^{(s)}_{321,4,5,6} = 0$.
Similarly, one may verify the Lie symmetries at higher multiplicities.
The superscript in $Z^{(s)}$ reminds of the presence of monomials in $s_{ij}$.

\newsubsec\WSdualCs Worldsheet dual expansions of BRST pseudo-invariants

In this section we will see the first non-trivial consequence
of the conjectural duality between worldsheet functions and kinematics: the
systematic construction of GEIs. This is done by exploiting the analogy between 
monodromy variations of $\cZ$-functions and the
BRST variations of Berends--Giele currents put forward in section~\DQdualsec.
The tentative idea is to assemble GEIs or ``worldsheet invariants'' following the same
combinatorics used in building kinematic BRST invariants $C_{1|A,B,C}$ and
$C^m_{1|A,B,C,D}$ in \scalarCs\ and \vectorCs\ from Berends--Giele currents. It turns
out that the worldsheet invariants constructed in this way give rise to GEIs as defined in
section~\GEIsec, i.e., their monodromy variations vanish.

At four and five points, the expressions for $C_{1|2,3,4} ,C_{1|23,4,5} $ and $C^m_{1|2,3,4,5} $
in \scalarCs\ and \vectorCs\ translate into
\eqnn\WSduals
$$\eqalignno{
E_{1|2,3,4}&=\cZ_{1,2,3,4}\,, &\WSduals\cr
E_{1|23,4,5} &= \cZ_{1,23,4,5} + \cZ_{12,3,4,5} - \cZ_{13,2,4,5}\,,\cr
E^m_{1|2,3,4,5} &= \cZ^m_{1,2,3,4,5} +  \big[ k_2^m \cZ_{12,3,4,5}
+ (2\leftrightarrow 3,4,5) \big]\,,
}$$
while at six points we have the unrefined GEIs,
\eqnn\SixElliptic
$$\eqalignno{
E_{1|234,5,6} & =
	 \cZ_{1,234,5,6}
        + \cZ_{12,34,5,6}
        + \cZ_{123,4,5,6}
        + \cZ_{412,3,5,6}
        - \cZ_{14,23,5,6}
        + \cZ_{143,2,5,6}\,,\cr
E_{1|23,45,6} & =
\cZ_{1,23,45,6} + \cZ_{12,45,3,6}
- \cZ_{13,45,2,6} + \cZ_{14,23,5,6}
- \cZ_{15,23,4,6}\cr
&\quad{}-\cZ_{412,3,5,6} + \cZ_{314,2,5,6}
+ \cZ_{215,3,4,6} - \cZ_{315,2,4,6}\,, &\SixElliptic\cr
E^m_{1|23,4,5,6} & = \cZ^m_{1,23,4,5,6} + \cZ^m_{12,3,4,5,6}
- \cZ^m_{13,2,4,5,6} +  k^m_3 \cZ_{123,4,5,6}
-   k^m_2 \cZ_{132,4,5,6}  \cr
&\quad{}+\big[ k^m_4 \cZ_{14,23,5,6} - k^m_4 \cZ_{214,3,5,6}
+ k^m_4 \cZ_{314,2,5,6}
+ (4\leftrightarrow 5,6)\bigr]\,, \cr
E^{mn}_{1|2,3,4,5,6} & =
\cZ^{mn}_{1,2,3,4,5,6} +  \big[ k_2^{m} \cZ^n_{12,3,4,5,6}+k_2^{n}
\cZ^m_{12,3,4,5,6} + (2\leftrightarrow 3,4,5,6) \big] \cr
&\quad{}-  \big[ (k_2^{m} k_3^{n} +k_2^{n} k_3^{m}) \cZ_{213,4,5,6}
+ (2,3|2,3,4,5,6) \big]\,,
}$$
see \tenssixpt\ for the superspace counterpart of the tensor. Moreover,
six points admit one instance of a refined GEI dual to the $P_{1|2|3,4,5,6}$
superfield \pseudoPsix,
\eqnn\covPSb
$$\eqalignno{
E_{1|2|3,4,5,6} &= \cZ_{2|1,3,4,5,6} + k_2^m \cZ^m_{12,3,4,5,6} + \big[
s_{23}\cZ_{123,4,5,6}
+(3\leftrightarrow4,5,6)\big]\,.   &\covPSb
}$$
The analogous seven-point expansions are displayed in Appendix~\someapp.

Based on the $D$-variations from section~\DQdualsec, it is straightforward to
verify that all of \WSduals\ and \SixElliptic\ are indeed GEIs upon using
momentum conservation\foot{In order to see that \covPSb\ defines a GEI as well,
one can either employ the explicit representation assembled in \SixErefined\ or
insert the integration-by-parts identity \covU\ among GEIs into the $D$
variation obtained from \JCmono.}. As we will see in the next section, the above
GEIs have obvious extensions by one extra word (slot) to match the slot
structure on the right-hand side of the above $D$-variations. For instance,
$E_{1|23,4,5,6}\equiv E_{1|23,4,5}$ will be needed for $D{\cal Z}_{123,4,5,6}=
\Omega_1 E_{1|23,4,5,6} - \Omega_3 E_{3|12,4,5,6}$.

\newsubsec\bootsec The bootstrap

At first glance, the discussion in sections~\DQdualsec\ and \WSdualCs\ seems to
suffer from a {\it chicken-and-egg\/} dilemma; in section~\DQdualsec, to obtain
the monodromy variations of the shuffle symmetric functions one needs the
associated GEIs from section~\WSdualCs, while the expressions of the GEIs
require the shuffle-symmetric functions from section~\DQdualsec.

The way out of this conundrum is to note that this self-recursive structure can
be exploited to {\it bootstrap} the shuffle-symmetric $\cZ$-functions order by
order in multiplicity, starting with the four-point solution which is taken to
be a constant. We will see how this works in practice in the following
subsections.

Note that the functions obtained below will be used inside one-loop correlators
of the open superstring, and as such, are considered to be multiplied by the
overall Koba--Nielsen factor \KNfactor. Therefore functions that differ by
derivatives of the Koba--Nielsen factor given in \zderivzero\ and \tauderivzero\
are considered equivalent as will be indicated by the symbol $\cong$.

\subsubsec Four-point worldsheet functions

From the computation of the four-point correlator in \refs{\GreenSW, \MPS}, it follows
that the four-point shuffle-symmetric worldsheet function is a constant.
Similarly, the expansion \WSduals\ implies that also its corresponding
GEI is a constant. Both are normalized to one,
\eqn\fourcte{
\cZ_{1,2,3,4} \equiv 1\,,\quad
E_{1|2,3,4} \equiv 1\, .
}
To proceed to the next level we define the slot extension of \fourcte\ as
$E_{1|2,3,4,5}\equiv 1$.

\newsubsubsec\fivewssec Five-point worldsheet functions

According to \protoC\ and \assumption, the monodromy variations of the shuffle-symmetric 
functions $\cZ_{12,3,4,5}$ and $\cZ^m_{1,2,3,4,5}$ at five points are given by
\eqnn\DZfive
$$\eqalignno{
D\cZ_{12,3,4,5} &= \Omega_1 E_{1|2,3,4,5} - \Omega_2E_{2|1,3,4,5} =
\Omega_{12} &\DZfive\cr
D\cZ^m_{1,2,3,4,5} &= \Omega_1 k^m_1 E_{1|2,3,4,5} + (1\leftrightarrow2,3,4,5) =
\sum_{j=1}^5 \Omega_j k_j^m\,.
}$$
A closer inspection of the linearized monodromies \covD\ naturally leads to
the following solutions of \DZfive,
\eqn\fiveSol{
\cZ_{12,3,4,5} = g^{(1)}_{12}\,,\qquad \cZ^m_{1,2,3,4,5} = \ell^m\,.
}
These expressions reproduce the desired singularity structure
$\cZ_{12,3,4,5} = z_{12}^{-1}+{\cal O}(z_{12})$
and regularity of $\cZ^m_{1,2,3,4,5}$, cf.\ section \Gnambiguitysec.

\newsubsubsubsec\FiveptGEI Assembling five-point GEIs

We can now assemble associated GEIs from the
expansions in \WSduals,
\eqnn\fiveEls
\eqnn\fiveElls
$$\eqalignno{
E_{1|23,4,5} &= \cZ_{1,23,4,5} + \cZ_{12,3,4,5} - \cZ_{13,2,4,5} &\fiveEls\cr
&= \gg1(2,3) + \gg1(1,2) - \gg1(1,3)\,,\cr
E^m_{1|2,3,4,5} &= \cZ^m_{1,2,3,4,5} +  \big[ k_2^m \cZ_{12,3,4,5}
+ (2\leftrightarrow 3,4,5) \big] &\fiveElls\cr
&= \ell^m + \bigl[ k_2^m \gg1(1,2) + (2\leftrightarrow3,4,5)\bigr]\,.
}$$
It is easy to check that \fiveEls\ and \fiveElls\ are indeed invariant under 
monodromy variations (using momentum conservation in the latter case).
Before proceeding to the next multiplicity, we define the slot extension of
\fiveEls\ and \fiveElls,
\eqn\sixEms{
E_{1|23,4,5,6}\equiv E_{1|23,4,5}\,,\qquad 
E^m_{1|2,3,4,5,6} \equiv
\ell^m + \bigl[ k_2^m \gg1(1,2) + (2\leftrightarrow3,4,5,6)\bigr]\,,
}
including an extra permutation $2\leftrightarrow6$ in the vector GEI. These extensions
are natural from the generating functions for GEIs to be given in a later work
and they will be used on the right-hand sides of the
monodromy variations of six-point $\cZ$-functions below.

\newsubsubsec\sixwssec Six-point worldsheet functions

According to \assumption\ and \JCmono, the six-point shuffle-symmetric
worldsheet functions satisfy the following monodromy variations:
\eqnn\sixMon
$$\eqalignno{
D\cZ_{123,4,5,6} &= \Omega_1 E_{1|23,4,5,6} - \Omega_3 E_{3|12,4,5,6}\,,
&\sixMon \cr
D\cZ_{12,34,5,6} &= \Omega_1 E_{1|2,34,5,6}-\Omega_2 E_{2|1,34,5,6}
+\Omega_3 E_{3|12,4,5,6}-\Omega_4 E_{4|12,3,5,6}\,,\cr
D\cZ^m_{12,3,4,5,6} &= \Omega_1 E^m_{1|2,3,4,5,6}-\Omega_2 E^m_{2|1,3,4,5,6}
+\big[ k_3^m \Omega_3 E_{3|12,4,5,6}+ (3\leftrightarrow 4,5,6) \big]\,,\cr
D\cZ^{mn}_{1,2,3,4,5,6} &= k^m_1\Omega_1 E^n_{1|2,3,4,5,6}
+k^n_1\Omega_1 E^m_{1|2,3,4,5,6}+ (1\leftrightarrow 2,3,4,5,6)\,,\cr
D {\cal Z}_{2|1,3,4,5,6} &= \Omega_2 k_2^m E^m_{2|1,3,4,5,6}\,.
}$$
In the appendix \appCone\ we will obtain the following solutions,
\eqnn\newgs
$$\eqalignno{
\cZ_{123,4,5,6}&= g^{(1)}_{12}g^{(1)}_{23} + g^{(2)}_{12} +
g^{(2)}_{23} - g^{(2)}_{13}\,, &\newgs\cr
\cZ_{12,34,5,6}&= g^{(1)}_{12}g^{(1)}_{34}
+ g^{(2)}_{13} + g^{(2)}_{24}
- g^{(2)}_{14} - g^{(2)}_{23}\,,\cr
\cZ^m_{12,3,4,5,6}&= \ell^m g^{(1)}_{12}
+ (k_2^m - k_1^m)g^{(2)}_{12}
+ \big[ k_3^m (g^{(2)}_{13} - g^{(2)}_{23}) + (3\leftrightarrow 4,5,6)\big]\,,\cr
\cZ^{mn}_{1,2,3,4,5,6}&= \ell^m\ell^n +
\bigl[( k_1^{m}k_2^{n}+k_1^{n}k_2^{m}) g^{(2)}_{12} + (1,2|1,2,3,4,5,6)\bigr]\,,\cr
{\cal Z}_{2|1,3,4,5,6}&=0\,.
}$$
In accordance with the discussion in section \shufflesec, their behavior as
the vertex insertions collide corresponds to their tree-level counterparts.
For instance, the short-distance behavior $\cZ_{123,4,5,6}\rightarrow
(z_{12}z_{23})^{-1}$ is the same as that of $\cZ^{\rm tree}_{123}$.

\newsubsubsubsec\SixptGEIsec Assembling six-point GEIs

Plugging the above solutions into the expansions \SixElliptic\ of six-point GEIs leads to
\eqnn\Eswithgs
$$\eqalignno{
E_{1|234,5,6} & =
 \gg1(1,2) \gg1(2,3)
       - \gg1(1,2) \gg1(2,4)
       + \gg1(1,2) \gg1(3,4)
       - \gg1(1,4) \gg1(2,3)
       + \gg1(1,4) \gg1(2,4)
       - \gg1(1,4) \gg1(3,4) &\Eswithgs\cr
&\quad{}      + \gg1(2,3) \gg1(3,4)
       + \gg2(2,3)
       - \gg2(2,4)
       + \gg2(3,4)\,,\cr
E_{1|23,45,6} & =\big(\gg1(1,2)+\gg1(2,3)+\gg1(3,1)\big)\big(\gg1(1,4)
+\gg1(4,5)+\gg1(5,1)\big)\,,\cr
E^m_{1|23,4,5,6} &=\big(
\ell^m + k_4^m g^{(1)}_{14} +  k_5^m g^{(1)}_{15}
+ k_6^m g^{(1)}_{16}\big)\big(\gg1(1,2)+\gg1(2,3)+\gg1(3,1)\big)\cr
&\quad{} + \big[k_2^m( \gg1(1,3)\gg1(2,3)
          + \gg2(1,2)
          - \gg2(1,3)
          - \gg2(2,3)) - (2\leftrightarrow 3)\big]\,,\cr
E^{mn}_{1|2,3,4,5,6} &= \ell^m\ell^n
+ \big[k_2^{(m}k_{3}^{n)}\gg1(1,2)\gg1(1,3)+ (2,3|2,3,4,5,6)\big]\cr
&\quad{} +\big[\ell^{(m}k_2^{n)}\gg1(1,2) + 2 k_2^m k_2^n \gg2(1,2)
 + (2\leftrightarrow3,4,5,6)\big]\,,
}$$ 
as well as
\eqnn\SixErefined
$$\eqalignno{
E_{1|2|3,4,5,6}
&= - 2 s_{12} g^{(2)}_{12} + g^{(1)}_{12} (\ell\cdot k_2 + s_{23}g^{(1)}_{23}
+ s_{24}g^{(1)}_{24}  + s_{25}g^{(1)}_{25}  + s_{26}g^{(1)}_{26})\cr
&\cong \p g^{(1)}_{12} + s_{12} (g_{12}^{(1)})^2 - 2 s_{12} g^{(2)}_{12}\,.&\SixErefined\cr
}$$
The second line follows from the first one via integration by parts according
to \zderiv.

The slot-extensions of the above GEIs are given by
\eqnn\sixExt
$$\eqalignno{
E_{1|234,5,6,7} &\equiv E_{1|234,5,6}\,,\quad
E_{1|23,45,6,7}  \equiv E_{1|23,45,6}\,,\cr
E^m_{1|23,4,5,6,7} & \equiv E^m_{1|23,4,5,6}
+ k_7^m g^{(1)}_{17}\big(\gg1(1,2)+\gg1(2,3)+\gg1(3,1)\big)\,,&\sixExt\cr
E^{mn}_{1|2,3,4,5,6,7} &\equiv \ell^m\ell^n
+ \big[k_2^{(m}k_{3}^{n)}\gg1(1,2)\gg1(1,3)+ (2,3|2,\ldots,7)\big]\cr
&\quad{} +\big[\ell^{(m}k_2^{n)}\gg1(1,2) + 2 k_2^m k_2^n \gg2(1,2)
 + (2\leftrightarrow3,4,5,6,7)\big]\,,
 \cr
 E_{1|2|3,4,5,6,7}&\equiv - 2 s_{12} g^{(2)}_{12} + g^{(1)}_{12} (\ell\cdot k_2 + s_{23}g^{(1)}_{23}
+ s_{24}g^{(1)}_{24}  +\ldots  + s_{27}g^{(1)}_{27}) \cr
&\cong \p g^{(1)}_{12} + s_{12} (g_{12}^{(1)})^2 - 2 s_{12} g^{(2)}_{12}\ \, ,
}$$
and they will be used to bootstrap the shuffle-symmetric
functions at seven points.

\newsubsubsec\sevenwssec Seven-point worldsheet functions

At seven points, the monodromy variations for the scalar shuffle-symmetric
functions following from \assumption\ and \sixExt\ are given by
\eqnn\DZsevens
$$\eqalignno{
D\cZ_{1234,5,6,7} &=\Omega_1E_{1|234,5,6,7} - \Omega_4E_{4|123,5,6,7}\,,
&\DZsevens\cr
D\cZ_{123,45,6,7} &=\Omega_1E_{1|23,45,6,7} - \Omega_3E_{3|12,45,6,7}
+ \Omega_4E_{4|123,5,6,7} - \Omega_5E_{5|123,4,6,7}\,,\cr
D\cZ_{12,34,56,7} &=\Omega_1E_{1|2,34,56,7} - \Omega_2E_{2|1,34,56,7} +
(12\leftrightarrow34,56)\,,
}$$
and admit the following solutions:
\eqnn\newsevengs
$$\eqalignno{
\cZ_{1234,5,6,7} &=
 \gg1(1,2)\gg1(2,3)\gg1(3,4)
	+ \gg3(1,2)+ \gg3(2,3) + \gg3(3,4) - 2 \gg3(4,1)  &\newsevengs\cr
&\quad{}+ \gg1(1,2)\bigl(\gg2(2,3) + \gg2(3,4) - \gg2(4,1) \bigr)
       + \gg1(2,3) \bigl(\gg2(1,2) + \gg2(3,4) - \gg2(4,1) \bigr)
       + \gg1(3,4) \bigl(\gg2(1,2) +\gg2(2,3) - \gg2(4,1) \bigr)\,,  \cr
\cZ_{123,45,6,7} &=\gg1(1,2)\gg1(2,3)\gg1(4,5)
+\gg1(4,5)\big(\gg2(1,2) + \gg2(2,3) - \gg2(1,3)\big)\cr
&\quad{}+\big(\gg1(1,2)+\gg1(2,3)+\gg1(3,1)\big)\big(\gg2(1,4) - \gg2(1,5) +
\gg2(3,5) - \gg2(3,4)\big)\,,
\cr
\cZ_{12,34,56,7}&= \gg1(1,2)\gg1(3,4)\gg1(5,6) + \gg1(1,2)\big(\gg2(3,5)
- \gg2(3,6) - \gg2(4,5) + \gg2(4,6)\big)\cr
&\quad{}+ \gg1(3,4)\big(\gg2(1,5)- \gg2(1,6) - \gg2(2,5)+ \gg2(2,6)\big)
+ \gg1(5,6)\big(\gg2(1,3) - \gg2(1,4) - \gg2(2,3)+ \gg2(2,4)\big)
          \cr
&\quad{}+ \gg1(1,5)\big(\gg2(1,3) - \gg2(1,4) - \gg2(3,5) + \gg2(4,5)\big)
+ \gg1(1,6)\big( \gg2(1,4)- \gg2(1,3) + \gg2(3,6)- \gg2(4,6)\big)
	\cr
&\quad{}+ \gg1(2,5)\big(\gg2(2,4)- \gg2(2,3)- \gg2(4,5) + \gg2(3,5)\big)
+\gg1(2,6)\big(\gg2(2,3) - \gg2(2,4)- \gg2(3,6)+ \gg2(4,6)\big)\,.
}$$
The solutions for the tensorial functions will be presented
in Appendix~\sevptapp, see in particular \SevVecOne, \SevTopTwo\
and \seVTens.

In addition to the above unrefined solutions,
the monodromy variations of the three seven-point topologies of
refined worldsheet functions following from \JCmono\ read
\eqnn\DZrefseven
$$\eqalignno{
D\cZ_{12|3,4,5,6,7} &= \Omega_1 E_{1|2|3,4,5,6,7} - \Omega_2 E_{2|1|3,4,5,6,7}\,,&\DZrefseven\cr
D\cZ_{1|23,4,5,6,7} &= \Omega_1 k_1^p E^p_{1|23,4,5,6,7}
+\Omega_2E_{2|1|3,4,5,6,7}
-\Omega_3E_{3|1|2,4,5,6,7}\,,\cr
D\cZ^m_{1|2,3,4,5,6,7} &=\Omega_1 k_1^p E^{pm}_{1|2,3,4,5,6,7}
+\big[\Omega_2 k_2^m E_{2|1|3,4,5,6,7} + (2\leftrightarrow3,4,5,6,7)\big]\,,
}$$
where the extended GEI above were defined in \sixExt, with solutions
\eqnn\sevNotorious
$$\eqalignno{
\cZ_{12|3,4,5,6,7} &= \p\gg2(1,2) + s_{12}\gg1(1,2)\gg2(1,2) - 3 s_{12}\gg3(1,2)
\,, &\sevNotorious\cr
\cZ_{1|23,4,5,6,7} &= \cZ_{13|2,4,5,6,7} - \cZ_{12|3,4,5,6,7}\,,\cr
\cZ^m_{1|2,3,4,5,6,7} &= -\big[ k_2^m \cZ_{12|3,4,5,6,7} +
(2\leftrightarrow3,4,5,6,7)\big]\,.
}$$
Although not manifest, the worldsheet singularities of the above functions are
the ones expected from their labeling according to the discussion in
section~\shufflesec. For instance, the function $\cZ_{12,34,56,7}$ can only have
singularities as $z_1\to z_2$ (corresponding to the word $12$) and similarly for
$34$ and $56$. However, its expansion contains certain factors of $g^{(1)}_{ij}$
that suggest the presence of ``forbidden'' singularities; like
$g^{(1)}_{15}\big(\gg2(1,3) - \gg2(1,4) - \gg2(3,5) + \gg2(4,5)\big)$ as $z_1\to
z_5$. But a careful analysis using the Laurent expansions \gnexpansion\ shows
that it is in fact non-singular as $z_1\to z_5$ (similar conclusions apply for
the other terms). Note that functions which involve Mandelstam variables such as
$s_{12}\gg1(1,2)$ are considered non-singular as they don't generate kinematic
poles when integrated along with the Koba--Nielsen factor. Therefore all
functions in \sevNotorious\ are in fact non-singular upon integration over
$z_j$.

Having the shuffle-symmetric worldsheet functions we can now assemble
seven-point GEIs as discussed in the previous section. The results are displayed in
Appendix~\bootstrapapp, see in particular \sevenM\ to \rewrtDalt. Also, the building
blocks of section \GenSeriessec\ turn out to admit the compact representations \rewrtD\ or \rewrtE.

\subsubsec Eight-point shuffle-symmetric worldsheet functions

The system of monodromy variations can be solved explicitly at eight points
following the bootstrap approach. This will be done in the appendix~\appCeight.

\newnewsec\newdualitysec Duality between worldsheet functions and kinematics

In this section, we will illustrate various further facets of the duality between 
worldsheet functions and kinematics. It will be exemplified that GEIs $E_{1|\ldots}^{\ldots}$ 
share the relations and symmetries of the kinematic factors $C^{\ldots}_{1|\ldots}$ and
$P^{\ldots}_{1|\ldots}$ discussed in part I. Some of these relations will be
shown to have an echo at the level of the ${\cal Z}$-functions. We spell out the
concrete evidence for the duality and formulate conjectures for the
all-multiplicity patterns. If these conjectures are correct, the kinematic and
worldsheet ingredients of the open-string correlators ${\cal K}_n$ in \fourcor\
to \sevencor\ enter on completely symmetric footing. Like this, we support the
double-copy structure of one-loop open-string amplitudes \MafraIOJ\ up to and
including seven points. At eight points we will sometimes encounter
terms proportional to the holomorphic Eisenstein series ${\rm G}_4$ that do not
have a corresponding kinematic companion. Accommodating these terms with the
duality between worldsheet functions and kinematics is left for a future work.

\newsubsec\thissubsec The GEI dual to BRST-cohomology identities

The appearance of the correlators ${\cal K}_n(\ell)$ in open- and closed-string
amplitudes is insensitive to BRST-exact terms. This has been exploited in
\partI\ to derive so-called Jacobi identities in the BRST cohomology that relate
momentum contractions $k_1^m C^{m\ldots}_{1|A,B,\ldots}$ and $k_A^m
C^{m\ldots}_{1|A,B,\ldots}$ to (pseudo-)invariants of lower tensor rank, see
section \CJACOBIsec. We will now exemplify that GEIs
$E^{m\ldots}_{1|A,B,\ldots}$ obey the same Jacobi identities between different
tensor rank and degree of refinement, where BRST-exact terms translate into
total derivatives.

\newsubsubsec\EjacA Five points

Based on momentum conservation, one can show that the following combinations
of five-point GEIs \fiveEls\ and \fiveElls\ conspire to total Koba--Nielsen derivatives \zderiv\
\eqnn\covU
$$\eqalignno{
k^m_1 E^m_{1|2,3,4,5} &= \partial_1 \log {\cal I}_5   &\covU \cr
k^m_2 E^m_{1|2,3,4,5} + \big[ s_{23} E_{1|23,4,5}
+ (3\leftrightarrow 4,5) \big] &= \partial_2 \log {\cal I}_5
}
$$
and can therefore be dropped from open- and closed-string amplitudes.
The first relation is in one-to-one correspondence with the cohomology
identity $Q {\cal J}_{1|2,3,4,5} = k_1^m C^m_{1|2,3,4,5}+\Delta_{1|2,3,4,5}$
after dropping the BRST-exact anomaly factor $\Delta_{1|2,3,4,5}$ (cf.\ section \Deltasec).
Similarly, the second line of \covU\ has the kinematic counterpart \deltasecE\ involving
$k^m_2 C^m_{1|2,3,4,5}$.

\newsubsubsec\EjacB Six points

Similarly, at six points we find Jacobi relations among the GEIs
in \Eswithgs\ and \SixErefined\ which exactly match the kinematic
identities listed in \sixptQDs\ (cf.\ section~$10$ of \partI),
\eqnn\covW
$$\eqalignno{
k^m_4 E^m_{1|23,4,5,6} &\cong -s_{24} E_{1|324,5,6} +s_{34} E_{1|234,5,6}
-s_{45} E_{1|23,45,6} -s_{46} E_{1|23,46,5}\,,\cr
k_{23}^m E^m_{1|23,4,5,6}  &\cong
\big[   s_{24} E_{1|324,5,6} -  s_{34} E_{1|234,5,6}
+ (4\leftrightarrow 5,6) \big] + E_{1|2|3,4,5,6} - E_{1|3|2,4,5,6}\, ,\cr
k^m_1 E^m_{1|23,4,5,6} &\cong E_{1|3|2,4,5,6} - E_{1|2|3,4,5,6}\, , &\covW\cr
k^m_2 E^{mn}_{1|2,3,4,5,6} &\cong k_2^n E_{1|2|3,4,5,6} - \big[
s_{23} E^n_{1|23,4,5,6} + (3\leftrightarrow 4,5,6)
\big]\,,\cr
k^m_1 E^{mn}_{1|2,3,4,5,6} &\cong - \big[ k_2^n E_{1|2|3,4,5,6}
+ (2\leftrightarrow 3,4,5,6) \big]\,.
}$$
As in \zderivzero, the $\cong$ notation is a reminder that $z_j$-derivatives have been
discarded in passing to the right-hand side.
Note that momentum conservation reduces the identities for contraction with $k_1$
to combinations of the remaining ones involving $k_{A}^m E^{m\ldots}_{1|A,\ldots}$.

\newsubsubsec\EjacC Higher multiplicity

More generally, the elliptic identities that are dual to the BRST-cohomology identities
in section~9 of \partI\ can be written as
\eqnn\morenotR
$$\eqalignno{
k_{A_1}^p E^p_{1|A_1, \ldots,A_4} &=
- \big[ E_{1| S[A_1,A_2],A_3, \ldots,A_4}
+ (A_2\leftrightarrow A_3, \ldots,A_4) \big]&\morenotR\cr
&\quad{}+\!\!\sum_{XY=A_1}\!\!\! \big[E_{1|X|Y,A_2,
\ldots,A_4}-(X\leftrightarrow Y) \big]\,,\cr
\noalign{\vskip4pt}
k_{A_1}^p E^{pm}_{1|A_1, \ldots,A_5} &= k_{A_1}^m E_{1|A_1|A_2, \ldots,A_5}
- \big[ E^m_{1| S[A_1,A_2],A_3, \ldots,A_5}+ (A_2\leftrightarrow A_3, \ldots,A_5) \big]\cr
&\quad{}+\!\! \sum_{XY=A_1}\!\!\! \big[E^m_{1|X|Y,A_2, \ldots,A_5}-
(X\leftrightarrow Y)\big]\,,
}$$
where the $S[A,B]$ map is defined in section \smapsubsec\ and yields $E_{1|S[2,3],4,5}=s_{23} E_{1|23,4,5}$
as well as $E_{1|S[23,4],5,6}=s_{34} E_{1|234,5,6} - s_{24} E_{1|324,5,6}$ in the simplest cases. Given a word
$A=a_1 a_2\ldots a_{|A|}$, the sum over deconcatenations $XY=A$ is understood to comprise all 
non-empty $X=a_1 a_2\ldots a_j$ and $Y= a_{j+1}\ldots a_{|A|}$ with $j=1,2,\ldots,|A|{-}1$.
We have verified all of \morenotR\ up to and including eight points, and
their higher-point generalizations are plausible by the dual kinematic identities given
in \JRal\ and \partI. Note the absence of elliptic-function duals to the
BRST-exact anomaly terms $\Delta^{m_1\ldots}_{1|A_1,\ldots}$ without refined slots.

Following the worldsheet duals of the higher-rank identities in section~9 of \partI,
one arrives at
\eqnn\morenotRR
$$\eqalignno{
k_{A_1}^p E^{pmn}_{1|A_1,\ldots,A_{6}} &=
k_{A_1}^{(m} E^{n)}_{1|A_1|A_2,\ldots,A_{6}}
- \big[ E^{mn}_{1| S[A_1,A_2],A_3,\ldots, A_{6}}
+ (A_2\leftrightarrow A_3,\ldots, A_{6}) \big]\cr
&\quad{}+\!\! \sum_{XY=A_1}\!\!\! \big[E^{mn}_{1|X|Y,A_2,\ldots,A_{6}}
-(X\leftrightarrow Y)\big] + \delta^{mn} G_{1|A_1|A_2,\ldots,A_{6}}
&\morenotRR\cr
k_{A_1}^p E^{pm_1\ldots m_r}_{1|A_1,\ldots,A_{r+4}} &=
k_{A_1}^{(m_1} E^{m_2\ldots m_r)}_{1|A_1|A_2,\ldots,A_{r+4}}
- \big[ E^{m_1\ldots m_r}_{1| S[A_1,A_2],A_3,\ldots, A_{r+4}}
+ (A_2\leftrightarrow A_3,\ldots, A_{r+4}) \big]\cr
&\quad{}+\!\! \sum_{XY=A_1}\!\!\! \big[E^{m_1\ldots m_r}_{1|X|Y,A_2,\ldots,A_{r+4}}
-(X\leftrightarrow Y)\big] + \delta^{(m_1 m_2}G^{m_3\ldots m_r)}_{1|A_1|A_2,\ldots,A_{r+4}}\,,\cr
}$$
for some a priori undetermined GEIs $G_{1|\ldots}$ in the trace component. The
latter can be thought of as a tentative GEI dual of the refined anomaly superfields
$\Delta^{m_1\ldots}_{1|A_1,\ldots,A_d|B,\ldots}$ that are no longer BRST-exact
if $d\geq 1$, see section \Deltasec. The representations of GEIs up to and
including eight points given in this work yield
$G^{m_1\ldots}_{1|A_1|A_2,\ldots}=0$, e.g. the seven-point GEIs in \sevenM\ to
\rewrtDalt\ can be checked to obey
\eqnn\EmoreB
$$\eqalignno{
k_2^p E^{mnp}_{1|2,3,\ldots,7}&=  k_2^{(m} E^{n)}_{1|2|3,\ldots,7}
- \big[ s_{23} E^{mn}_{1|23,4,\ldots,7} + (3\leftrightarrow 4,\ldots,7) \big]
&\EmoreB\cr
}$$
which amounts to $G_{1|2|3,4,5,6,7} = 0$ in \morenotRR. Still, it is worthwhile
to keep in mind that non-zero choices of $G^{m_1\ldots}_{1|A_1|A_2,\ldots}$ are
still compatible with the duality between kinematics and worldsheet functions.

Finally, the above identities generalize straightforwardly to slot-extensions of GEIs such as
$E_{1|A,B,C} \rightarrow E_{1|A,B,C,D}$ and its generalizations in the $D{\cal Z}$-variations, namely
\eqnn\morenotRext
$$\eqalignno{
k_{A_1}^p E^{pm_1\ldots m_r}_{1|A_1,\ldots,A_{r+5}} &=
k_{A_1}^{(m_1} E^{m_2\ldots m_r)}_{1|A_1|A_2,\ldots,A_{r+5}}
- \big[ E^{m_1\ldots m_r}_{1| S[A_1,A_2],A_3,\ldots, A_{r+5}}
+ (A_2\leftrightarrow A_3,\ldots, A_{r+5}) \big]\cr
&\quad{}+\!\! \sum_{XY=A_1}\!\!\! \big[E^{m_1\ldots m_r}_{1|X|Y,A_2,\ldots,A_{r+5}}
-(X\leftrightarrow Y)\big]  + \delta^{(m_1 m_2}G^{m_3\ldots m_r)}_{1|A_1|A_2,\ldots,A_{r+5}} \,. &\morenotRext 
}$$
At higher degree of refinement, appropriate choices of GEIs
should obey the dual of the most general Jacobi identity \allJacv\ on
the kinematic side
\eqnn\allJacvE
$$\eqalignno{
0&= \big[ G^{m_1\ldots m_r}_{1|A_2,\ldots,A_d|A_1,B_1,\ldots,B_{r+d+2}}
+ (A_1\leftrightarrow A_2,\ldots,A_d) \big]
+ \delta^{(m_1 m_2} G^{m_3\ldots m_r)}_{1|A_1,\ldots,A_d|B_1,\ldots,B_{r+d+2}} \cr
&+\big[k_{A_1}^p E^{pm_1\ldots m_r}_{1|A_2,\ldots,A_d|A_1,B_1,\ldots,B_{r+d+2}}
+(A_1\leftrightarrow A_2,\ldots,A_d) \big]
- k^{(m_1}_{A_1A_2\ldots A_d} E^{m_2\ldots m_r)}_{1|A_1,\ldots,A_d|B_1,\ldots,B_{r+d+2}}
 \cr
&+ \Big(  \big[  E^{m_1\ldots m_r}_{1|A_2,\ldots,A_d| S[A_1,B_1] , B_2,\ldots,B_{r+d+2} }
+ (B_1 \leftrightarrow B_2,\ldots,B_{r+d+2}) \big]  &\allJacvE \cr
&- \sum_{XY=A_1} ( E^{m_1\ldots m_r}_{1|X,A_2,\ldots,A_d | Y,B_1,\ldots,B_{r+d+2}}
-E^{m_1\ldots m_r}_{1|Y,A_2,\ldots,A_d| X,B_1,\ldots,B_{r+d+2}} )
+(A_1\leftrightarrow A_2,\ldots,A_d) \Big) \, ,
}$$
that are checked up to and including eight points. These proposals will serve 
as a key input for the all-multiplicity construction of GEIs from generating series. 
Note that the first term $G^{m_1\ldots m_r}_{1|A_2,\ldots,A_d|A_1,B_1,\ldots,B_{r+d+2}}$ 
does not have any refined slots at $d=1$ and should vanish by the duality with the 
BRST-exact unrefined anomaly superfields. In fact, we even observe stronger identities
among seven- and eight-point GEIs such as
\eqnn\dreamII
$$\eqalignno{
0&\cong k_3^m E^m_{1|2|3,4,5,6,7} - s_{23} E_{1|23|4,5,6,7} + \big[ s_{34} E_{1|2|34,5,6,7} + (4\leftrightarrow 5,6,7) \big]
&\dreamII
}$$
with a single momentum contraction, which implies \allJacvE\ upon symmetrization 
in $2\leftrightarrow3$. The kinematic dual of \dreamII\ involving $k_3^m P^m_{1|2|3,4,5,6,7} 
+ \Delta_{1|3|2,4,5,6,7}$ can be found in \deltasecF. As detailed in section \closedLEdd, 
identities like \dreamII\ that involve just a single momentum contraction $k_{A_1}^p 
E^{pm_1\ldots }_{1|A_2,\ldots,A_d|A_1,B_1,\ldots}$ play a key role for the path towards
local and BRST-invariant $n$-point correlators in future work.

\newsubsec\dualbasissec The GEI dual to BRST change-of-basis identities

In section~11 of \partI\ several identities among (pseudo-)invariants were derived
using BRST-cohomology manipulations that implement a {\it change of basis}\foot{They were
referred to as ``BRST-canonicalization'' identities in \partI.}.
The simplest examples are
\eqnn\CdualOne
$$\eqalignno{
C_{3|12,4,5} &= C_{1|23,4,5} + Q(\ldots)\,, &\CdualOne\cr
 C_{2|1,34,5} &= C_{1|2,34,5}+C_{1|23,4,5} - C_{1|24,3,5} + Q(\ldots)\,, \cr
C^m_{2|1,3,4,5} &= C^m_{1|2,3,4,5} + \big[k_3^m C_{1|23,4,5}
+(3\leftrightarrow 4,5)\big] + Q(\ldots)\,, \cr
P_{2|1|3,4,5,6}&= P_{1|2|3,4,5,6} + \cY_{12,3,4,5,6} + Q(\ldots)\,,
}$$
where the right-hand side is written in terms of the canonical basis of
$C_{1|A,B,C}$ and $P_{1|A|B,C,D,E}$ with leg 1 in the first position of the
subscript. The BRST-exact terms in the ellipses are spelled out in \partI.
Naturally, these identities have an elliptic dual under $C\rightarrow E$ as well
as its ``refined'' version\foot{The pseudo-invariant $P_{i|A|B, \ldots}$ should
really be denoted $C_{i|A|B, \ldots}$, as it would unify this and countless
other formulas.} $P\rightarrow E$.

\newsubsubsec\EjacG Five points

It is straightforward to show that the five-point GEIs in \fiveEls\
obey change-of-basis identities dual to \CdualOne,
\eqnn\unhatfive
$$\eqalignno{
E_{3|12,4,5} &= E_{1|23,4,5}\,, &\unhatfive \cr
 E_{2|1,34,5} &= E_{1|2,34,5}+E_{1|23,4,5} - E_{1|24,3,5}\,,   \cr
E^m_{2|1,3,4,5} &= E^m_{1|2,3,4,5} + \big[k_3^m E_{1|23,4,5}
+(3\leftrightarrow 4,5) \big]\,.
}$$
As detailed in appendix~\bootstrapapp, similar change-of-basis identities involving GEIs play
a major role in the solution of the monodromy-variation equations. On the right-hand sides
of the monodromy variations $D\cZ$ in section \DQdualsec, however, the GEIs are ``extended'' to
have one additional word. While the scalar identities in \unhatfive\ hold in identical form for
$E_{1|23,4,5,6}=E_{1|23,4,5}$, the vector identity is extended by an obvious extra permutation
involving leg $6$, i.e.\  $E^m_{2|1,3,4,5,6} = E^m_{1|2,3,4,5,6} + \big[k_3^m E_{1|23,4,5,6}
+(3\leftrightarrow 4,5,6) \big]$.

\newsubsubsec\EjacH Six points

Change-of-basis identities among six-point GEIs take the identical form as compared to the
relations among (pseudo-)invariants in section 11 and appendix F of \partI,
\eqnn\cobA
$$\eqalignno{
E_{2|134,5,6}&=   E_{1|342,5,6}   &\cobA \cr
E_{2|13,45,6}&= E_{1|32,45,6} + E_{1|324,5,6} - E_{1|325,4,6}   \cr
E_{2|1,345,6}&=  E_{1|2,345,6} +  E_{1|234,5,6}+E_{1|254,3,6}
+ E_{1|325,4,6} + E_{1|23,45,6} + E_{1|25,43,6} \cr
 E_{2| {1}, {3 4}, {5 6}} &= E_{1|{2}, {3 4}, {5 6}} + E_{1| {2 3}, {5 6}, {4}}  -
 E_{1|{2 4}, {5 6}, {3}} + E_{1| {2 5}, {3 4}, {6}}  -
 E_{1|{2 6}, {3 4}, {5}}   \cr
 &\quad{} - E_{1| { 325}, {6}, {4}}  +
 E_{1| {326}, {5}, {4}}+ E_{1|{425}, {6}, {3}}  -
 E_{1| {426}, {5}, {3}}
 \cr
E^m_{2| {1 3}, {4}, {5},6}&= E^m_{1|32,4,5,6} + \big[ k_4^m E_{1|324,5,6} +(4\leftrightarrow 5,6) \big]\cr
 E^m_{2| {1 }, {34}, {5},6} &=  E^m_{1|2,34,5,6}+E^m_{1|23,4,5,6}-E^m_{1|24,3,5,6}+k_4^m E_{1|234,5,6}-k_3^m E_{1|243,5,6}  \cr
 &\quad{}+ \big[ k_5^m (E_{1| {2 5}, {3 4}, {6}}-E_{1| {325}, {4}, {6}}  +E_{1|{42 5}, {3}, {6}})   +(5\leftrightarrow 6) \big]\, .
}$$
Similar to the translation of BRST variations to $D{\cal Z}$-variations in the
previous section, the ${\cal Y}$-superfield in the pseudo-invariant identity
of \CdualOne\ has no GEI analogue,
\eqnn\cobB
$$\eqalignno{
E_{2|1|3,4,5,6} &=  E_{1|2|3,4,5,6} &\cobB\cr
E_{2|3|1,4,5,6} &=  E_{1|3|2,4,5,6} + k_3^m E_{1|23,4,5,6}^m
+ \big[ s_{34} E_{1|234,5,6} + (4\leftrightarrow 5,6) \big]
\cr
E^{mn}_{2|1,3,4,5,6}&=    E^{mn}_{1|2,3,4,5,6}
+ \big[k_3^{(m} E^{n)}_{1|23,4,5,6} + (3\leftrightarrow 4,5,6) \big] \cr
 & \quad{} - \big[  k^{(m}_3 k^{n)}_4 E_{1|324,5,6}  + (3,4|3,4,5,6) \big]\,.
}$$
More general cases such as expanding $E_{2|13,45,67,89}$ in terms of
$E_{1|\ldots}$ have no matching analogous identities in terms of $C_{1|A,B,C}$,
so the required change-of-basis identities are not readily available from
\partI. These identities can, however, be generated using the general algorithm
described in the appendix~\changebasisapp.

\newsubsec\tracesZsec The worldsheet analogue of kinematic trace relations

We have seen in section~\tracesT\ that the kinematic building blocks satisfy
certain identities that relate traces of tensorial building blocks at refinement
$d$ to sums of building blocks\foot{ Note that the building blocks with $d=0$
are denoted by $M$ rather than $\cJ$.} of refinement $d{+}1$. For instance, \HOc\
at the level of Berends--Giele currents reads \partI
\eqn\HOcAgain{
{1\over 2}\delta_{np}\cJ^{np m_1\ldots m_r}_{A_1,\ldots,A_d|B_1,\ldots,B_{d+r+5}} =
  \cJ^{m_1\ldots m_r}_{A_1,\ldots,A_d,B_1|B_2,\ldots,B_{d+r+5}}
+(B_1{\leftrightarrow} B_2,\ldots,B_{d+r+5}) \,,
}
and it is natural to ask what is the corresponding statement in terms of
worldsheet functions. Given that this identity relates BRST-covariant Berends--Giele
superfields rather than (pseudo-)invariants, their worldsheet analogues should
concern the ${\cal Z}$-functions subject to non-vanishing $D$-variations.
Note that the worldsheet functions depend on one
additional word when compared to their kinematic counterpart
($\cZ_{A,B,C,D}\leftrightarrow M_{A,B,C}$), therefore their trace relations will
also have one extra permutation.

\newsubsubsec\somessssecA Six points

At six points one can show from the explicit solutions \newgs\ for the $\cZ$-functions that
the following trace relation is satisfied up to a total derivative \tauderivzero\ in $\tau$:
\eqnn\tracZex
$$\eqalignno{
\half\d_{mn}\cZ^{mn}_{1,2,3,4,5,6} &\cong \cZ_{1|2,3,4,5,6}
+ (1\leftrightarrow2,3,4,5,6)\,.&\tracZex\cr
}$$
In order so see this, we note that the functions $\cZ_{1|2,3,4,5,6}$ on the right-hand side vanish
(see \newgs\ and appendix \appCone), and the trace of the tensor in \newgs\ yields the 
$\tau$-derivative \tauderiv\ of the Koba--Nielsen factor,
\eqn\traceB{
{1\over 2} \delta_{mn} \cZ^{mn}_{1,2,3,4,5,6} =
{1\over 2}\ell^2 + \big[ s_{12} g^{(2)}_{12}
+(1,2|1,2,3,4,5,6) \big] = 2\pi i{\p\over\p\tau}\log{\cal I}_6(\ell)\,.
}

\newsubsubsec\somessssecD Seven points

Similarly, the solutions of the seven-point monodromy variations in section \sevenwssec\ satisfy
\eqnn\sevTraces
$$\eqalignno{
\half \cZ^{mpp}_{1,2,3,4,5,6,7} - \big[\cZ^m_{2|1,3,4,5,6,7} +
\,(2\leftrightarrow3, \ldots,7)\big]&\cong\cZ^m_{1|2,3,4,5,6,7}\,,&\sevTraces\cr
\half \cZ^{pp}_{12,3,4,5,6,7} - \big[\cZ_{3|12,4,5,6,7} +
\,(3\;\leftrightarrow4,5,6,7)\big]&\cong\cZ_{12|3,4,5,6,7}\,,\cr
}$$
in accordance with the expectation from the analogy with kinematic building
blocks \HOcAgain. Note that $\tau$-derivatives acting on both the Koba--Nielsen
factor and $\ell^m$ or $g^{(1)}_{12}$ have been discarded in \sevTraces, using
the mixed heat equation \mixedheat\ for the latter.

\newsubsubsec\somessssecC Eight points

From the discussion in section~\Gnambiguitysec\ we know that the solutions of
the eight-point monodromy variations are slightly ambiguous due to the
Eisenstein series ${\rm G}_4$. This freedom can be exploited to yield two sets
of solutions differing by terms proportional to ${\rm G}_4$, depending on
whether they satisfy the trace relations or not. On the one hand, the {\it
naive\/} solutions to the monodromy equations in the appendix~\appCeight\ fail
to satisfy all but one of the dual trace relations,
\eqnn\failedtraces
$$\eqalignno{
\half \cZ^{pp}_{12,34,5, \ldots,8} - \big[\cZ_{12|34,5, \ldots,8}
+ (12\leftrightarrow34,5, \ldots,8)\big]&\cong
- R_{12,34,5,6,7,8}\,,&\failedtraces\cr
\half \cZ^{pp}_{123,4,5,6,7,8} - \big[\cZ_{123|4,5,6, \ldots,8}
+ (123\leftrightarrow4, \ldots,8)\big]&\cong
- R_{123,4,5,6,7,8}\,,\cr
\half \cZ^{mpp}_{12,3,4,5,6,7,8} - \big[\cZ^m_{12|3,4,5,6, \ldots,8}
+ (12\leftrightarrow3, \ldots,8)\big]&\cong
- R^m_{12,3,4,5,6,7,8}\,,\cr
\half \cZ^{mnpp}_{1,2,3,4,5,6,7,8} - \big[\cZ^{mn}_{1|2,4,5,6,7,8}
+ (1\leftrightarrow2,3, \ldots,8)\big]&\cong
- R^{mn}_{1,2,3,4,5,6,7,8}\,,\cr
\half\cZ^{pp}_{1|2,3,4,5,6,7,8} - \big[ \cZ_{1,2|3,4,5,6,7,8}
+ (2\leftrightarrow3,\ldots,8) \big]&\cong0\,,
}$$
where
\eqnn\RemainGfour
$$\eqalignno{
R_{12,34,5,6,7,8} & = 3{\rm G}_4\big(
	 s_{13}
        - s_{14}
        - s_{23}
        + s_{24}\big)\,,&\RemainGfour\cr
R_{123,4,5,6,7,8} & = 3{\rm G}_4\big(s_{12}-2s_{13}+s_{23}\big)\,,\cr
R^m_{12,3,4,5,6,7,8} &= 3{\rm G}_4\big(
s_{12} (k_2^m - k_1^m)
+ \big[ k_3^m (s_{13}-s_{23}) + (3\leftrightarrow4,5,6,7,8)\big]\big)\,,\cr
R^{mn}_{1,2,3,4,5,6,7,8} &=3{\rm G}_4 k_1^{(m} k_2^{n)} s_{12} +(1,2|1,2, \ldots,8)\,.
}$$
But note that these failed trace relations are a peculiarity of certain
eight-point $\cZ$-functions that will be used in the eight-point
correlator in section~\EightPointsec. Since these functions will be multiplying
local kinematic building blocks, one may exploit the kinematic trace relations
reviewed in section~\tracesT\ to add {\it deformations\/}
\eqn\deformdef{
\hat\cZ \equiv \cZ + \d\cZ\,,
}
while keeping the overall eight-point correlator unchanged. Starting from the
naive solutions $\cZ$ of the monodromy variations in the appendix~\appCeight,
the deformed functions $\hat\cZ$ in \deformdef\ can be made to satisfy all
trace relations by adding\foot{Beware of the definition \controversy, in particular,
$\d^{(mn}\d^{pq)}=\d^{mn}\d^{pq} +\d^{mp}\d^{nq} + \d^{mq}\d^{pn}$.}
\eqnn\GfourUnref
$$\eqalignno{
\d\cZ^{mn}_{12,34,5,6,7,8} &= - \d^{mn}R_{12,34,5,6,7,8}\, , \ \ \ \ \ \ 
\d\cZ^{mn}_{123,4,5,6,7,8} = - \d^{mn}R_{123,4,5,6,7,8}\,,\cr
\d\cZ^{mnp}_{12,3,4,5,6,7,8} &=
-\d^{(mn}R^{p)}_{12,3,4,5,6,7,8}\,, &\GfourUnref\cr
\d\cZ^{mnpq}_{1,2,3,4,5,6,7,8} &=
-\d^{(mn}R^{pq)}_{1,2,3,4,5,6,7,8}
+{1\over4}\d^{(mn}\d^{pq)} R^{aa}_{1,2,3,4,5,6,7,8}\,,
}
$$
in the unrefined cases,
while the deformations of the refined functions read,
\eqnn\GfourRef
$$\eqalignno{
\eqalign{
\d\cZ_{123|4,5,6,7,8} &= - R_{123,4,5,6,7,8}\,,\cr
\d\cZ^m_{1|23,4,5,6,7,8} &= -R^m_{23,1,4,5,6,7,8}\,,\cr
\d\cZ^m_{12|3,4,5,6,7,8} &= -R^m_{12,3,4,5,6,7,8}\,,\cr
\d\cZ^{mn}_{1|2,3,4,5,6,7,8} &= -R^{mn}_{1,2,3,4,5,6,7,8}
-{1\over4}\d^{mn}R^{aa}_{1,2,3,4,5,6,7,8}\,.
}\hskip-15pt\eqalign{
\d\cZ_{1|23,45,6,7,8} &= - R_{23,45,1,6,7,8}\,,\cr
\d\cZ_{1|234,5,6,7,8} &= - R_{234,1,5,6,7,8}\,,\cr
\d\cZ_{12|34,5,6,7,8} &= - R_{12,34,5,6,7,8}\,,\cr
\phantom{{1\over4}R^{aa}_1}&\phantom{M}\hskip70pt\GfourRef
}}$$
In addition, in order to preserve the last trace relation of \failedtraces, we have
\eqn\doublyDeform{
\d\cZ_{1,2|3,4,5,6,7,8} =-{1\over4}R^{aa}_{1,2,3,4,5,6,7,8}\,,
}
where the shorthands $R$ proportional to ${\rm G}_4$ were defined in
\RemainGfour.  Once we present the eight-point correlator in
section~\EightPointsec, it will be straightforward to verify that the above
deformations \deformdef\ keep it invariant.

\newsubsec\DeltaZsec The worldsheet analogue of kinematic anomaly invariants

The vanishing of the six-point function $\cZ_{1|2,3,4,5,6}$ can be understood as
a correspondence between refined worldsheet functions at multiplicity $n$ and unrefined
$\cY$ superfields at multiplicity $n{-}1$. More precisely, the BRST-exact linear
combinations $\Delta_{1| \ldots}$ of unrefined anomaly superfields \partI\ reviewed
in section \Deltasec\ are observed to match the vanishing of the corresponding linear
combinations of refined worldsheet functions under the map
\eqn\ZDeltamap{
\cY^{m \ldots}_{1A,B_1, \ldots}\leftrightarrow \cZ^{m
\ldots}_{1A|B_1, \ldots}\,.
}
In the following we will use the notation $\cZ^\Delta_{1|A,B,C,D,E}$
to denote the worldsheet counterpart of $\Delta_{1|A,B,C,D,E}$ that follows the
same combinatorics (with obvious generalizations to tensors and refined cases).
We will see that the six- and seven-point $\cZ^\Delta$
vanish up to total derivatives (confirming the suggested duality) whereas
subtle contributions $\sim {\rm G}_4$ may arise at eight points.

\newsubsubsec\sixptDeltaZ Six points

At six points, the vanishing of the components
$\langle\Delta_{1|2,3,4,5}\rangle=\langle\cY_{1,2,3,4,5}\rangle$ suggests that
its worldsheet analogue under the map \ZDeltamap\ also vanishes. Indeed, as
anticipated in \newgs\ and detailed in appendix~\appCone,
\eqn\cYZdual{
\cZ^\Delta_{1|2,3,4,5,6}\equiv
\cZ_{1|2,3,4,5,6} \cong 0\,.
}

\newsubsubsec\sevptDeltaZ Seven points

The natural next step is to check whether the seven-point refined functions
following from the combinatorics of the six-point BRST-exact superfields \partI,
\eqnn\EightDeltas
$$\eqalignno{
\Delta_{1|23,4,5,6} &= \cY_{1,23,4,5,6}
+ \cY_{12,3,4,5,6}
- \cY_{13,2,4,5,6}\,,&\EightDeltas\cr
\Delta^m_{1|2,3,4,5,6} &= \cY^m_{1,2,3,4,5,6}
+ \big[ k_2^m \cY_{12,3,4,5,6} + (2 \leftrightarrow3,\ldots,6)\big]\,,
}$$
also vanish. This is indeed the case, as the solutions \sevNotorious\ for
the refined ${\cal Z}$-functions imply the vanishing of
\eqnn\constS
$$\eqalignno{
\cZ^\Delta_{1|23,4,5,6,7} &\equiv \cZ_{1|23,4,5,6,7}  + \cZ_{12|3,4,5,6,7} -
\cZ_{13|2,4,5,6,7}\cong0\,,&\constS \cr
\cZ^{\Delta,m}_{1|2,3,4,5,6,7} &\equiv \cZ^m_{1|2,3,4,5,6,7}+ \big[ k_2^m \cZ_{12|3,4,5,6,7}
+ (2\leftrightarrow 3,4,\ldots,7)\big]\cong0\,.
}$$
Therefore, the pattern established in the six-point vanishing
of $\cZ_{1|2,3,4,5,6}$ in \cYZdual\ extends to seven points; worldsheet functions that correspond
to BRST-exact superfields $\Delta_{1|\ldots}$ vanish up to total derivatives.

\newsubsubsec\eightptDeltaZ Eight points

However, at eight points something peculiar happens. From the superfield
expansions of the BRST-exact anomaly building blocks, the map \ZDeltamap\ leads to
\eqnn\DelZs
$$\eqalignno{
\cZ^\Delta_{1|234,5,6,7,8} &=
\cZ_{1|234,5,6,7,8} + \cZ_{12|34,5,6,7,8}
+ \cZ_{123|4,5,6,7,8} - \cZ_{124|3,5,6,7,8}\cr
&\quad{}- \cZ_{14|23,5,6,7,8} - \cZ_{142|3,5,6,7,8} + \cZ_{143|2,5,6,7,8}\,, &\DelZs\cr
\cZ^\Delta_{1|23,45,6,7,8}
&= \cZ_{1|23,45,6,7,8} + \cZ_{12|45,3,6,7,8} - \cZ_{13|45,2,6,7,8}
+ \cZ_{14|23,5,6,7,8} - \cZ_{15|23,4,6,7,8}\cr
&\quad{}-\cZ_{412|3,5,6,7,8} + \cZ_{314|2,5,6,7,8}
+ \cZ_{215|3,4,6,7,8}- \cZ_{315|2,4,6,7,8}\,, \cr
\cZ^{\Delta,m}_{1|23,4,5,6,7,8}
&= \cZ^m_{1|23,4, \ldots,8} + \cZ^m_{12|3, \ldots,8} - \cZ^m_{13|2,4, \ldots,8}
+ k^m_3 \cZ_{123|4,5,6,7,8} - k^m_2 \cZ_{132|4,5,6,7,8}\cr
&\quad{}+\big[ k^m_4 \cZ_{14|23,5,6,7,8} - k^m_4
\cZ_{214|3,5,6,7,8} + k^m_4 \cZ_{314|2,5,6,7,8} + (4\leftrightarrow
5,6,7,8)\bigr]\,,\cr
\cZ^{\Delta, mn}_{1|2,3,4,5,6,7,8} & =
\cZ^{mn}_{1|2,3,4,5,6,7,8} +  \big[ k_2^{m} \cZ^n_{12|3,4,5,6,7,8}
+k_2^{n} \cZ^m_{12|3,4,5,6,7,8} + (2\leftrightarrow 3, \ldots,8) \big] \cr
&\quad{} -  \big[ (k_2^{m} k_3^{n} + k_2^{n}k_3^{m})
\cZ_{213|4,5,6,7,8} + (2,3|2, \ldots,8) \big]\,.
}$$
In addition, the worldsheet analogue of the non-BRST-exact building block
$\Delta_{1|2|3,4,5,6,7}$ in \deltasecC\ gives rise to
\eqn\PZdelta{
\cZ^\Delta_{1|2|3,\ldots,8} = \cZ_{1,2|3,\ldots,8}  +  k_2^m \cZ^{m}_{12|3,\ldots,8}
+ \big[ s_{23} \cZ_{123|4,\ldots,8} +
(3\leftrightarrow 4,5,\ldots,8) \big]\,.
}
Given that the monodromy variations used to obtain the eight-point functions
$\cZ$ cannot detect
explicit appearances of the modular form ${\rm G}_4$, we have
two possible scenarios:
\medskip
\item{$i)$} use $\cZ$-functions without ${\rm G}_4$ corrections that do not satisfy
the trace relations;
\item{$ii)$} use $\hat \cZ$-functions in \deformdef\ with ${\rm G}_4$ corrections
that satisfy the trace relations.
\medskip
\noindent It turns out that the functions from option $i)$ lead to vanishing $\cZ^\Delta$,
including \PZdelta:
\eqn\noGfour{
\eqalign{
\cZ^\Delta_{1|234,5,6,7,8} &\cong 0\,,\cr
\cZ^\Delta_{1|23,45,6,7,8}&\cong0\,,
}\qquad\eqalign{
\cZ^{\Delta\,m}_{1|23,4,5,6,7,8}&\cong0\,,\cr
\cZ^{\Delta\, mn}_{1|2,3,4,5,6,7,8} &\cong0\,.
}\qquad\eqalign{
\cZ^\Delta_{1|2|3,\ldots,8} &\cong0\,,\cr
{}&{}\cr
}}
The trace-satisfying functions $\hat \cZ$ from option $ii)$, however, lead to non-vanishing
analogues $\hat \cZ^\Delta$ that are defined by replying $ \cZ \rightarrow \hat \cZ$ in \DelZs\ and \PZdelta,
\eqnn\ZDeltafail
$$\eqalignno{
\hat \cZ^\Delta_{1|234,5,6,7,8} &= 3{\rm G}_4\big(
        2\ss(1,3)
       - \ss(1,2)
       - \ss(1,4)
       + 2\ss(2,4)
       - \ss(2,3)
       - \ss(3,4)
       \big)\,, &\ZDeltafail\cr
\hat \cZ^\Delta_{1|23,45,6,7,8}&= 3{\rm G}_4\big(
        \ss(2,5)
       + \ss(3,4)
       - \ss(2,4)
       - \ss(3,5)\big)\,,\cr
\hat \cZ^{\Delta,m}_{1|23,4,5,6,7,8}&= 3{\rm G}_4\Big[
s_{23}k_2^m
- \ss(1,2)(2 k_2^m + k_3^m)
-\big[k_4^m s_{24} + (4\leftrightarrow5,6,7,8)\big]
       - (2\leftrightarrow3)\Big]\,,\cr
\hat \cZ^{\Delta, mn}_{1|2,3,4,5,6,7,8} & = 3{\rm G}_4 s_{23}\Big(k_2^{(m} k_2^{n)} +
k_3^{(m}k_3^{n)} - k_2^{(m}k_3^{n)} + \half\d^{mn}(s_{12}+s_{13}-s_{23})\Big) +
(2,3|2, \ldots,8)\,,\cr
\hat \cZ^\Delta_{1|2|3,4,5,6,7,8} &= 3{\rm G}_4\big(
       3 \ss(2,3)\ss(2,4)
       + \ss(1,3)\ss(1,4)
       - \ss(3,4)(
        \ss(2,3)
       + \ss(2,4)
       + \half\ss(3,4)
       + \half\ss(1,2)
       )
	\big) + (3,4|3, \ldots,8)\,.
}$$
As will become clear in the discussion of the eight-point correlator in
section~\EightPointsec, the subtleties associated to the presence or absence
of ${\rm G}_4$ terms are responsible for the difficulties in obtaining a BRST-closed
eight-point correlator.

\newsubsec\seconeeightanymore The GEI dual to trace relations

Also the trace relations among pseudo-invariants such as ${1\over
2}\delta_{mn}C^{mn}_{1|2,3,4,5,6}= P_{1|2|3,\ldots,6}$ $+(2\leftrightarrow 3,4,5,6)$
and its generalizations in \pseudoTRA\ have an echo at the level of GEIs.

\newsubsubsec\EjacD Six points

At six points, the GEIs \Eswithgs\ and \SixErefined\ are related by
\eqnn\traceA
$$\eqalignno{
{1\over 2} \delta_{mn} E^{mn}_{1|2,3,4,5,6} &=
{1\over 2}\ell^2 + \big[ s_{12} g^{(2)}_{12}
+(1,2|1,2,3,4,5,6) \big] + \big[ E_{1|2|3,4,5,6}
+ (2\leftrightarrow 3,4,5,6) \big] \cr
&=  \big[ E_{1|2|3,4,5,6} + (2\leftrightarrow 3,4,5,6) \big]
+ 2\pi i{\partial \over \partial \tau} \log{\cal I}_6(\ell)\,, &\traceA
}$$
where we have used  \tauderiv\ to identify ${1\over 2}\ell^2+\sum_{i<j}s_{ij}g^{(2)}_{ij}$
as a $\tau$-derivative of the Koba--Nielsen factor.
Note that this trace relation has a $\cZ$-function counterpart given in \traceB.

\newsubsubsec\EjacE Seven points

Similarly, we have checked that the seven-point tensor traces of GEIs
obey relations analogous to the dual (pseudo-)invariants,
\eqnn\tracesevenA
$$\eqalignno{
{1\over 2} \delta_{mn} E^{mn}_{1|23,4,5,6,7} &\cong E_{1|23|4,5,6,7}
+ \big[ E_{1|4|23,5,6,7} + (4\leftrightarrow 5,6,7) \big]
\cr
{1\over 2} \delta_{np} E^{mnp}_{1|2,3,4,5,6,7} &\cong
\big[ E^m_{1|2|3,4,5,6,7} + (2\leftrightarrow 3,\ldots,7) \big]\,. &\tracesevenA
}$$
Similar to the $\cZ$-function counterparts \sevTraces, the equivalence $\cong$
refers to $\tau$-derivatives that have been discarded.

\newsubsubsec\EjacF Eight points

At eight points, however, the GEI-duals of the kinematic trace relations
\pseudoTRA\ exhibit deviations proportional to ${\rm G}_4$.
After expanding the GEIs in terms of $\hat\cZ$-functions (obtained from the
Berends--Giele expansion of their corresponding pseudo BRST invariants, see
appendix~\bootstrapapp) one can show that
\eqnn\traceeightA
$$\eqalignno{
{1\over 2} \delta_{mn} E^{mn}_{1|234,5,6,7,8}
- \big[ E_{1|234|5,6,7,8} + (234\leftrightarrow 5,6,7,8) \big]
&\cong \hat\cZ^\Delta_{1|234,5,6,7,8}
&\traceeightA\cr
{1\over 2} \delta_{mn} E^{mn}_{1|23,45,6,7,8}
- \big[ E_{1|23|45,6,7,8} + (23\leftrightarrow 45,6,7,8)\big]
&\cong \hat\cZ^\Delta_{1|23,45,6,7,8} \cr
{1\over 2} \delta_{np} E^{mnp}_{1|23,4,5,6,7,8} - \big[
E^m_{1|23|4,5,6,7,8} + (23\leftrightarrow 4,5,6,7,8) \big]
&\cong \hat\cZ^{\Delta,m}_{1|23,4,5,6,7,8} \cr
{1\over 2} \delta_{pq} E^{mnpq}_{1|2,3,\ldots,8}
- \big[ E^{mn}_{1|2|3,4,\ldots,8} + (2\leftrightarrow 3,\ldots,8) \big]
&\cong \hat\cZ^{\Delta,mn}_{1|2,3,4,5,6,7,8}  \cr
{1\over 2} \delta_{mn} E^{mn}_{1|2|3,\ldots,8}
- \big[ E_{1|2,3|4,\ldots,8} + (3\leftrightarrow 4,\ldots,8) \big]
&\cong \hat\cZ^\Delta_{1|2|3,4,5,6,7,8}\,,
}$$
where the various functions $\hat\cZ^{\Delta}$ are described in
section~\eightptDeltaZ\ and defined in \ZDeltafail. The above results were
obtained using the trace-satisfying representation $\hat\cZ$ in the expansions
of the GEIs. We know from \noGfour\ that all $\cZ^\Delta$-functions vanish if we
use the representation of shuffle-symmetric functions that do not satisfy the
trace relations, so one could wonder if the above elliptic traces would vanish
in that case. Unfortunately, this does not happen. In fact, the first three
relations of \traceeightA\ are independent on the choice of $\cZ$ or $\hat\cZ$,
while the other two change (but do not vanish in either case).

\newsubsubsec\EjacMany Higher multiplicities

At higher multiplicity, suitable choices of the GEIs are expected to admit the dual
of the kinematic relation \pseudoTRA,
\eqn\Etracesecond{
\delta_{np}\widehat E^{npm_1\ldots m_r}_{1|B_1,\ldots,B_{r+5}}
 = 2 \widehat E^{m_1 \ldots m_r}_{1|B_1|B_2, \ldots, B_{r+5}}
 + (B_1 \leftrightarrow B_2,\ldots,B_{r+5})\,,
}
or more generally, the dual of the higher-refinement relation \pseudoTRB,
\eqn\Etracethird{
\delta_{np} \widehat E^{npm_1\ldots m_r}_{1|A_1,\ldots,A_{d} | B_1,\ldots,B_{d+r+5}} =
2\widehat E^{m_1 \ldots m_r}_{1|A_1, \ldots,A_{d},B_1|B_2, \ldots, B_{d+r+5}}
+ (B_1 \leftrightarrow B_2,\ldots, B_{d+r+5}) \, .
}
The hat notation in \Etracesecond\ and \Etracethird\ is used to indicate that,
beyond seven points, the expressions for $E$ presented in this work do not
necessarily match the trace-satisfying GEIs $\widehat E$. We leave it to the 
future to identify the missing redefinitions by ${\rm G}_{k\geq 4}$ relating the 
GEIs $E$ of this work to the trace-satisfying GEIs $\widehat E$ in \Etracesecond\ 
and \Etracethird.

\newnewsec\GenSeriessec Simplified representations of GEIs

In this section, we review and extend the construction of elliptic functions
from the Kronecker--Eisenstein series \refs{\DolanEH, \TsuchiyaNF} and identify
ubiquitous building blocks for GEIs. These building blocks turn out to yield
compact expressions for the GEIs in section \bootsec\ and will be used to
present explicit all-multiplicity formulae for unrefined GEI of tensor rank
$r\leq 2$.

\newsubsec\copypasteA Elliptic functions and their extensions

One can show via \shift\ that the cyclic product $F(z_{12},\alpha) F(z_{23},\alpha) \ldots
F(z_{n-1,n},\alpha) F(z_{n,1},\alpha) $ of Kronecker--Eisenstein series \EisKron\ is an elliptic function
of the punctures $z_1,z_2,\ldots,z_n$~\DolanEH,
\eqn\ellrepD{
F(z_{12},\alpha) F(z_{23},\alpha) \ldots F(z_{n-1,n},\alpha) F(z_{n,1},\alpha)
= \sum_{w=0}^{\infty} \a^{-n+w} V_w(1,2,\ldots,n)\, ,
}
where the dependence on $\tau$ is kept implicit for ease of notation. Since this
property is independent on $\alpha$, each term on the right-hand side of
\ellrepD\ is an elliptic function $V_w$ in $n$ punctures $z_1,z_2,\ldots,z_n$ by
itself. At the level of linearized monodromies \covC, we have
$DF(z_{ij},\alpha) =\alpha \Omega_{ij}F(z_{ij},\alpha) $ and therefore
\eqn\VWellipt{
DV_w(1,2,\ldots,n)  = 0\,.
}
The simplest examples of the elliptic functions $V_w$ in \ellrepD\ are $V_0(1,2,\ldots,n)=1$ and
\eqn\ellrepDex{
V_1(1,2,\ldots,n) = \sum_{j=1}^n g^{(1)}_{j,j+1}\,, \qquad
V_2(1,2,\ldots,n) = \sum_{j=1}^n g^{(2)}_{j,j+1} + \sum_{1\leq i<j}^n g^{(1)}_{i,i+1} g^{(1)}_{j,j+1}\,,
}
subject to cyclic identification $z_{n+1}\equiv z_1$. Their generating series in
\ellrepD\ and the reflection properties
\eqn\reflKron{
F(-z,-\alpha,\tau) = - F(z,\alpha,\tau)\,, \qquad
g^{(n)}(-z,\tau) = (-1)^n g^{(n)}(z,\tau)
}
imply cyclicity and reflection (anti-)symmetry for the functions $V_w$,
\eqn\dihed{
V_w(1,2,3,\ldots,n) = V_w(2,3,\ldots,n,1) = (-1)^w V_w(1,n,\ldots,3,2)\,.
}
Moreover, one can show via Fay relations \FayKron\ or \Faygn\ that the functions $V_w(1,2,\ldots,n)$
with $w=n{-}2$ obey the shuffle symmetry
\eqn\VWshuff{
V_{n-2}(1,(2,3,\ldots,j) \shuffle (j{+}1,\ldots,n))=0\,, \qquad j=2,3,\ldots,n{-}1 \, .
}
Given that shuffle symmetry is shared by Berends--Giele currents and (pseudo-)invariant
kinematic factors, the $V_w(1,2,\ldots,n)$ with $w=n{-}2$ will play a key role for
the duality between worldsheet functions and kinematics.

\subsubsec Derivative extension of elliptic functions

Compact representations of vectorial and tensorial GEIs will require
extensions of the set of $V_w$-functions \ellrepD\ that are covariant rather than
invariant under linearized monodromies. Functions with these properties can be
constructed by inserting a derivative with respect to the bookkeeping variable
$\alpha$ into their generating series:
\eqn\ellrepF{
 F(z_{12},\alpha) F(z_{23},\alpha) \ldots F(z_{n-1,n},\alpha) \partial_\a F(z_{n,1},\alpha)
  \equiv \sum_{w=-1}^{\infty} \a^{-n+w} \partial V_w(1,2,\ldots,n)\,.
}
The notation $\partial V_w$ for the functions on the right-hand side reminds of the
$\alpha$-derivative on the left-hand side and should not be confused with ${\partial \over \partial z_j}$.
Based on $D \partial_\a F(z_{ij},\alpha)= \Omega_{ij} \big[ \a \partial_\a F(z_{ij},\alpha) + F(z_{ij},\alpha)  \big]$,
the monodromy variations of the $\partial V_w$-functions in \ellrepF\ can be written as
\eqn\ellrepH{
D\partial V_w(1,2,\ldots,n) = \Omega_{n1}  V_w(1,2,\ldots,n)\,.
}
Given that their $D$-variation is expressible in terms of the elliptic $V_w$-functions
of \ellrepD, the $\partial V_w$-functions are said to be {\it monodromy-covariant}.

The desired expressions for the ${\cal Z}$-functions and GEIs turn out to only involve
$\partial V_w(1,2,\ldots,n)$ with $w=n{-}2$. The simplest examples admit the
following expansions
\eqnn\delVs
$$\eqalignno{
\partial V_0(1,2)&=\gg1(2,1) &\delVs\cr
\partial V_1(1,2,3) &=  \gg2(3,1) -  \gg1(1,2) \gg1(2,3)- \gg2(1,2)- \gg2(2,3)
= \gg1(3,1)\Big(\gg1(1,2)  + \gg1(2,3)\Big) + 2 \gg2(3,1)\cr
\partial V_2(1,2,3,4) &=  \gg1(4,1)\Big(\gg1(1,2) \gg1(2,3)
+ \gg1(1,2) \gg1(3,4)
+ \gg1(2,3) \gg1(3,4)
+ \gg2(1,2)
+ \gg2(2,3)
+ \gg2(3,4)\Big)\cr
&\quad{}+ 2\gg2(4,1)\Big( \gg1(1,2)  + \gg1(2,3) + \gg1(3,4)\Big)
+ 3  g^{(3)}_{41}\,,\cr
\p V_3(1,2,3,4,5) &=
       \gg1(5,1)\Big(\gg1(1,2)\gg1(2,3)\gg1(3,4)
       + \gg1(1,2)\gg1(2,3)\gg1(4,5)
       + \gg1(1,2)\gg1(3,4)\gg1(4,5)
       + \gg1(2,3)\gg1(3,4)\gg1(4,5)
      + \gg1(1,2)\gg2(2,3)\cr
&\qquad{}+ \gg1(1,2)\gg2(3,4)
       + \gg1(1,2)\gg2(4,5)
      +\gg1(2,3)\gg2(1,2)
     + \gg1(2,3)\gg2(3,4)
      +\gg1(2,3)\gg2(4,5)
      +\gg1(3,4)\gg2(1,2)
           + \gg1(3,4)\gg2(2,3)\cr
&\qquad{}
      + \gg1(3,4)\gg2(4,5)
      + \gg1(4,5)\gg2(1,2)
     + \gg1(4,5)\gg2(2,3)
      + \gg1(4,5)\gg2(3,4)
      + \gg3(1,2)
      + \gg3(2,3)
      + \gg3(3,4)
     + \gg3(4,5)\Big)\cr
&\quad{}      + 2\gg2(5,1)\Big(\gg1(1,2)\gg1(2,3)
       + \gg1(1,2)\gg1(3,4)
       + \gg1(1,2)\gg1(4,5)
       + \gg1(2,3)\gg1(3,4)
       + \gg1(2,3)\gg1(4,5)
       + \gg1(3,4)\gg1(4,5)\cr
&\qquad{}  + \gg2(1,2)
       + \gg2(2,3)
       + \gg2(3,4)
      + \gg2(4,5)\Big)\cr
&\quad{}     + 3\gg3(5,1)\Big(\gg1(1,2)
       + \gg1(2,3)
       + \gg1(3,4)
       + \gg1(4,5)\Big)
       + 4\gg4(5,1)\,,
}$$
as one can check via  $\partial_\a F(z,\alpha) =-{1\over \a^2}+ \sum_{n=1}^\infty n\a^{n-1} g^{(n+1)}(z)$.
Alternatively, the expansions \delVs\ can be written using the definition
$V_p(\slash1,2,3, \ldots,\slash{q}) \equiv V_p(1,2,3,
\ldots,q)\big|_{g^{(k)}_{1q}\rightarrow0 }$
as $\p V_w(1, \ldots, n) = \sum_{p=1}^{w+1}
p\gg{p}(1,n)V_{w+1-p}(\slash1,2,3, \ldots,\slash{n})$ when $w=n{-}2$.

Note that the cyclicity of $V_w$ does not extend to the $\partial V_w$, but the
shuffle symmetry \VWshuff\ at $w=n{-}2$ reappears in a modified form:
\eqn\VWshuffA{
\partial V_{n-2}((1,2,\ldots,j) \shuffle (j{+}1,\ldots,n))=0\,, \qquad j=1,2,\ldots,n{-}1\,.
}
Also, the generating series \ellrepF\ immediately implies the reflection property
(valid for general $w$ and $n$)
\eqn\VWshuffnote{
\partial V_w(1,2,\ldots,n) =  (-1)^{w+1} \p V_w(n,\ldots,2,1)\,.
}

\subsubsec Higher-derivative extension of elliptic functions

By extending \ellrepF\ to involve higher derivatives in $\alpha$, we are led to
monodromy covariant functions $\partial^M V_w$ in
\eqn\highDA{
 F(z_{12},\alpha ) F(z_{23},\alpha ) \ldots F(z_{n-1,n},\alpha ) \partial^M_\a F(z_{n,1},\alpha )
 \equiv \sum_{w=-M}^{\infty} \a^{-n+w}  \partial^M V_w(1,2,\ldots,n)  \ ,
}
where $D \partial_\alpha^M F(z_{ij},\alpha) = \Omega_{ij} [\alpha \partial_\alpha^M F(z_{ij},\alpha) 
+ M  \partial_\alpha^{M-1} F(z_{ij},\alpha) ]$ implies that
\eqnn\highDB
$$\eqalignno{
D\partial^M V_w(1,2,\ldots,n) &= M \Omega_{n1} \partial^{M-1} V_w(1,2,\ldots,n)\,.&\highDB
}$$
The expansion of $\partial^M_\a F(z_{n,1},\alpha )$ in terms of $g^{(n)}_{ij}$ gives rise to
expressions such as
\eqnn\highDD
$$\eqalignno{
\partial^M V_0(1,2) &= M!  \, g^{(M)}_{21}\,&\highDD\cr
\p^2 V_1(1,2,3) & = 2 \gg2(3,1)\big(\gg1(1,2) + \gg1(2,3)\big) + 6\gg3(3,1)\cr
\p^2 V_2(1,2,3,4) &=
2\gg2(4,1)\big(\gg1(1,2)\gg1(2,3)
       + \gg1(1,2)\gg1(3,4)
       + \gg1(2,3)\gg1(3,4)
       + \gg2(1,2)
       + \gg2(2,3)
       + \gg2(3,4)
       \big)\cr
&\quad{}       + 6\gg3(4,1)\big(\gg1(1,2)
       + \gg1(2,3)
       + \gg1(3,4)\big)
       + 12\gg4(4,1)\,.
}$$
Again, the cyclic symmetry of $V_w$ is lost for $\partial^M V_w$ with $M\geq 1$, and
there is no analogue of the shuffle symmetries \VWshuff\ and \VWshuffA\ at $M\geq 2$.
Still, the reflection property in \dihed\ generalizes to
\eqn\dihedpV{
\partial^MV_w(1,2,\ldots,n) =  (-1)^{w+M} \p^M V_w(n,\ldots,2,1) \, .
}

\subsec Explicit examples of GEIs
\par\subseclab\seconeeight

In this section, we apply the elliptic functions $V_w$ and their derivative-extensions
$\partial^M V_w$ to cast the GEIs from the bootstrap procedure into compact form.
Given the trivial GEI $E_{1|2,3,4}=1$ at four points, the simplest example of the $V_w$-functions 
occurs at five points, where the GEIs \fiveEls\ and \fiveElls\ can be rewritten as
\eqn\startSix{
E_{1|23,4,\ldots}=V_1(1,2,3)\,,\qquad E^m_{1|2,3,4,\ldots}=\ell^m + \sum_{j\geq 2} k_j^m g^{(1)}_{1j} \,,
}
see \ellrepDex\ for $V_1$. Here and in the following, the number of
slots (i.e.\ the upper bound on the summation range for $j\geq 2$) is kept unspecified in order to
account for the extensions as in \sixEms.

\newsubsubsec\seconeeightA Six points

At six points, the definitions in \ellrepDex\ and \delVs\ can be used to condense the
scalars and the vector GEI in \Eswithgs\ to
\eqnn\rewrtB
$$\eqalignno{
E_{1|234,5,\ldots} &= V_2(1,2,3,4)\,,  &\rewrtB \cr
E_{1|23,45,\ldots} &= V_1(1,2,3)V_1(1,4,5)\,, \cr
E^m_{1|23,4,5,\ldots} &= \Big( \ell^m +\sum_{j\geq 4} k_j^m g^{(1)}_{1j}
 \Big)V_1(1,2,3)+ k_2^m \partial V_1(2,3,1) - k_3^m \partial V_1(3,2,1)\,, \cr
E^{mn}_{1|2,3,4,5,\ldots} &= \ell^m \ell^n + \sum_{j\geq 2} \ell^{(m} k_j^{n)} g_{1j}^{(1)}
 +2 \sum_{j\geq 2} k_j^m k_j^{n} g_{1j}^{(2)} + \sum_{2\leq i<j} k_{i}^{(m} k_j^{n)} g_{1i}^{(1)}g_{1j}^{(1)}\, ,
}$$
where the unspecified summation range automatically accounts for the extensions
in \sixExt. With the covariant monodromy variation \ellrepH\ of $\partial V_1$ at hand, it is
easy to verify that $DE^m_{1|23,4,5,\ldots} =0$. One may identify the above $g_{1j}^{(1)}$
and $g_{1j}^{(2)}$ as $-\partial V_0(1,j)$ and ${1\over 2}\partial^2 V_0(1,j)$, respectively,
to arrive at a uniform presentation for the coefficients of $k_j^m$.

In view of $\partial_1 g^{(1)}_{12} =V_2(1,2) -  {\rm G}_2$, the
refined GEI \SixErefined\ is also expressible in terms of elliptic functions
\eqnn\GtwoA
$$\eqalignno{
E_{1|2|3,4,5,\ldots} &=  (1-s_{12}) V_2(1,2) - {\rm G}_2 &\GtwoA  \cr
&\cong - 2 s_{12} g^{(2)}_{12} + g^{(1)}_{12} \Big(\ell\cdot k_2 + \sum_{j\geq 3} s_{2j}g^{(1)}_{2j} \Big)  \, ,
}$$
where the last line again follows from integration by parts.

\newsubsubsec\seconeeightB Seven points

At seven points, the scalar GEIs in \sevenM\ can be compactly written as
\eqnn\muchsimpler
$$\eqalignno{
E_{1|2345,6,7,\ldots} &=V_3(1,2,3,4,5)\,, &\muchsimpler\cr
E_{1|234,56,7,\ldots} & = V_2(1,2,3,4)V_1(1,5,6)\,,\cr
E_{1|23,45,67,\ldots} &= V_1(1,2,3)V_1(1,4,5)V_1(1,6,7)\,,
}$$
and the vectors \sevenP\ simplify as well when expressed in
terms of $V_w$- and $\partial V_w$-functions,
\eqnn\muchsh
$$\eqalignno{
E^m_{1|234,5,6,\ldots}&=\Big(\ell^m+  \sum_{j\geq 5} g^{(1)}_{1j} k^m_j \Big) V_2(1, 2, 3, 4)  + 
 k_2^m \p V_2(2, 3, 4,1) &\muchsh\cr
 &+ k_4^m \p V_2( 4, 3, 2,1) - 
 k_3^m \big[ \p V_2( 3, 2, 4,1) + \p V_2(3, 4, 2,1)\big]\,,\cr
E^m_{1|23,45,6,\ldots}&= \Big(\ell^m+  \sum_{j\geq6} g^{(1)}_{1j} k^m_j \Big)  V_1(1, 2, 3) V_1(1, 4, 5) \cr
&  + V_1(1, 4, 5)  \big[k_2^m \p V_1( 2, 3,1) - k_3^m \p V_1( 3, 2,1) \big] \cr
&  + V_1(1, 2, 3)  \big[ k_4^m \p V_1( 4, 5,1) - k_5^m \p V_1( 5, 4,1) \big] \, .\cr
}$$
Similarly, the $\partial^2 V_w$-functions in \highDA\ allow for compact representations
of the two- and three-tensors in \sevenS\foot{Note that
our conventions lead to $\ell^{(m} \ell^n k_j^{p)} = \ell^{m} \ell^n k_j^{p}
+\ell^{m} \ell^p k_j^{n}+\ell^{n} \ell^p k_j^{m}$.},
\eqnn\rewrtC
$$\eqalignno{
E^{mn}_{1|23,4,5,\ldots}&=\Big( \ell^m\ell^n
+ \sum_{j \geq 4} g^{(1)}_{1j} \ell^{(m} k^{n)}_j \Big) V_1(1, 2, 3)  &\rewrtC\cr
&+   \p V_1( 2, 3,1) \Big( \ell^{(m} k^{n)}_2 + \sum_{j\geq4} k_2^{(m} k_j^{n)} 
g^{(1)}_{1j} \Big)
+ \p^2 V_1( 2, 3,1) k_2^m k_2^n \cr
&-   \p V_1(3, 2,1) \Big( \ell^{(m} k^{n)}_3 + 
\sum_{j\geq4} k_3^{(m} k_j^{n)}  g^{(1)}_{1j} \Big)  
 - \p^2 V_1( 3,2,1) k_3^m k_3^n \cr
&+ 2V_1(1,2,3) \sum_{j\geq4} k_j^m k_j^n  g^{(2)}_{1j} 
+ V_1(1,2,3) \sum_{4\leq i<j} k_i^{(m} k_j^{n)}  g^{(1)}_{1i} g^{(1)}_{1j} \cr
&+ {1\over 2} k_2^{(m} k_3^{n)} \big[\p^2 V_1(1, 2, 3) - \p^2 V_1(1, 3, 2)
+ \p^2 V_1(2, 1, 3) \big] \, ,\cr
E^{mnp}_{1|2,3,4,5,\ldots}&=\ell^m\ell^n\ell^p+ \sum_{j\geq2} \ell^{(m} \ell^n k_j^{p)}g^{(1)}_{1j}  
+\sum_{2\leq i<j} \ell^{(m} k_i^n k_j^{p)}  g^{(1)}_{1i} g^{(1)}_{1j} \cr
&+2 \sum_{j\geq2}
\ell^{(m} k_j^n k_j^{p)} g^{(2)}_{1j}  +\sum_{2\leq i<j<l} k_i^{(m} k_j^n k_l^{p)} 
g^{(1)}_{1i} g^{(1)}_{1j}  g^{(1)}_{1l}  \cr
&+6 \sum_{j\geq2} k_j^m k_j^n k_j^p  g^{(3)}_{1j}
+2\sum_{2\leq i<j} \big[ k_i^{(m} k_i^n k_j^{p)}  g^{(2)}_{1i} g^{(1)}_{1j}  +(i\leftrightarrow j) \big]\,.
}$$
Using the monodromy variation \highDB\ of $\partial^M V_w$ and the shuffle
symmetries \VWshuff\ and \VWshuffA\ of $V_{n-2}$ and $\partial V_{n-2}$, all the
above $E^{\ldots}_{1|\ldots}$ can be verified to be GEIs with pen-and-paper
effort. Similarly, one can show that the refined GEIs \rewrtDalt\ whose combinatorics mimic
the Berends--Giele expansion of the refined superfields $P$ from \partI\ can be
rewritten more compactly as
\eqnn\rewrtD
$$\eqalignno{
E_{1|23|4,5,\ldots}  &= - s_{123} V_3(1,2,3) +(g_{12}^{(1)}+g_{31}^{(1)}) \p g^{(1)}_{23}
+ \p g^{(2)}_{23} &\rewrtD\cr
E_{1|4|23,5,\ldots}  &= \big[\p g^{(1)}_{14} - s_{14} V_2(1,4) \big]V_1(1,2,3)
- s_{24} V_3(1,2,4) + s_{34} V_3(1,3,4)
\cr
E^m_{1|2|3,4,5,\ldots}  &= \big[\p g^{(1)}_{12} - s_{12} V_2(1,2) \big]
\Big(\ell^m + \sum_{j \geq 3} k_j^m g^{(1)}_{1j} \Big)
+ \sum_{j \geq 3} k_j^m s_{2j} V_3(1,2,j)\cr
& \ \ \ \ + k_2^m \big[ \p g^{(2)}_{12} + s_{12}(g_{12}^{(1)} g_{12}^{(2)} - 3 g^{(3)}_{12})\big]\,.
}$$
Alternatively, using integration-by-parts identities leads to
\eqnn\rewrtE
$$\eqalignno{
E_{1|23|4,5,\ldots}  &=  \Big(g^{(1)}_{12}g^{(1)}_{23}+ \tfrac12 (g^{(2)}_{12}+g^{(2)}_{23}) \Big)
\Big(\ell\cdot k_3 + \sum_{j\geq 4} s_{3j} g^{(1)}_{3j} \Big)&\rewrtE  \cr
&- \Big(g^{(1)}_{13}g^{(1)}_{32}+ \tfrac12 (g^{(2)}_{13}+g^{(2)}_{32}) \Big)
\Big(\ell\cdot k_2 + \sum_{j \geq 4} s_{2j} g^{(1)}_{2j} \Big) \cr
&- \Big( s_{23} \big[3 g^{(3)}_{23} +  2  g^{(2)}_{23} (g^{(1)}_{12}+g^{(1)}_{31}) 
+ \tfrac12 g^{(1)}_{23}(g_{12}^{(2)} + g_{13}^{(2)})\big]
   + {\rm cyc}(1,2,3) \Big)\,,\cr
E_{1|4|23,5,\ldots}  &=
V_1(1,2,3) \Big[ g_{14}^{(1)} \big(\ell \cdot k_4 - s_{24} g^{(1)}_{24} 
- s_{34} g^{(1)}_{34}
+ \sum_{j \geq 5}s_{4j} g^{(1)}_{4j} \big)  - 2s_{14} g^{(2)}_{14} \Big]  \cr
& - s_{24} V_3(1,2,4) + s_{34} V_3(1,3,4)\,,\cr
E^m_{1|2|3,4,5,\ldots}  &= \Big(\ell^m + \sum_{j\geq 3} k_j^m g_{1j}^{(1)} \Big)
\Big[g_{12}^{(1)} \big(k_2\cdot \ell + \sum_{l \geq 3} s_{2l}g^{(1)}_{2l} \big)
-2 s_{12}g^{(2)}_{12} \Big] \cr
&+ \sum_{j\geq3} k_j^m s_{2j} V_3(1,2,j)  + k_2^m
\Big[g_{12}^{(2)}\big( k_2\cdot \ell +  \sum_{l\geq3} s_{2l}g^{(1)}_{2l}\big)
-3 s_{12}g^{(3)}_{12} \Big]\,.
}$$
Instead of \rewrtD, one can also use $\partial_1 g^{(1)}_{12} =V_2(1,2) -  {\rm G}_2$ and
$\partial_1 g^{(2)}_{12} = 3 g^{(3)}_{12} - g^{(1)}_{12}g^{(2)}_{12} - {\rm G}_2 g^{(1)}_{12}$ to write
\eqnn\GtwoB
$$\eqalignno{
E_{1|23|4,5,\ldots} &=  (1-s_{123}) V_3(1,2,3)- {\rm G}_2 V_1(1,2,3) \, , &\GtwoB
}$$
and the analogous identities for $z_j$-derivatives of general $g^{(n)}_{ij}$-functions 
read\foot{This follows from the expansion of
$(\p_z-\p_\alpha) F(z,\alpha,\tau)= (g^{(1)}(\alpha,\tau ) - g^{(1)}(z,\tau ) )F(z,\alpha,\tau)$.}
\eqn\zderi{
\p_z g^{(n)}(z,\tau) = (n{+}1) g^{(n+1)}(z,\tau)
- g^{(1)}(z,\tau) g^{(n)}(z,\tau)  - \sum_{k=2}^{n+1} {\rm G}_k
\,g^{(n+1-k)}(z,\tau)\,.
}

\newsubsec\seconeclosed Closed all-multiplicity formulae for GEIs

\newsubsubsec\seconeeightC Scalars at all multiplicities

The above examples of scalar GEIs in  \startSix, \rewrtB\ and \muchsimpler\ line up with
\eqn\newEA{
E_{1|A,B,C} = V_{|A|-1}(1,A)V_{|B|-1}(1,B)V_{|C|-1}(1,C)\, .
}
Given that all the $V_w(1,2,\ldots,n)$-functions on the right-hand side have $w=n{-}2$,
the GEIs in \newEA\ exhibit the desired shuffle symmetry in each slot by \VWshuff. Although
only the functions \newEA\ with three multiparticle slots enter open-string amplitudes,
the later discussion will benefit from
an extension to unspecified numbers of slots,
\eqn\morenewEA{
E_{1|A_1,A_2,\ldots} = \prod_{j\geq 1} V_{|A_j|-1}(1,A_j)\,.
}

\newsubsubsec\seconeeightE Closed formulae for vectors and two-tensors

The above examples of vector and two-tensor GEIs can be lined up with the closed formulae
\eqnn\newEOcl
$$\eqalignno{ 
 E^m_{1|A,B,C,\ldots} &=\ell^m 
V_{|A|-1}(1,A)V_{|B|-1}(1,B)V_{|C|-1}(1,C) \ldots  \cr
&+ \Big[ 
\sum_{j=1}^{|A|} (-1)^{j-1} k^m_{a_j}   \p V_{|A|-1}(a_j,( a_{j-1} \ldots a_2a_1 \shuffle a_{j+1} \ldots a_{|A|} ) ,1) \cr
& \ \ \ \ \ \times
V_{|B|-1}(1,B)V_{|C|-1}(1,C)\ldots +(A\leftrightarrow B,C,\ldots) \Big]
 &\newEOcl
    }$$
as well as
\eqnn\newEOtcl
$$\eqalignno{ 
& E^{mn}_{1|A,B,C,\ldots} =  \ell^m  \ell^n
V_{|A|-1}(1,A)V_{|B|-1}(1,B)V_{|C|-1}(1,C)\ldots  \cr
&\ \ + \Big[ 
\sum_{j=1}^{|A|} (-1)^{j-1}  \ell^{(m}k_{a_j}^{n)}  \p V_{|A|-1}(a_j,( a_{j-1} \ldots a_2a_1 \shuffle a_{j+1} \ldots a_{|A|} ) ,1) \cr
& \ \ \ \ \ \times
V_{|B|-1}(1,B)V_{|C|-1}(1,C)\ldots +(A\leftrightarrow B,C,\ldots) \Big]
 \cr
& \ \ +   \Big[
\sum_{i=1}^{|A|} \sum_{j=1}^{|B|}   (-1)^{i+j}  k_{a_i}^{(m}k_{b_j}^{n)}\p V_{|A|-1}(a_i,( a_{i-1} \ldots a_2a_1 \shuffle a_{i+1} \ldots a_{|A|} ) ,1) \cr
&  \ \ \ \ \ \times \p V_{|B|-1}(b_j,( b_{j-1} \ldots b_1 \shuffle b_{j+1} \ldots b_{|B|} ) ,1) \cr
&  \ \ \ \ \ \times V_{|C|-1}(1,C)V_{|D|-1}(1,D)\ldots +(A,B| A,B,C,D\ldots) \Big]  &\newEOtcl \cr
& \ \ +{1\over 2} \Big[  k_A^{(m}
\sum_{j=1}^{|A|} k_{a_j}^{n)}  (-1)^{j-1}     \p^2 V_{|A|-1}(a_j,( a_{j-1} \ldots a_2a_1 \shuffle a_{j+1} \ldots a_{|A|} ) ,1) \cr
&  \ \ \ \ \ \times
V_{|B|-1}(1,B)V_{|C|-1}(1,C) \ldots +(A\leftrightarrow B,C,\ldots) \Big]
\cr
&\ \ -{1\over 2} \Big[  V_{|B|-1}(1,B)V_{|C|-1}(1,C)\ldots
\sum_{1=i<j}^{|A|} (-1)^{i+j+|A|}  k_{a_i}^{(m} k_{a_j}^{n)}   \cr
&  \ \ \ \ \ \times \p^2 V_{|A|-1}(a_i,( a_{i-1} \ldots a_2a_11 a_{|A|} \ldots a_{j+1} \shuffle a_{i+1}a_{i+2} \ldots a_{j-1} ) ,a_j)
+(A\leftrightarrow B,C,\ldots) \Big] \, .
}$$
Up to multiplicity seven, the complete set of unrefined GEIs is accessible from
the above closed formulae and \rewrtC. At higher tensor rank, the system of $\p^M
V_w(1,2,\ldots,n) $-functions in \highDA\ is no longer sufficient to represent
the coefficients of $k_i^{(m} k_j^{\phantom{(}\!\! \!n} \,k_l^{p)}$ and
higher-rank terms. This shortcoming motivates the
development of more powerful tools for all-multiplicity and all-rank
constructions of GEIs, which we leave for a future work.

\newnewsec\secthree Integrating the loop momentum and modular invariance

The purpose of this section is to set the stage for integrating the one-loop correlators
of part III over the loop momentum. We will see below that loop-integrated GEIs 
yield manifestly single-valued worldsheet functions that largely conspire to modular
weight $(n{-}4,0)$. The loop integrals of individual GEIs at $(n{\geq} 6)$ points also 
feature terms of different modular weights that (as will be shown in part III) cancel 
from the amplitude by kinematic identities among their coefficients. Such 
modular anomalies will be illustrated to follow the patterns of BRST anomalies of 
pseudo-invariants. Like this, we extend the duality between worldsheet functions and kinematics to anomalies.

\subsec The non-holomorphic Kronecker--Eisenstein series

As detailed in section \EKsec, the meromorphic constituents $g^{(n)}(z,\tau)$ 
of the chirally-split open-string correlators ${\cal K}_n(\ell)$ descend from
the Kronecker--Eisenstein series \EisKron. The doubly-periodic counterparts
of $g^{(n)}(z,\tau)$ that will result from loop integration can be generated
from the non-holomorphic completion \BrownLevin,
\eqnn\NHKron
$$\eqalignno{
\Omega(z,\alpha,\tau) &\equiv 
e^{ 2\pi i  \alpha{\Im z \over \Im \tau} } F(z,\alpha,\tau)
 \equiv \sum_{n=0}^{\infty} \alpha^{n-1} f^{(n)}(z,\tau)\,, &\NHKron
}$$
where the exponential factor is tailored to cancel the $B$-cycle monodromies \shift,
\eqnn\doubleper
$$\eqalignno{
f^{(n)}(z,\tau)&=f^{(n)}(z+1,\tau) =f^{(n)}(z+\tau,\tau)  &\doubleper\cr
  \Omega(z,\alpha,\tau) &=\Omega(z+1,\alpha,\tau) =\Omega(z+\tau,\alpha,\tau)\,.
}$$
 The doubly-periodic but non-holomorphic functions $f^{(n)}$ in \NHKron\ are related to the
holomorphic $g^{(n)}$ with $B$-cycle monodromies \gnmonodromy\ via \BroedelVLA
\eqn\lopF{
f^{(n)}(z,\tau) \equiv \sum_{k=0}^n {\nu^k \over k!}
g^{(n-k)}(z,\tau)\,,\quad
\nu \equiv 2\pi i {\Im z \over \Im \tau}\,,
}
where the simplest examples are $f^{(0)}=1$ and
\eqn\lopH{
f^{(1)}(z,\tau) = g^{(1)}(z,\tau) + \nu\,, \quad
f^{(2)}(z,\tau) = g^{(2)}(z,\tau) + \nu g^{(1)}(z,\tau) + \half\nu^2\,.
}
Apart from double-periodicity, the non-holomorphic Kronecker--Eisenstein series
and the functions $f^{(n)}$ exhibit covariant modular transformations with holomorphic
weights $(1,0)$ and $(n,0)$, respectively, \Zagier
\eqnn\modKron
$$\eqalignno{
\Omega\left({z\over c\tau+d},{\alpha \over c\tau+d},{a\tau + b\over c\tau + d}\right)
& = (c\tau+d) \, \Omega(z,\alpha,\tau)\,, &\modKron \cr
f^{(n)}\left({z\over c\tau+d},{a\tau + b\over c\tau + d}\right)
&= (c\tau+d)^n \,  f^{(n)}(z,\tau)\,,
}$$
where $a,b,c,d$ form an ${\rm SL}_2(\Bbb Z)$ matrix. Similarly,
each holomorphic derivative in $z$ adds holomorphic weight $(1,0)$ to the $f^{(n)}$.
However, meromorphicity of the $g^{(n)}$ is replaced by the condition
\eqn\quasihol{
\Big( {\p \over \p \bar \tau} + {\Im z \over \Im \tau }
{\p \over \p \bar z} \Big) f^{(n)}(z,\tau) = 0
}
following from
\eqn\nonhol{
{\p \over  \p \bar z} f^{(n)}(z,\tau) =
-{\pi \over \Im\tau} f^{(n-1)}(z,\tau)\,,
\qquad {\p\over\p\bar \tau} f^{(n)}(z,\tau) =
{\pi \Im z \over (\Im \tau)^2}  f^{(n-1)}(z,\tau)\,.
}
It will be convenient to extend the shorthand notation \gij\ 
for $g^{(n)}_{ij}$ to their doubly-periodic counterparts,
\eqn\fij{
f^{(n)}_{ij} \equiv f^{(n)}(z_i-z_j,\tau)\,,
}
which we will use from now on. The Fay identity \FayKron\ of the Kronecker--Eisenstein series is
unchanged when replacing $F(\ldots) \rightarrow \Omega(\ldots)$.
Accordingly, the relations \Faygn\ to rearrange products $g^{(n)}_{12} g^{(m)}_{23}$
also hold when globally trading $g^{(n)}_{ij} \rightarrow f^{(n)}_{ij}$. For instance, the
simplest examples \lopH\ of $f^{(n)}$ satisfy the analogue
$f^{(1)}_{12} f^{(1)}_{23}+f^{(2)}_{12}+{\rm cyc}(1,2,3)=0$ of \lopEa.

\newsubsec\intellmom Integrating out the loop momentum

In this section, we set the stage for loop integrals over both the Koba--Nielsen
factor
\eqnn\modsqnew
$$\eqalignno{
\big|{\cal I}_n(\ell) \big|^2 = \exp \Big( \sum^n_{i<j}s_{ij}  \Big\{ &\log\big| \theta_1(z_{ij},\tau)  \big|^2  - {i\pi \over \tau {-} \bar \tau} \Big[ \sum_{j=1}^n k_j(z_j{-}\bar z_j) \Big]^2\cr
& \ \ \ + {\tau{ -} \bar \tau \over 4\pi i} \Big[
\ell  + 2\pi i \sum_{j=1}^{n} k_j {z_j {-} \bar z_j \over \tau {-} \bar \tau}
\Big]^2  \Big\}\Big) &\modsqnew
}$$
and $\ell$-dependent open- and closed-string correlators in
the amplitudes \theamphere\ and \theclosedamphere. For closed-string correlators independent
on $\ell$, the result of the Gaussian loop integral
\eqn\notationellA{
\hat {\cal I}_n \equiv \int d^D \ell \ \big|{\cal I}_n(\ell) \big|^2 
 = {(2\pi i)^D\over (2 \Im \tau)^{{D\over 2}}}\exp \Big(
 \sum^n_{i<j}s_{ij} \Big[ \log\big| \theta_1(z_{ij},\tau)  \big|^2
 - {2\pi \over \Im \tau} (\Im z_{ij})^2 \Big]\Big)
}
has already been spelled out in \KNKNB. Zero-mode integration at
$n\geq 5$ points, however, requires generalizations of \notationellA\ to additional
polynomials $p(\ell)$ in the loop momentum besides $|{\cal I}_n(\ell) |^2$. We will 
use the square-bracket notation
\eqn\notationellB{
\int d^D \ell \ \big|{\cal I}_n(\ell)  \big|^2 \, p(\ell) =  \hat {\cal I}_n \, [[ p(\ell) ]]
}
to compactly address the net effect $[[ p(\ell) ]]$ of the shifts in the Gaussian
integration variable in \modsqnew. The right-hand side of \notationellB\ is normalized to $[[1]]=1$,
and the loop integrals over polynomials in $\ell$ are most
conveniently written in terms of the shorthands
\eqn\earlynu{
\nu_{ij} \equiv 2\pi i {\Im z_{ij} \over \Im \tau}\, ,\quad
L_{0}^m\equiv  - \sum_{j=1}^n k_j^m \nu_j = \sum_{j=2}^n k_j^m\nu_{1j} \, ,
}
where momentum conservation has been used to 
eliminate $k^m_1 = - k^m_2 - \cdots - k^m_n$ from the definition of $L_0^m$.
As a result of straightforward Gaussian integration, we have (recall the
convention \controversy\ where all terms generated by (anti)symmetrization
of indices have unit coefficient, e.g., $\d^{(mn}k^{p)}\equiv \d^{mn}k^p + \d^{mp}k^n + \d^{np}k^m$)
\eqnn\KNKNC
$$\eqalignno{
 [[\ell^m]] &= L_{0}^m\,, &\KNKNC \cr
 [[\ell^m \ell^n ]] &=L_{0}^m L_{0}^n -{\pi\over \Im \tau} \delta^{mn}\,, \cr
 [[\ell^m \ell^n \ell^p ]] &=L_{0}^m L_{0}^n L_0^p -{ \pi\over \Im \tau}
 \delta^{(mn} L_0^{p)}\,, \cr
 [[\ell^m \ell^n \ell^p \ell^q ]] &=L_{0}^m L_{0}^n L_0^p L_0^q
  -{ \pi\over \Im \tau} \delta^{(mn} L_0^{p} L_0^{q)}
  + \Big( { \pi \over \Im \tau} \Big)^2 \delta^{m(n} \delta^{pq)}\,,
}$$
which are sufficient to integrate open-string correlators at $n\leq 8$ points
and closed-string correlators at $n\leq 6$ points. In general, following standard Gaussian
integration rules, one has to sum over all possibilities to perform pairwise contractions
$\ell^m \ell^n \rightarrow -{\pi \over \Im \tau} \delta^{mn}$ on a subset of the loop momenta
in the integrand while setting the others to $\ell^m \rightarrow L_0^m$.

The open-string analogue of \notationellB\ reads
\eqn\notationopen{
\int d^D \ell \ \big|{\cal I}_n(\ell)\big| \, p(\ell) = \hat {\cI}^{\rm open}_n \, [[ p(\ell) ]]\,,
}
where $\hat {\cal I}^{\rm open}_n$ is defined in \KNKNBop, and one can take advantage of 
the same expressions \KNKNC\ for $[[ p(\ell) ]]$ that apply to the closed string. The imaginary parts
in \earlynu\ then ensure that the results \KNKNC\ can be specialized to all the open-string
topologies by suitable choices of the integration domains for $z_j$ and $\tau$.

In summary, \notationellB\ and \notationopen\ are tailored to express the open-
and closed-string amplitudes \theamphere\ and \theclosedamphere\ in the following form
\eqnn\newamprep
$$\eqalignno{
{\cal A}_n  &=
\sum_{\rm top} C_{\rm top} \int_{D_{\rm top}}\!\!\!\! 
d\tau \, dz_2 \, d z_3 \, \ldots \, d z_{n} \,  \hat {\cal I}^{\rm open}_n \, [[ \, \langle {\cal K}_n(\ell)  \rangle \, ]] \, ,
&\newamprep
\cr
{\cal M}_n  &=
 \int_{{\cal F}}
d^2\tau \, d^2z_2 \, d^2 z_3 \, \ldots \, d^2 z_{n} \, \hat {\cal I}_n \, [[ \,
\langle {\cal K}_n(\ell)\rangle \, \langle\tilde{\cal K}_n(-\ell)\rangle \, ]]
\,,
}$$
where all the remnants of the loop momenta in the correlators are captured by the
Gaussian brackets $[[\ldots]]$ exemplified in \KNKNC. In the remainder of this section, 
we will evaluate $[[E_{1|\ldots}]]$ for various GEIs and elaborate on the modified 
integration-by-parts rules adapted to \notationellA\ instead of $|{\cal I}_n(\ell)|^2$.
This will be applied in section \loopintsec\ to provide manifestly single-valued expressions 
for open- and closed-string correlators $[[ {\cal K}_n(\ell)  ]]$ and $[[ {\cal K}_n(\ell) 
\tilde{\cal K}_n(-\ell)]]$.

\subsubsec Integrating unrefined GEIs

In section \GEIsec, GEIs $E^{\ldots}_{1|\ldots}$ have been introduced as meromorphic
functions that are doubly-periodic up to shifts of the loop momentum. Hence, upon
integration over $\ell$, GEIs are guaranteed to become doubly-periodic, and the
functions $f^{(n)}$ in \NHKron\ turn out to be the natural framework to represent the
dependence of $[[E^{\ldots}_{1|\ldots}]]$ on $z_j$.

Unrefined scalar GEIs $E_{1|A,B,C}$ were found to be elliptic functions in the conventional
sense and expressible in terms of the $V_w$-functions of \ellrepD, see e.g.\ \startSix\ and \rewrtB.
Given that the generating series \ellrepD\ of $V_w$ are unchanged when the Kronecker--Eisenstein
series are replaced by their doubly-periodic completions \NHKron, one can globally replace
$g^{(n)} \rightarrow f^{(n)}$ in any $V_w$, and in fact, in any $E_{1|A,B,C}$. For instance, all
the imaginary parts $\nu_{ij}$ of \earlynu\ cancel out from
\eqn\Vwithfs{
V_1(1,2,\ldots,n) = \sum_{j=1}^n f^{(1)}_{j,j+1}\,, \qquad
V_2(1,2,\ldots,n) = \sum_{j=1}^n f^{(2)}_{j,j+1} + \sum_{1\leq i<j}^n f^{(1)}_{i,i+1} f^{(1)}_{j,j+1}\,,
}
which gives rise to $[[E_{1|2,3,4}]]=1$ and
\eqnn\scalarGEIf
$$\eqalignno{
[[E_{1|23,4,5}]] &= V_1(1,2,3)= f^{(1)}_{12}+ f^{(1)}_{23}+ f^{(1)}_{31}  &\scalarGEIf\cr
[[E_{1|234,5,6}]] &= V_2(1,2,3,4)= f^{(1)}_{12} f^{(1)}_{34}+ f^{(1)}_{23} f^{(1)}_{41}+ \big[ f^{(1)}_{12} f^{(1)}_{23}+ f^{(2)}_{12} + {\rm cyc}(1,2,3,4) \big]   \cr
[[E_{1|23,45,6}]] &= V_1(1,2,3) V_1(1,4,5)= (f^{(1)}_{12}+ f^{(1)}_{23}+ f^{(1)}_{31})(f^{(1)}_{14}+ f^{(1)}_{45}+ f^{(1)}_{51})  \, .\cr
}$$
Similarly, the all-multiplicity formula \newEA\ for unrefined scalar GEIs generalizes to 
\eqn\newEAagain{
[[E_{1|A,B,C}]] = V_{|A|-1}(1,A)V_{|B|-1}(1,B)V_{|C|-1}(1,C)\, .
}
The $[[\ldots]]$ have no effect on these $\ell$-independent functions but have been included
into \scalarGEIf\ and \newEAagain\ to harmonize with the examples below. 

For vectorial and tensorial GEIs, the loop momenta integrate to polynomials in $\nu_{ij}$ as a result
of the Gaussian brackets in \KNKNC. In order to manifest the double-periodicity of $[[E^{m_1\ldots}_{1|A,B,\ldots}]]$,
these factors of $\nu_{ij}$ can be combined with the meromorphic functions $g^{(n)}_{ij}$ to obtain their
doubly-periodic completion $f^{(n)}_{ij}$. Based on the conversion \lopF\ between $g^{(n)}_{ij}$ and $f^{(n)}_{ij}$
as well as the expressions for the GEIs in \fiveElls\ and \Eswithgs, we find
\eqnn\intellA
$$\eqalignno{
[[E^m_{1|2,3,4,5}]] &= k^m_2 f^{(1)}_{12} + (2\leftrightarrow 3,4,5)\,,&\intellA\cr
[[E^m_{1|23,4,5,6}]] &= k^m_3 f^{(1)}_{12} f^{(1)}_{23} +k^m_2 f^{(1)}_{13} f^{(1)}_{23} + \big[
k_4^m f^{(1)}_{14}(f^{(1)}_{23} + f^{(1)}_{12} + f^{(1)}_{31})
+(4\leftrightarrow 5,6) \big]\cr
&\quad{} + k_{23}^m (f^{(2)}_{12} - f^{(2)}_{13}) + (k_3^m - k_2^m)f^{(2)}_{23}\,, \cr
[[E^{mn}_{1|2,3,4,5,6}]] &= - {\pi\over\Im\tau}\delta^{mn}
+ 2 \big[ k_2^m k_2^n f^{(2)}_{12} + (2\leftrightarrow 3,4,5,6) \big]\cr
&\quad{} + \big[ (k_2^{m} k_3^{n}{+}k_2^{m} k_3^{n}) f^{(1)}_{12} f^{(1)}_{13} + (2,3|2,3,4,5,6)\big]\,,
}$$
and higher-multiplicity results will be given below.

\subsubsec Modular anomalies

By the modular weight $(w,0)$ of $f^{(w)}_{ij}$, see \modKron, almost all of the
examples \scalarGEIf\ to \intellA\ of integrated $n$-point GEIs are modular
forms of weight $(n{-}4,0)$. The only exception is the first term $-
{\pi\over\Im\tau}\delta^{mn}$ of modular weight $(1,1)$ in the expression
\intellA\ for the tensor $[[E^{mn}_{1|2,3,4,5,6}]]$ whose remaining
terms $f^{(2)}_{ij}$ and $f^{(1)}_{ij} f^{(1)}_{kl}$ carry weight $(2,0)$.
Accordingly, contributions to $[[E^{m_1m_2\ldots}_{1|A,B,\ldots}]]$ at $n$
points that depart from modular weight $(n{-}4,0)$ are referred to as a {\it
modular anomalies}, the simplest example being the above $-
{\pi\over\Im\tau}\delta^{mn}$.

For unrefined GEIs, modular anomalies can be conveniently traced back to
contractions $\ell^m \ell^n \rightarrow -{\pi \over \Im \tau} \delta^{mn}$, so
they only arise at tensor rank $r{\geq} 2$ (the situation for refined GEIs
is different, see section \moreanom). Scalar GEIs $[[E_{1|A,B,C}]] = E_{1|A,B,C}$ 
reduce to elliptic $V_w$-functions of weight $(w,0)$, and the integral $[[E^m_{1|A,B,C,D}]] $ over vector
GEIs follows from setting $\ell^m \rightarrow 0$ and $g^{(n)}_{ij} \rightarrow
f^{(n)}_{ij}$, see e.g.\ \intellA. The modular anomalies of the tensorial
seven-points GEIs \rewrtC\ are the contributions $\sim  {\pi\over\Im\tau}$ in
\eqnn\newEOx
$$\eqalignno{
[[E^{mn}_{1|23,4,5,6,7}]] &=  - {\pi\over\Im\tau}\delta^{mn} V_{1}(1,2,3)
+ 2  V_1(1,2,3) \big[ k_4^m k_4^n  f^{(2)}_{14} + (4\leftrightarrow 5,6,7)\big] &\newEOx \cr
&+ V_1(1,2,3) \big[ k_4^{(m} k_5^{n)}  f^{(1)}_{14} f^{(1)}_{15} + (4,5|4,5,6,7) \big] \cr
&+
\Big( \big[  k_2^{(m} k_4^{n)}  f^{(1)}_{14}+(4\leftrightarrow 5,6,7)\big] 
\big[ 2 f^{(2)}_{12} + f^{(1)}_{12}(f^{(1)}_{23}  {+} f^{(1)}_{31} )  \big]- (2\leftrightarrow 3) \Big) \cr
&+\Big( k_2^m k_2^n \big[ 6 f^{(3)}_{12} + 2 f^{(2)}_{12}(f^{(1)}_{23}{+}f^{(1)}_{31}) \big]
 -  (2\leftrightarrow 3)  \Big)
\cr
&+   k_2^{(m} k_3^{n)} \big[ 2 f^{(3)}_{12} + 2 f^{(3)}_{31} - f^{(3)}_{23} + f^{(1)}_{23} (f^{(2)}_{12}{+}f^{(2)}_{13}) \big] \ .
 \cr
[[E^{mnp}_{1|2,3,4,\ldots,7}]] &= - {\pi\over\Im\tau}\delta^{(mn} \big[ k_2^{p)} f^{(1)}_{12} +(2\leftrightarrow3,4,\ldots,7)\big]
+ 6 \big[ k_2^m k_2^n k_2^p f^{(3)}_{12} + (2\leftrightarrow3,\ldots,7)\big] \cr
&  + 2 \big[ k_2^{(m} k_2^n k_3^{p)} f^{(2)}_{12} f^{(1)}_{13}
+k_2^{(m} k_3^n k_3^{p)} f^{(1)}_{12} f^{(2)}_{13}
+(2,3|2,3,\ldots,7)\big] 
 \cr
 & +  \big[ k_2^{(m}k_{3}^{n}k_4^{p)}f^{(1)}_{12}f^{(1)}_{13}f^{(1)}_{14}+
(2,3,4|2,3,\ldots,7)\big] \, .
}$$
Before pointing out analogous modular anomalies in the loop
integrals of refined GEIs, we shall elaborate on the integration-by-parts
relations relevant to the results for $[[E^{m_1\ldots}_{1|A|B,\ldots}]]$.

\subsubsec Integration by parts

The integration-by-parts relations of meromorphic correlators ${\cal K}_n(\ell)$ were
governed by the derivatives of the $\ell$-dependent Koba--Nielsen factor ${\cal I}_n(\ell)$,
see section \totalderivsec. Accordingly, the loop-integrated Koba--Nielsen factor $\hat {\cal I}_n$ 
in \notationellA\ gives rise to a modified set of integration-by-parts relations. The 
$z_j$-derivatives \zderivzero\ straightforwardly generalize to
\eqn\newzderiv{
{\p\over\p z_i} \hat {\cal I}_n =\big(
 \sum_{j\neq i}^n s_{ij} f^{(1)}_{ij}\big)\hat {\cal I}_n\, ,
}
while the $\tau$-derivative \tauderivzero\ requires more adjustments after integration over $\ell$.
After momentum conservation, the Koba--Nielsen exponent in \notationellA\
has the following $\tau$-derivative
\eqn\stepone{
{\p \over \p \tau} \sum^n_{i<j}s_{ij} \Big[ \log\big| \theta_1(z_{ij},\tau)  \big|^2 - {2\pi \over \Im \tau} (\Im z_{ij})^2 \Big] 
= \sum^n_{i<j} s_{ij} \Big[ { 1\over 2\pi i } f^{(2)}_{ij}  -  { \Im z_{ij} \over \Im \tau } f^{(1)}_{ij} \Big] \, ,
}
where the admixtures of $f^{(1)}_{ij}$ cancel from the action of the differential operator
\eqn\stepfour{
\nabla_{\tau} \equiv {\p \over \p   \tau} + \sum_{j=2}^n {\Im z_{j1} \over \Im \tau }
{\p \over \p   z_j}  
}
depending on $n$ punctures $z_j$. The operator $\nabla_{\tau}$ obeys the usual Leibniz property
and appears naturally in the following generalization of the mixed heat equation \mixedheat,
\eqn\newmixedht{
\nabla_{\tau} f^{(w)}_{ij} = {w\over 2\pi i } \p  f^{(w+1)}_{ij} - {w\over 2i \, \Im \tau}  f^{(w)}_{ij} \, .
}
Then, after taking the prefactor of $\hat {\cal I}_n\sim (\Im \tau)^{-D/2}$ in \notationellA\ into account, 
a convenient analogue of the $\tau$-derivative \tauderivzero\ after loop integration reads
\eqn\steptwo{
\nabla_\tau \hat {\cal I}_n = \hat{\cal I}_n \Big\{
{1\over 2\pi i } \sum_{i<j}^n s_{ij} f^{(2)}_{ij} + {i D \over 4 \Im \tau}
\Big\} \, ,
}
where we will set the number of spacetime dimensions to $D=10$ henceforth.
The operator \stepfour\ can be aligned into the following boundary term
\eqnn\stepthree
$$\eqalignno{
& {\p \over \p  \tau} \big(  h(z,\tau) \hat {\cal I}_n  \big) + \sum_{p=2}^n {\p \over \p   z_p}  \Big( {\Im z_{p1} \over \Im \tau } h(z,\tau) \hat {\cal I}_n  \Big)
&\stepthree  \cr
& \ \ \ =  h(z,\tau)  \hat{\cal I}_n \Big\{
{1\over 2\pi i } \sum_{i<j}^n s_{ij} f^{(2)}_{ij} + {n{-}6 \over 2i\, \Im \tau}
\Big\}
+\hat{\cal I}_n \nabla_\tau h(z,\tau) \, , 
}$$
with $h(z,\tau)$ denoting an arbitrary function on the worldsheet.
Since both of \newzderiv\ and \stepthree\ integrate to zero within string amplitudes, 
we conclude the following equivalence classes of integrated correlators $[[\ldots]]$,
\eqnn\newzderivzero
\eqnn\newtauderivzero
$$\eqalignno{
\Big( \sum_{j\neq i}^n s_{ij} f^{(1)}_{ij} \Big) h(z,\tau)
+   {\p h(z,\tau)\over\p z_i} &\cong 0
\, , \ \ \ \ \forall \,h(z,\tau)\,,&\newzderivzero\cr
\Big(\sum_{ i<j}^n s_{ij} f^{(2)}_{ij} \Big) h(z,\tau)
+ 2\pi i \Big( {n{-}6 \over 2i\, \Im \tau} + \nabla_\tau \Big) h(z,\tau)
&\cong 0\, , \ \ \ \ \forall \,h(z,\tau)\,,&\newtauderivzero
}$$
see \zderivzero\ and \tauderivzero\ for their chirally-split analogues. The simplest example
of \newtauderivzero\ with $h(z,\tau)=1$ has been used in \MafraNWR\ to identify 
the BRST variation of the $(n=6)$-point closed-string amplitude as a boundary term.

Note that the holomorphic derivative ${\p \over\p z_i} $ in \newzderivzero\ acts non-trivially
on the contributions $\bar f^{(w)}_{ij}$ from the opposite chiral half in closed-string amplitudes. 
This follows from the complex conjugate
\eqn\ccnonhol{
{\p \over  \p  z} \bar f^{(n)}(z,\tau) =
-{\pi \over \Im\tau} \bar f^{(n-1)}(z,\tau)\,,
\qquad {\p\over\p \tau} \bar f^{(n)}(z,\tau) =
{\pi \Im z \over (\Im \tau)^2} \bar f^{(n-1)}(z,\tau)  
}
of \nonhol\ and gives rise to examples such as \refs{\Richards, \oneloopMichael}
\eqn\exLR{
f^{(1)}_{12} \bar f^{(1)}_{23}  \cong {1\over s_{12}}
\Big( \bar f^{(1)}_{23}  \sum_{j=3}^n s_{2j} f_{2j}^{(1)}
- {\pi \over \Im \tau} \Big)\, .
}
The differential operator \stepfour\ in turn annihilates undifferentiated $\bar f^{(w)}_{ij}$
and only acts on $\bar z$-derivatives of the $ \bar f^{(w)}_{ij} $ from the opposite chiral half
in closed-string amplitudes
\eqn\onlydiffs{
\nabla_{\tau} \bar f^{(w)}_{ij} =0\, , 
\ \ \ \ \ \ \nabla_\tau \Big( {\p \bar f^{(w)}_{ij}  \over \p \bar z} \Big) = -{\pi \bar f^{(w-1)}_{ij}  \over 2i \,(\Im \tau)^2} \, .
}
On these grounds, the analysis of boundary terms in $\tau$ is facilitated when
loop-integrated GEIs $[[E^{m_1\ldots}_{1|A,\ldots}]]$ are expressed in terms of 
undifferentiated $f^{(w)}_{ij}$.

\newsubsubsec\moreanom Integrating refined GEIs

After loop integration, the integration-by-parts equivalent representations
of the simplest refined GEI $E_{1|2|3,4,5,6}$ in \SixErefined\ translate into
\eqnn\intellC
$$\eqalignno{
[[E_{1|2|3,4,5,6}]] &=  - {\pi\over\Im\tau} +  \p f^{(1)}_{12} + s_{12} (f_{12}^{(1)})^2
- 2 s_{12} f^{(2)}_{12} &\intellC \cr
&\cong - {\pi\over\Im\tau} - 2 s_{12} f^{(2)}_{12} + f^{(1)}_{12} (s_{23}f^{(1)}_{23}
+ s_{24}f^{(1)}_{24}  + s_{25}f^{(1)}_{25}  + s_{26}f^{(1)}_{26})  \cr
&\cong - 2 s_{12} f^{(2)}_{12} + f^{(1)}_{12} \big[s_{23}f^{(1)}_{23}
+ (3\leftrightarrow 4,5,6) \big]+ \nu_{12}\big[s_{12}f^{(1)}_{12}+ (1\leftrightarrow 3,4,5,6)\big]\,.\cr
}$$
The first line follows from inserting $\p g^{(1)}_{12} = \p f^{(1)}_{12} - {\pi
\over \Im \tau}$ into \SixErefined, and the second and third line result from
the integration-by-parts relation \newzderivzero\ after discarding $\p_2(
f^{(1)}_{12} \hat {\cal I}_6)$ and $\p_2( \nu_{12} \hat {\cal I}_6)$,
respectively. One can also arrive at the last line by inserting $[[\ell^m]] =
\sum_{j=2}^6 k_j^m \nu_{1j}$ into the first line of \SixErefined\ and expressing
all the $g^{(n)}_{ij}$ in terms of $f^{(n)}_{ij}$ and $\nu_{ij}$.

At seven points, the refined GEIs \rewrtD\ integrate to
\eqnn\newrewrtD
$$\eqalignno{
[[E_{1|23|4,5,6,7} ]] &= - { \pi \over \Im \tau} V_1(1,2,3) - s_{123} V_3(1,2,3) +(f_{12}^{(1)}+f_{31}^{(1)}) \p f^{(1)}_{23} + \p f^{(2)}_{23} &\newrewrtD
\cr
[[E_{1|4|23,5,6,7}]]  &= - { \pi \over \Im \tau} V_1(1,2,3) {+} \big[\p f^{(1)}_{14} {-} s_{14} V_2(1,4) \big]V_1(1,2,3) {-} s_{24} V_3(1,2,4) {+} s_{34} V_3(1,3,4)
\cr
[[E^m_{1|2|3,4,5,6,7} ]] &= - { \pi \over \Im \tau} \big[ k_2^m f^{(1)}_{12} +(2\leftrightarrow 3,\ldots,7)  \big]  
+ k_2^m \big[ \p f^{(2)}_{12} + s_{12}(f_{12}^{(1)} f_{12}^{(2)} - 3 f^{(3)}_{12}) \big] 
\cr
& \! \! \! \! \!  \! \! \! \! \! + \big[\p f^{(1)}_{12} {-} s_{12} V_2(1,2) \big]  \big[ k_3^m f^{(1)}_{13} {+}(3\leftrightarrow 4,\ldots,7)  \big]  {+} \big[   k_3^m s_{23} V_3(1,2,3){+}(3\leftrightarrow 4,\ldots,7)  \big]\, ,
}$$
where we reiterate that the elliptic $V_w$-functions are unchanged under the global replacement
of $g^{(n)}_{ij} \rightarrow f^{(n)}_{ij}$. One can perform integrations by parts \newzderivzero\ 
similar to \intellC\ to avoid the appearance of $\p f^{(n)}_{ij}$ on the right-hand side.
Similar to \intellA\ and \newEOx, the factors of ${ \pi \over \Im \tau}$ on the right-hand sides
of \intellC\ and \newrewrtD\ signal a modular anomaly: They depart from the purely holomorphic
modular weights $(n,0)$ and $(n{+}1,0)$ of the $f^{(n)}_{ij}$ and $\p f^{(n)}_{ij}$.

Note that the trace relations \traceA\ and \tracesevenA\ of six- and seven-point GEIs 
can be verified at the level of the above expressions for the $[[E_{1|\ldots}]]$: While
\eqnn\newtrex
$$\eqalignno{
&{1\over 2} \delta_{mn} [[E^{mn}_{1|2,3,4,5,6}]] \hat {\cal I}_6  
+ \big( [[E_{1|2|3,4,5,6}]]  + (2\leftrightarrow 3,4,5,6) \big)  \hat {\cal I}_6  \cr
& \ \ \ \ = 2\pi i  {\p \over \p  \tau} \hat {\cal I}_6  + 
2\pi i \sum_{p=2}^6 {\p \over \p   z_p}  \Big( {\Im z_{p1} \over \Im \tau } \hat {\cal I}_6 \Big)
&\newtrex
}$$
is a consequence of \stepthree\ at $n=6$ and $h(z,\tau)=1$, the seven-point
analogues require a specialization of \newtauderivzero\ to\foot{More generally,
the choices of $h(z,\tau)$ in \newtauderivzero\ relevant to integrated $n$-point
closed-string correlators $[[{\cal K}_n(\ell) \tilde {\cal K}_n(-\ell)]]$ have
the form ${1\over (\Im \tau)^{m}}\prod_k f_{a_kb_k}^{(w_k)}\bar
f_{c_kd_k}^{(\bar w_k)}$ with $m{+}\sum_k w_k = n{-}6$. In these cases, the
second line of the following equivalence relation \newtauderivzero\ vanishes (we
are suppressing the $\bar f_{c_kd_k}^{(\bar w_k)}$ they are annihilated by
$\nabla_\tau$),
$$\eqalignno{
0 &\cong   \bigg( {1 \over (\Im \tau)^m} \prod_k f_{a_kb_k}^{(w_k)}
\Big( \sum_{i<j}^n s_{ij} f^{(2)}_{ij} \Big)
+\sum_{r} {w_r\p f_{a_rb_r}^{(w_r+1)} \over (\Im \tau)^m} \prod_{k\neq r} f_{a_kb_k}^{(w_k)} \bigg) \cr
& \ \ \ +2\pi i\bigg( {n-6-m-w_1-w_2-\ldots \over 2i \, \Im \tau} \bigg)\,
{1 \over (\Im \tau)^m} \prod_k f_{a_kb_k}^{(w_k)}\, .
}$$}
\eqn\specseven{
\Big( {n{-}6 \over 2i\, \Im \tau} + \nabla_\tau \Big) f^{(1)}_{ij} \, \Big|_{n=7}
= { \p f^{(2)}_{ij} \over 2\pi i}\,,
}
see \newmixedht\ for the action of $\nabla_\tau$ on $f^{(w)}_{ij}$.

\subsubsec Modular anomalies versus BRST anomalies

The above instances of modular anomalies furnish another incarnation of the
duality between kinematics and worldsheet functions. Modular anomalies are proposed to be
the worldsheet counterpart of anomalous BRST variations such as
\eqnn\qanom
$$\eqalignno{
Q C^{mn}_{1|2,3,4,5,6} &= - \delta^{mn}  \Gamma_{1|2,3,4,5,6}  \, ,  \ \ \ \ \ \, Q P_{1|2|3,4,5,6} = -  \Gamma_{1|2,3,4,5,6} &\qanom\cr
Q C^{mn}_{1|23,4,\ldots,7} &= - \delta^{mn}  \Gamma_{1|23,4,\ldots,7} \,  ,  \ \ \ \
Q P_{1|23|4,5,6,7} = Q P_{1|4|23,5,6,7} = -  \Gamma_{1|23,4,5,6,7} \cr
Q C^{mnp}_{1|2,3,\ldots,7} &= - \delta^{(mn}  \Gamma^{p)}_{1|2,3,\ldots,7} \,  ,  \ \ \ \ \ \ \,
Q P^m_{1|2|3,\ldots,7} =  -  \Gamma^m_{1|2,3,\ldots,7} \, ,
}$$
where the anomaly invariants $\Gamma_{1|\ldots}$ are defined in section \symmCPtwo, and generalizations
of \qanom\ can be found in \GAMb. The idea is to associate the anomaly invariants with the slot extensions
$[[E_{1|2,3,4,5,6}]] = 1$ and
\eqnn\slotexten
$$\eqalignno{
[[E_{1|23,4,5,6,7}]] &=  V_1(1,2,3) \, , \ \ \ \
[[E^m_{1|2,3,\ldots,7}]]= k_2^m f^{(1)}_{12} + (2\leftrightarrow 3,\ldots,7) 
 &\slotexten \cr
}$$
of earlier results according to the general dictionary
\eqn\anomdict{
\Gamma^{m_1\ldots m_r}_{1|A_1,\ldots,A_d|B_1,\ldots,B_{d+r+5}}\leftrightarrow {\pi \over \Im \tau}
[[E^{m_1\ldots m_r}_{1|A_1,\ldots,A_d|B_1,\ldots,B_{d+r+5}}]] \, .
}
Under these identifications, the combinatorics of \qanom\ literally 
translates into the following modular anomalies at six points
\eqnn\dualmodanomalysix
$$\eqalignno{
[[E^{mn}_{1|2,3,4,5,6} ]] &= - { \pi \over \Im \tau} \delta^{mn}+ {\rm modular} \ {\rm weight} \ (2,0) &\dualmodanomalysix
\cr
[[ E_{1|2|3,4,5,6}]]  &= - { \pi \over \Im \tau} + {\rm modular} \ {\rm weight} \ (2,0)
}$$
and at seven points
\eqnn\dualmodanomaly
$$\eqalignno{
[[E^{mn}_{1|23,4,5,6,7} ]] &= - { \pi \over \Im \tau} \delta^{mn} V_1(1,2,3) + {\rm modular} \ {\rm weight} \ (3,0) &\dualmodanomaly
\cr
[[E^{mnp}_{1|2,3,4,5,6,7} ]] &= - { \pi \over \Im \tau} \delta^{(mn}  \big[ k_2^{p)} f^{(1)}_{12} +(2\leftrightarrow 3,\ldots,7)  \big]  
+ {\rm modular} \ {\rm weight} \ (3,0)
\cr
[[E^m_{1|2|3,4,5,6,7} ]] &= - { \pi \over \Im \tau} \big[ k_2^m f^{(1)}_{12} +(2\leftrightarrow 3,\ldots,7)  \big]  
+ {\rm modular} \ {\rm weight} \ (3,0) 
\cr
[[E_{1|23|4,5,6,7} ]] &= - { \pi \over \Im \tau} V_1(1,2,3) + {\rm modular} \ {\rm weight} \ (3,0)
\cr
[[E_{1|4|23,5,6,7}]]  &= - { \pi \over \Im \tau} V_1(1,2,3) + {\rm modular} \ {\rm weight} \ (3,0)
\, ,
}$$
where the weight-$(n{-}4,0)$ parts can be found in \intellA, \newEOx,
\intellC\ and \newrewrtD. As we will see in section \SixLoopsec,
the above instances of modular anomalies drop out from the
integrated six-point correlator $[[{\cal K}_6(\ell)]]$. The
cancellation of modular anomalies will be shown to furnish a dual to the
localization of BRST anomalies $Q{\cal K}_n(\ell)$ on the boundary of moduli
space.

While the dictionary \anomdict\ is expected to extend to higher multiplicity, it
is not clear whether it applies to higher powers $({\pi \over \Im \tau})^m$ with
$m\geq 2$. It remains to clarify whether the absence of tensor structures
$\delta^{m(n} \delta^{pq)}$ in $Q C^{mnpq}_{1|2,3,\ldots,8} = - \delta^{(mn}
\Gamma^{pq)}_{1|2,\ldots,8}$ can be reconciled with the contribution $[[\ell^m
\ell^n \ell^p \ell^q]]=({\pi \over \Im \tau})^2 \delta^{m(n}
\delta^{pq)}+\ldots$ to $[[E^{mnpq}_{1|2,\ldots,8}]]$.

\newsec Conclusions

In this paper we continued setting up the ingredients that will be needed to
build up one-loop correlators for massless open- and closed-string amplitudes
in the pure-spinor formalism. We have introduced two classes of worldsheet 
functions that will manifest different aspects of the correlators to be assembled in part III.
Both of them are constructed from loop momenta and combinations of Jacobi theta 
functions $g^{(n)}_{ij}=g^{(n)}(z_i{-}z_j,\tau)$ that are the coefficients in the Laurent
expansion of the Kronecker--Eisenstein series \BrownLevin.

The first class of worldsheet functions, denoted by $\cZ$, is designed to
capture the worldsheet singularities arising when the vertex operators approach
each other on a genus-one surface. These singularities are straightforward to
handle via an OPE analysis, and their behavior when the vertices are close
together is the same as products of $1/z_{ij}=1/(z_i{-}z_j)$ functions well-known from the
tree-level correlators.

However, the OPE analysis is not enough to completely determine the one-loop
${\cal Z}$-functions as there can be non-singular pieces that do not vanish on a genus-one
surface\foot{These non-singular parts are absent at tree level where the
knowledge of the singular behavior is enough to fix the whole function.}. Instead,
our starting point to constrain the non-singular pieces is the following observation on
tree-level correlators: The products of singular functions $1/z_{ij}$ at genus zero end up assembling chains
$1/(z_{12}z_{23} \ldots z_{p-1,p})$ \nptString\ that obey shuffle symmetries among their
labels $1,2, \ldots,p$. By imposing the same shuffle symmetries among the 
labels of their one-loop counterparts $\cZ$ and using Fay identities one proves the existence of 
non-singular pieces in the one-loop worldsheet functions. 

The algorithmic determination of
these non-singular pieces follows from another surprising feature of these functions; their
properties mimic those of superfield building blocks discussed in part I. 
More precisely, the role of the pure-spinor BRST charge
acting on the superfields is replaced by a monodromy operator acting on the
genus-one functions and the loop momentum. This observation, among others 
along the same lines, has been interpreted as a duality between worldsheet functions and kinematics.

The second class of worldsheet functions discussed in this paper concerns the
generalized elliptic integrands (GEIs) briefly introduced in \MafraIOJ.
GEIs are monodromy-invariant combinations of $\cZ$-functions, and already their
very construction is driven by the duality between worldsheet functions and kinematics: 
GEIs can be assembled from the monodromy-covariant functions $\cZ$ in exactly the 
same combinatorial manner as kinematic BRST invariants are assembled from 
Berends--Giele superfield building blocks (reviewed in part I). These definitions
lead to a plethora of relations that apply in similar if not identical form to the
superfield building blocks, manifesting various further incarnations of the duality 
between worldsheet functions and kinematics.

A multitude of identities among ${\cal Z}$-functions and GEIs has been discussed 
in this paper that support their duality connection with superfield building blocks. 
However, we observed that holomorphic Eisenstein series lead to departures from 
a strict duality between functions and kinematics starting at eight points. The solution 
to this puzzling behavior, for instance through systematic redefinitions of $\cZ$-functions 
and GEIs via Eisenstein series, will be left for the future. Furthermore, a preliminary analysis
indicates that the functions considered in this paper admit compact generating-series
representations whose detailed presentation we also leave for future work.

The relevance of both the ${\cal Z}$-functions as well as
GEIs for the assembly of one-loop correlators will become apparent in the sequel
part III of this series of papers.

\bigskip \noindent{\bf Acknowledgements:} We are indebted to the IAS Princeton
and to Nima Arkani-Hamed for kind hospitality during an inspiring visit which initiated this project.
This research was supported by the Munich Institute for Astro- and Particle
Physics (MIAPP) of the DFG cluster of excellence ``Origin and
Structure of the Universe'', and we are grateful to the organizers for creating
a stimulating atmosphere. CRM is supported by a University 
Research Fellowship from the Royal Society. The research of OS was supported in part by 
Perimeter Institute for Theoretical Physics. Research at Perimeter Institute is supported by the Government of
Canada through the Department of Innovation, Science and Economic Development
Canada and by the Province of Ontario through the Ministry of Research,
Innovation and Science.

\appendix{A}{Bootstrapping the shuffle-symmetric worldsheet functions}
\applab\bootstrapapp

\noindent
This appendix complements the results of the bootstrap techniques for ${\cal Z}$-functions
outlined in section~\bootsec\ with derivations based on the system of
monodromy variations. The key steps will be presented in detail for six points
and for some selected seven- and eight-point functions; the results from
the omitted derivations can be obtained with
reasonable effort \FORM\ and do not require any new methods.

In the derivations below we will use the representations of GEIs obtained in
section~\GenSeriessec\ as they lead to considerably shorter results;
in some cases, they even suggest pattern-driven general closed formul\ae.

\newsubsec\appCone Six points

The starting point at six points is given by the extended GEIs \sixEms\ from the
five-point results \fiveEls\ and \fiveElls, namely
\eqn\startSix{
E_{1|23,4,5,6}=V_1(1,2,3)\,,\qquad E^m_{1|2,3,4,5,6}=\ell^m - \big[\p V_0(1,2) +
(2\leftrightarrow3,4,5,6)\big]\,.
}
They are written in terms of the $V_w$- and $\p V_w$-functions with
generating series in \ellrepD, \ellrepF\ and \highDA\ for convenience.
As we have seen in section~\bootsec, the monodromy variations of the
six-point shuffle-symmetric worldsheet functions are given by
\eqnn\sixMonApp
$$\eqalignno{
D\cZ_{123,4,5,6} &= \Omega_1 E_{1|23,4,5,6} - \Omega_3 E_{3|12,4,5,6}\,,
&\sixMonApp \cr
D\cZ_{12,34,5,6} &= \Omega_1 E_{1|2,34,5,6}-\Omega_2 E_{2|1,34,5,6}
+\Omega_3 E_{3|12,4,5,6}-\Omega_4 E_{4|12,3,5,6}\,,\cr
D\cZ^m_{12,3,4,5,6} &= \Omega_1 E^m_{1|2,3,4,5,6}-\Omega_2 E^m_{2|1,3,4,5,6}
+\big[ k_3^m \Omega_3 E_{3|12,4,5,6}+ (3\leftrightarrow 4,5,6) \big]\,,
\cr
D\cZ^{mn}_{1,2,3,4,5,6} &= k^m_1\Omega_1 E^n_{1|2,3,4,5,6}
+k^n_1\Omega_1 E^m_{1|2,3,4,5,6}+ (1\leftrightarrow 2,3,4,5,6)\,,\cr
D {\cal Z}_{2|1,3,4,5,6} &= \Omega_2 k_2^m E^m_{2|1,3,4,5,6}\,.
}$$
To solve these equations using the generating-series techniques of 
section~\GenSeriessec\ it will be convenient to rewrite the above GEIs
in a basis where leg $1$ is in the special slot. This can be
done by exploiting the duality with the BRST invariants and using the identities
of section~\EjacH. In this new basis we have:
\eqnn\sixMonAppBa
\eqnn\secMon
\eqnn\vecMon
\eqnn\tenMon
\eqnn\refMon
$$\eqalignno{
D\cZ_{123,4,5,6} &= \Omega_{13} E_{1|23,4,5,6}\,,
&\sixMonAppBa \cr
D\cZ_{12,34,5,6} &= \Omega_{12} E_{1|2,34,5,6}
+ \Omega_{32} E_{1|23,4,5,6}
+ \Omega_{24} E_{1|24,3,5,6}\,, &\secMon\cr
D\cZ^m_{12,3,4,5,6} &= \Omega_{12}E^m_{1|2,3,4,5,6}
- \big[\Omega_{23}k_3^m E_{1|23,4,5,6}
+ (3\leftrightarrow 4,5,6)\big]\,,&\vecMon\cr
D\cZ^{mn}_{1,2,3,4,5,6} &= \big[\Omega_{21}
(k_2^m E^n_{1|2,3,4,5,6}+k_2^n E^m_{1|2,3,4,5,6})
+(2\leftrightarrow 3,4,5,6) \big] &\tenMon\cr
& + \big[
(k_2^m k_3^n + k_2^n k_3^m)\Omega_{23} E_{1|23,4,5,6}
+(2,3|2,3,4,5,6) \big]\,,\cr
D\cZ_{2|1,3,4,5,6} &= \Omega_2k_2^m \big(E^m_{1|2,3,4,5,6} + \big[k_3^m
E_{1|23,4,5,6} + (3\leftrightarrow4,5,6)\big]\,.&\refMon
}$$
The scalar equations are easily solved using cyclic symmetry of $V_w(1,2, \ldots,n)$
and the monodromy variations $D\p V_w(1,2, \ldots,n) = -\Omega_{1n}V_w(1,2,
\ldots,n)$. We get,
\eqnn\scalarSix
$$\eqalignno{
\cZ_{123,4,5,6} &=-\p V_1(1,2,3)\,,&\scalarSix\cr
\cZ_{12,34,5,6} &=-\p V_0(1,2)V_1(1,3,4) + \p V_1(4,1,2) - \p V_1(3,1,2)\,,
}$$
whose equivalence with the solutions presented in \newgs\
is easily established using Fay identities.
Let us now solve the monodromy variation \vecMon\ of the vectorial function
\eqnn\monSix
\eqnn\Dsix
$$\eqalignno{
D\cZ^m_{12,3,4,5,6} &= \Omega_{12}E^m_{1|2,3,4,5,6}
- \big[\Omega_{23}k_3^m E_{1|23,4,5,6}
+ (3\leftrightarrow 4,5,6)\big]&\monSix\cr
&= \Omega_{12}\ell^m
       + \Omega_{12} k_2^m g^{(1)}_{12}
       + \big[
       k_3^m(\Omega_{12}g^{(1)}_{13} - \Omega_{23}V_1(1,2,3))
       + (3\leftrightarrow 4,5,6)\big]\,,\quad\qquad{} &\Dsix
}$$
where the second line follows from \startSix.
Noting that
$D(g^{(1)}_{12}\ell^m) = \Omega_{12}\ell^m - \gg1(1,2)\sum_{j=2}^6\Omega_{1j}k_j^m$
one can rewrite \Dsix\ as follows
\eqnn\sixexp
$$\eqalignno{
D\cZ^m_{12,3,4,5,6} &= \big(\Omega_{12}\ell^m -
\gg1(1,2)\sum_{j=2}^6\Omega_{1j}k_j^m\big)
       + 2 k_2^m\Omega_{12} g^{(1)}_{12} &\sixexp\cr
 &      + \big[
       k_3^m(\Omega_{12}g^{(1)}_{13} + \Omega_{13}g^{(1)}_{12}
       - \Omega_{23}V_1(1,2,3)
       )
       + (3\leftrightarrow 4,5,6)\big]\,.
}$$
The solution to \sixexp\ can be obtained by inspection and is given by
\eqnn\integralSix
$$\eqalignno{
\cZ^m_{12,3,4,5,6} &= \gg1(1,2)\ell^m + 2k_2^m \gg2(1,2)
+ \big[k_3^m\big(
\gg1(1,2)\gg1(1,3) - \p V_1(3,1,2)\big)
+ (3\leftrightarrow 4,5,6) \big] \cr
&= \ell^m g^{(1)}_{12} 
+ (k_2^m - k_1^m)g^{(2)}_{12}
+ \big[ k_3^m (g^{(2)}_{13} - g^{(2)}_{23}) + (3\leftrightarrow 4,5,6)\big] 
\, , &\integralSix 
}$$
see \newgs. The equality in the last line 
follows from momentum conservation and
$\gg1(1,2)\gg1(1,3) - \DV1(3,1,2) - \gg2(1,2) = \gg2(1,3) - \gg2(2,3)$, which
can be shown using Fay identities. As a side remark,
note that one can arrive at \integralSix\ from \Dsix\ using an
effective ``integration'' rule $\int\Omega_{ij}g^{(n)}_{ij} = (n{+}1)g^{(n+1)}_{ij}$, $\forall \ n\in{\Bbb N}$
to ``invert'' the $D$ operator.

The solution to the tensorial monodromy variation \tenMon,
$$\eqalignno{
D\cZ^{mn}_{1,2,3,4,5,6} &= \big[\Omega_{21}
k_2^{(m} E^{n)}_{1|2,3,4,5,6} 
+(2\leftrightarrow 3,4,5,6) \big] \cr
& + \big[
k_2^{(m} k_3^{n)}  \Omega_{23} E_{1|23,4,5,6}
+(2,3|2,3,4,5,6) \big]
}$$
can be found similarly. First one plugs in the vectorial extended GEI from
\startSix\ to obtain
\eqnn\tensEx
$$\eqalignno{
D\cZ^{mn}_{1,2,3,4,5,6} &= \big[\Omega_{21}k_2^{(m}\ell^{n)}_{\phantom{2}}
+ 2k_2^m k_2^n \Omega_{21}\gg1(1,2) +
(2\leftrightarrow3,4,5,6)\big]&\tensEx\cr
&+\big[k_2^{(m}k_3^{n)}\big(\Omega_{21}\gg1(1,3) + \Omega_{31}\gg1(1,2) -
\Omega_{32}V_1(3,1,2) \big)+ (2,3|2,3,4,5,6)\big]\,,
}$$
whose solution is easily found after noticing that $D(\ell^m \ell^n) =
\Omega_{21}k_2^{(m}\ell_{\phantom{2}}^{n)} + (2\leftrightarrow3,4,5,6)$,
\eqnn\tensSol
$$\eqalignno{
\cZ^{mn}_{1,2,3,4,5,6} & = \ell^m \ell^n - 2\big[k_2^m k_2^n
\gg2(1,2)+(2\leftrightarrow3,4,5,6)\big] &\tensSol\cr
&- \big[k_2^{(m}k_3^{n)}\big(\gg1(1,2)\gg1(1,3)-\p V_1(3,1,2)\big)
+ (2,3|2,3,4,5,6)\big]\,.
}$$
Fay identities imply that \tensSol\ is equivalent to the
expression given in \newgs,
\eqn\newgsTensr{
\cZ^{mn}_{1,2,3,4,5,6}= \ell^m\ell^n +
\bigl[ k_1^{(m}k_2^{n)} g^{(2)}_{12} + (1,2|1,2,3,4,5,6)
\bigr]\,.
}
The solution to the refined worldsheet function can be easily found.
After plugging in the extended GEIs on the right-hand side of \refMon, 
straightforward algebra leads to
\eqn\refVan{
D\cZ_{2|1,3,4,5,6} = \Omega_2\big((\ell\cdot k_2) + s_{21}\gg1(2,1)
+ s_{23}\gg1(2,3)
+ s_{24}\gg1(2,4)
+ s_{25}\gg1(2,5)
+ s_{26}\gg1(2,6)\big)\cong0\,,
}
which vanishes in view of the total-derivative relation \zderiv. Therefore,
one can choose
\eqn\refinVan{
\cZ_{2|1,3,4,5,6}=0\,,
}
see \newgs.
As mentioned in section~\sixwssec, this vanishing is compatible with
a duality between refined worldsheet functions and BRST-exact superfields,
see \cYZdual.

Plugging the results above in the expressions \SixElliptic\ and \covPSb\ leads to the 
expressions \Eswithgs, \SixErefined\ and \rewrtB\ for GEIs. Their seven-point 
extensions \sixExt\ will be used in the next step of the bootstrap procedure.

\newsubsec\sevptapp Seven points

The solution to the scalar monodromy variations
\eqnn\DZsevensAgain
$$\eqalignno{
D\cZ_{1234,5,6,7} &=\Omega_1E_{1|234,5,6,7} - \Omega_4E_{4|123,5,6,7}\,,
&\DZsevensAgain\cr
D\cZ_{123,45,6,7} &=\Omega_1E_{1|23,45,6,7} - \Omega_3E_{3|12,45,6,7}\,,
+ \Omega_4E_{4|123,5,6,7} - \Omega_5E_{5|123,4,6,7}\cr
D\cZ_{12,34,56,7} &=\Omega_1E_{1|2,34,56,7} - \Omega_2E_{2|1,34,56,7} +
(12\leftrightarrow34,56)\,,
}$$
is easily obtained after rewriting the GEIs in the canonical basis and
using \rewrtB,
\eqnn\scalarSev
$$\eqalignno{
      D\cZ_{1234,5,6,7} &= \Omega_{14}V_2(1,2,3,4) &\scalarSev\cr
   D\cZ_{123,45,6,7} &=
         \Omega_{13} V_1(1,2,3)V_1(1,4,5)
       - \Omega_{34} V_2(1,2,3,4)
       + \Omega_{35} V_2(1,2,3,5) \cr
   D\cZ_{12,34,56,7} &=
        \Omega_{12} V_1(1,3,4)V_1(1,5,6) \cr
&       + \big[\Omega_{23}\big(V_2(1,2,3,6) - V_1(1,2,3)V_1(1,5,6) - V_2(1,2,3,5)\big)
       -(3\leftrightarrow4)\big]\cr
&	+  \big[\Omega_{25}\big(V_2(1,2,5,4) - V_1(1,2,5)V_1(1,3,4) - V_2(1,2,5,3)\big)
       - (5\leftrightarrow6)\big]\,.
}$$
Noting the fundamental equation \ellrepH\ and cyclicity of $V_w(1, \ldots,n)$ we
arrive at the following solutions
\eqnn\borSev
$$\eqalignno{
\cZ_{1234,5,6,7} &= - \p V_2(1,2,3,4)\,,&\borSev\cr
\cZ_{123,45,6,7} &=
- \p V_1(1,2,3) V_1(1,4,5)
+ \p V_2(5,1,2,3)
-\p V_2(4,1,2,3) \cr
\cZ_{12,34,56,7} &= -\p V_0(1,2) V_1(1, 3, 4) V_1(1, 5, 6)
+ \p V_1(4, 1,2) V_1(1, 5, 6) \cr
&- \p V_1( 5, 1,2) V_1(1, 3, 4)
- \p V_1( 3, 1,2) V_1(1, 5, 6)
+ \p V_1( 6, 1,2) V_1(1, 3, 4) \cr
 &+ \p V_2( 3, 6, 1,2) - \p V_2( 3, 5, 1,2)
 + \p V_2( 4, 5, 1,2) - \p V_2( 4, 6, 1,2)  \cr
 &+ \p V_2(6, 3, 1,2)- \p V_2( 5, 3, 1,2)
 + \p V_2( 5, 4, 1,2) - \p V_2( 6, 4, 1,2)\,.
}$$
A long but straightforward application of Fay identities demonstrates the
equivalence between the above solutions and the ones presented in the main text, \newsevengs.
While the above form of the functions is easy to derive from the monodromy
variations, it does not expose the singularity structure as the vertex positions approach each other.
This constitutes a drawback of the representation in \borSev\ and motivates the rewriting in
\newsevengs.

\subsubsec Vectorial seven-point functions

The monodromy variation \assumption\ of the vectorial seven-point function
$\cZ^m_{123,4,5,6,7}$ can be written in a basis of GEIs as
\eqnn\sevenDA
$$\eqalignno{
D\cZ^m_{123,4,5,6,7} &= \Omega_{13}E^m_{1|23,4,5,6,7}
+ \big[k_4^m \Omega_{43}E_{1|234,5,6,7} + (4\leftrightarrow
5,6,7)\big]\cr
&=\Omega_{13}V_1(1,2,3)\ell^m
+ k_2^m\Omega_{13}\p V_1(2,3,1)
- k_3^m\Omega_{13}\p V_1(3,2,1)\cr
&+\big[
k_4^m
\big(\Omega_{13}\gg1(1,4)V_1(1,2,3)
- \Omega_{34}V_2(1,2,3,4)\big)
+ (4\leftrightarrow 5,6,7)\big] \, .&\sevenDA\cr
}$$
Similarly as before, in order to integrate the
term containing $\ell^m$ in the above variation, we add and subtract
$ \p V_1(1,2,3)\sum_{j=2}^7 \Omega_{1j}k_j^m$
to obtain
\eqnn\trickedS
$$\eqalignno{
D\cZ^m_{123,4,5,6,7} &=
\Omega_{13}V_1(1,2,3)\ell^m + \p V_1(1,2,3) \big[  \Omega_{12}k_2^m +(2\leftrightarrow 3,4,5,6,7)\big]  &\trickedS\cr
& + k_2^m \big(\Omega_{13}\p V_1(2,3,1) {-} \Omega_{12}\p V_1(1,2,3) \big)
 - k_3^m \big(\Omega_{13}\p V_1(3,2,1) {+} \Omega_{13}\p V_1(1,2,3) \big)\cr
& +\big[
k_4^m
\big(\Omega_{13}\gg1(1,4)V_1(1,2,3)
- \Omega_{14}\p V_1(1,2,3)
- \Omega_{34}V_2(1,2,3,4)
\big)
+ (4\leftrightarrow 5,6,7)\big] \, .
}$$
One can then show that \trickedS\ integrates to
\eqnn\intSevenA
$$\eqalignno{
\cZ^m_{123,4,5,6,7} &= -\ell^m \p V_1(1,2,3)
+ k_3^m \p^2 V_1(1,2,3) &\intSevenA\cr
&+ \half k_2^m\big[\p^2 V_1(1,2,3) + \p^2 V_1(2,3,1) + \p^2
V_1(2,1,3)\big]\cr
& -\big[k_4^m\big(
\p V_2(4,1,2,3) + \gg1(1,4)\p V_1(1,2,3)
\big)
+ (4\leftrightarrow 5,6,7)
\big]\,,
}$$
which can be rewritten as
\eqnn\intSevenB
$$\eqalignno{
\cZ^m_{123,4,5,6,7} &= -\ell^m \p V_1(1,2,3)
-\half k_1^m \p^2 V_1(1,2,3)
+\half k_3^m \p^2 V_1(1,2,3) &\intSevenB\cr
&+ \half k_2^m\big[\p^2 V_1(2,3,1) + \p^2
V_1(2,1,3)\big]\cr
& +\big[k_4^m
V_1(1,2,3)(\gg2(1,4)-\gg2(3,4))
+ (4\leftrightarrow 5,6,7)
\big]\,.
}$$
To see this one uses momentum conservation and the identity
\eqn\surpr{
\p V_2(4,1,2,3) + \gg1(1,4)\p V_1(1,2,3)=
-V_1(1,2,3)(\gg2(1,4)-\gg2(3,4)) - \half\p^2 V_1(1,2,3)\,.
}
Alternatively, the expression \intSevenB\ can be rewritten in
terms of $g^{(n)}_{ij}$-functions in order to make its singularity structure
more evident,
\eqnn\SevVecOne
$$\eqalignno{
\cZ^m_{123,4,5,6,7} &=
\ell^m\big[\gg1(1,2)\gg1(2,3) + \gg2(1,2)-\gg2(1,3)+\gg2(2,3)\big]
+ (k_3^m - k_1^m)\bigl(\gg1(1,2)\gg2(1,3) +\gg1(2,3)\gg2(1,3)
-3\gg3(1,3)\bigr)\cr
&+ k_2^m \bigl(\gg1(1,3) (\gg2(1,2)-\gg2(2,3))+\gg2(1,3) (\gg1(2,3)-\gg1(1,2)) \bigr)&\SevVecOne\cr
&+ \big[k^m_4 \big(\gg1(1,2)+\gg1(2,3)+\gg1(3,1)\big)\big( \gg2(1,4)-\gg2(3,4)\big)
+ (4\leftrightarrow 5,6,7)\big]\,.
}$$
Similarly, the monodromy variation of $\cZ^m_{12,34,5,6,7}$,
\eqnn\anotherDZ
$$\eqalignno{
D\cZ^m_{12,34,5,6,7} &= \Omega_{12}E^m_{1|2,34,5,6,7} -
\Omega_{23}E^m_{1|23,4,5,6,7}
+ \Omega_{24}E^m_{1|24,3,5,6,7} &\anotherDZ\cr
&+ k_3^m \Omega_{24}E_{1|243,5,6,7}
- k_4^m \Omega_{23}E_{1|234,5,6,7}\cr
& +\Big[ k_5^m\big(
\Omega_{25}E_{1|254,3,6,7}
- \Omega_{25}E_{1|253,4,6,7}
-\Omega_{25}E_{1|25,34,6,7}\cr
&\quad{}+ \Omega_{24}E_{1|245,3,6,7}
- \Omega_{23}E_{1|235,4,6,7}
+ (5\leftrightarrow 6,7)\Big]\,,
}$$
is readily integrated and yields, after using identities similar to \surpr\ and momentum conservation,
the following result:
\eqnn\NotAmess
$$\eqalignno{
\cZ^m_{12,34,5,6,7} &= \ell^m\Big[
	\gg1(1,2)\V1(1,3,4)
          - \DV1(3,1,2)
          + \DV1(4,1,2)\Big] &\NotAmess\cr
	  &- \Big[k_1^m\big(
	\gg2(1,2)\V1(2,3,4) + \half\D2V1(1,2,4) - \half\D2V1(1,2,3)
	  \big) - (1\leftrightarrow 2)\Big]\cr
	  & - \Big[k_3^m\big(
          \gg2(3,4)\V1(4,1,2) + \half\D2V1(3,4,2) - \half\D2V1(3,4,1)
	  \big) - (3\leftrightarrow4)\Big]\cr
	  & + \Big[
	   k_5^m\big(
	   \gg2(5,1)\V1(1,3,4)
          - \gg2(5,2)\V1(2,3,4)
          + \gg2(5,3)\V1(3,1,2)
          - \gg2(5,4)\V1(4,1,2)
	  \big)
	  + (5\leftrightarrow 6,7)
	  \Big]\,.
}$$
Its expansion in terms of $g^{(n)}_{ij}$-functions can be shown to read,
\eqnn\SevTopTwo
$$\eqalignno{
\cZ^m_{12,34,5,6,7} & =
\ell^m\bigl(\gg1(1,2)\gg1(3,4) + \gg2(1,3) - \gg2(1,4) + \gg2(2,4)
-\gg2(2,3)\bigr) +(k_{12}^m - k_{34}^m) \big( \gg3(1,4) -\gg3(1,3) + \gg3(2,3) -\gg3(2,4) \big) \cr
&+ \big[ \gg1(3,4)\gg2(1,2) (k_2^m-k_1^m)
+k_1^m  \gg1(1,2) \big(\gg2(2,3)-\gg2(2,4) \big)
+k_2^m  \gg1(1,2) \big(\gg2(1,3)-\gg2(1,4)\big) + (12\leftrightarrow 34) \big]    \cr
&+ \Big\{ k_5^m\bigl[\gg1(1,5)(\gg2(1,3)-\gg2(1,4)+\gg2(4,5)-\gg2(3,5))
+ \gg1(2,5)(\gg2(2,4)-\gg2(2,3)+\gg2(3,5)-\gg2(4,5))    &\SevTopTwo \cr
&\qquad{}+\gg1(1,2)(\gg2(3,5){-}\gg2(4,5))
+\gg1(3,4)(\gg2(1,5){-}\gg2(2,5))+\gg3(1,4)-\gg3(1,3)+\gg3(2,3)-\gg3(2,4)\bigr] + (5\leftrightarrow 6,7)
\Big\}\,.
}$$
This completes the bootstrapping of
the vectorial shuffle-symmetric functions for seven points.

\subsubsec Tensorial functions

An analogous procedure can be used for solving the tensorial seven-point
functions starting from their monodromy variations given in \assumption. The
outcome can be written as
\eqnn\seVTens
$$\eqalignno{
\cZ^{mn}_{12,3,4,5,6,7} &= \ell^m \ell^n \gg1(1,2) 
+ \big[ \ell^{(m} k_3^{n)}  (\gg2(1,3)-\gg2(2,3)) + 2  k_3^m k_3^n
 (\gg3(1,3)-\gg3(2,3) ) + (3\leftrightarrow 4,5,6,7) \big] \cr
&+ \gg2(1,2) (\ell^{(m} k_2^{n)} -\ell^{(m} k_1^{n)} ) + \gg3(1,2)(2 k_1^m k_1^n+2 k_2^m k_2^n - k_1^m k_2^n - k_2^m k_1^n) &\seVTens \cr
&+ \big[ k_3^{(m} k_1^{n)} (\gg1(1,2) \gg2(2,3) {-} \gg3(1,3){+}\gg3(2,3))
+k_3^{(m} k_2^{n)}  (\gg1(1,2) \gg2(3,1) {-} \gg3(1,3){+}\gg3(2,3))
+ (3\leftrightarrow 4,5,6,7) \big] \cr 
&+ \big[ k_3^{(m} k_4^{n)} 
(\gg1(1,2) \gg2(3,4) + \gg1(3,4)(\gg2(1,3)-\gg2(2,3)-\gg2(1,4) + \gg2(2,4))\cr
&\quad{}+\gg3(1,3)-\gg3(2,3)+\gg3(1,4)-\gg3(2,4))
+(3,4|3,4,5,6,7) \big]\,,\cr
\cZ^{mnp}_{1,2,3,4,5,6,7} &= \ell^m \ell^n\ell^p
+ \big[k_1^{(m}k_2^n\ell^{p)}\gg2(1,2) -  k_1^{(m} (k_1^n-k_2^n)k_2^{p)}
\gg3(1,2) + (1,2|1,2,3,4,5,6,7)\big]\cr
&+\big[k_1^{(m}k_2^nk_3^{p)} \big(
g^{(1)}_{ 23} (g^{(2)}_{  1 2} - g^{(2)}_{  1 3})
+ g^{(3)}_{  1 2} + g^{(3)}_{  1 3}
\big)
+ (1,2,3|1,2,3,4,5,6,7)\big]\,.
}$$
Note that the coefficient of $k_1^{m}k_2^nk_3^{p}$ in the last line is totally
symmetric in $1,2,3$. Again, their singularity structure within a given word
is the same as in their tree-level counterparts, see section \shufflesec.

\newsubsubsec\someapp Assembling seven-point GEIs

Now that the shuffle-symmetric ${\cal Z}$-functions at seven points are known, one can 
assemble the GEIs as described in section~\WSdualCs. The scalar GEIs
follow from the replacement rule $M_A M_{B,C,D}\rightarrow \cZ_{A,B,C,D}$ applied
to the Berends--Giele expansion of the BRST invariants
$C_{1|2345,6,7}$, $C_{1|234,56,7}$ and $C_{1|23,45,67}$ from \partI,
\eqnn\sevenM
$$\eqalignno{
E_{1|2345,6,7} &=
\cZ_{1,2345,6,7}  +  \cZ_{512,34,6,7}+ \cZ_{12,345,6,7}
+ \cZ_{123,45,6,7} +\cZ_{1234,5,6,7}   \cr
&+  \cZ_{5123,4,6,7} +\cZ_{51,234,6,7}  +  \cZ_{451,23,6,7}
+ \cZ_{3451,2,6,7} +  \cZ_{4512,3,6,7}\,,\cr
E_{1|234,56,7} &= \cZ_{1,234,56,7}  +  \cZ_{214,3,56,7}
+ \cZ_{15,234,6,7} - \cZ_{16,234,5,7}  +  \cZ_{12,34,56,7} &\sevenM\cr
& + \cZ_{123,4,56,7} + \cZ_{14,32,56,7} + \cZ_{143,2,56,7} + \cZ_{612,34,5,7} + \cZ_{6123,4,5,7}\cr
& + \cZ_{5124,3,6,7} + \cZ_{614,32,5,7} + \cZ_{6143,2,5,7} + \cZ_{5142,3,6,7} - \cZ_{512,34,6,7}\cr
& - \cZ_{5123,4,6,7} - \cZ_{6124,3,5,7} - \cZ_{514,32,6,7} - \cZ_{5143,2,6,7} - \cZ_{6142,3,5,7}\,,\cr
E_{1|23,45,67} &= \cZ_{1,23,45,67}+\cZ_{12,3,45,67} - \cZ_{13,2,45,67}+\cZ_{14,5,23,67}  -  \cZ_{15,4,23,67}   \cr
&
 +\cZ_{16,7,23,45}- \cZ_{17,6,23,45}
 + \cZ_{217,3,45,6} - \cZ_{317,2,45,6}  -\cZ_{216,3,45,7}\cr
& + \cZ_{316,2,45,7} + \cZ_{413,5,67,2} - \cZ_{513,4,67,2}  -\cZ_{412,5,67,3} + \cZ_{512,4,67,3}\cr
&  + \cZ_{615,7,23,4} - \cZ_{715,6,23,4}  -\cZ_{614,7,23,5} + \cZ_{714,6,23,5} 
 +\cZ_{7135,2,4,6} \cr
& + \cZ_{7153,2,4,6}-\cZ_{7125,3,4,6} - \cZ_{7152,3,4,6} -\cZ_{7134,2,5,6} - \cZ_{7143,2,5,6}\cr
& +  \cZ_{7124,3,5,6} + \cZ_{7142,3,5,6} -\cZ_{6135,2,4,7} - \cZ_{6153,2,4,7}+\cZ_{6125,3,4,7}\cr
& + \cZ_{6152,3,4,7} +\cZ_{6134,2,5,7} + \cZ_{6143,2,5,7} - \cZ_{6124,3,5,7} -\cZ_{6142,3,5,7}\,,\cr
}$$
and read as in \muchsimpler\ after the solutions for $\cZ$ obtained above are plugged in.
Similarly, the lengthy expansions of the vectorial GEIs
\eqnn\sevenP
$$\eqalignno{
E^m_{1|234,5,6,7} &= \cZ_{1,234,5,6,7}^m  +  \cZ_{123,4,5,6,7}^m
+  \cZ_{412,3,5,6,7}^m   +  \cZ_{341,2,5,6,7}^m   \cr
&+ \cZ_{12,34,5,6,7}^m  + \cZ_{41,23,5,6,7}^m
+ k_2^m \cZ_{1432,5,6,7} + k_4^m \cZ_{1234,5,6,7}\cr
& - k_3^m(\cZ_{1423,5,6,7}+ \cZ_{1243,5,6,7})
- \big[ k_5^m  (\cZ_{51,234,6,7} + \cZ_{512,34,6,7} + \cZ_{514,32,6,7}\cr
& \ \ \ \  + \cZ_{5123,4,6,7}  + \cZ_{5143,2,6,7}
- \cZ_{5124,3,6,7} + \cZ_{5142,3,6,7}  )  + (5\leftrightarrow 6,7) \big]\,, &\sevenP\cr
E^m_{1|23,45,6,7} &= \cZ_{1,23,45,6,7}^m
+  \cZ_{12,3,45,6,7}^m   - \cZ_{13,2,45,6,7}^m 
+ \cZ_{14,23,5,6,7}^m - \cZ_{15,23,4,6,7}^m   \cr
&+ \cZ_{413,2,5,6,7}^m  + \cZ_{512,3,4,6,7}^m 
- \cZ_{412,3,5,6,7}^m  -  \cZ_{513,2,4,6,7}^m \cr
&+ \big[k_3^m (\cZ_{123,45,6,7}- \cZ_{4123,5,6,7} + \cZ_{5123,4,6,7} ) - (2\leftrightarrow3)\big]
\cr
&+ \big[k_5^m (\cZ_{145,23,6,7} - \cZ_{2145,3,6,7}  + \cZ_{3145,2,6,7} ) - (4\leftrightarrow5)\big]
\cr
& - \big[ k_6^m (\cZ_{61,23,45,7} + \cZ_{612,3,45,7} - \cZ_{613,2,45,7} + \cZ_{614,23,5,7} - \cZ_{615,23,4,7} \cr
& \ \ \ \   - (\cZ_{6134,2,5,7}+\cZ_{6143,2,5,7}) - (\cZ_{6125,3,4,7}+\cZ_{6152,3,4,7}) \cr
& \ \ \ \   + (\cZ_{6135,2,4,7}+\cZ_{6153,2,4,7})
+ (\cZ_{6124,3,5,7}+\cZ_{6142,3,5,7}) ) + (6 \leftrightarrow 7) \big]\,,\cr
}$$
collapse to a few terms \muchsh\ when rewritten in terms of $V_w$- and $\p V_w$-functions. Similarly,
the tensorial GEIs
\eqnn\sevenS
$$\eqalignno{
E^{mn}_{1|23,4,5,6,7} &= \cZ^{mn}_{1,23,4,5,6,7}  + \cZ^{mn}_{12,3,4,5,6,7}
- \cZ^{mn}_{13,2,4,5,6,7} + k_3^{(m} \cZ^{n)}_{123,4,5,6,7}
- k_2^{(m}  \cZ^{n)}_{132,4,5,6,7}\cr
&+  \big[ k_4^{(m} k_5^{n)} \big\{  -\cZ_{514,23,6,7} + (\cZ_{1245,3,6,7}+  {\rm symm}(2,4,5))\cr
& \ \ \ \ -(\cZ_{1345,2,6,7}+ {\rm symm}(3,4,5))   \big\}  + (4,5|4,5,6,7) \big] \cr
&+ \big[ k_4^{(m} \big\{ \cZ^{n)}_{14,23,5,6,7}  -\cZ_{214,3,5,6,7}^{n)}  + \cZ_{314,2,5,6,7}^{n)} &\sevenS \cr
& \ \ \ \  + k_2^{n)} \cZ_{4132,5,6,7}  - k_3^{n)} \cZ_{4123,5,6,7}\big\}
+ (4 \leftrightarrow 5,6,7) \big]\,,\cr
%
E^{mnp}_{1|2,3,4,5,6,7} &= \cZ^{mnp}_{1,2,3,4,5,6,7}
+ \big[ k_2^{(m} \cZ_{12,3,4,5,6,7}^{np)}  + ( 2 \leftrightarrow 3,4,5,6,7) \big] \cr
&-  \big[ k_2^{(m} k_3^{n} \cZ_{213,4,5,6,7}^{p)} + ( 2,3| 2,3,4,5,6,7) \big]  \cr
&+  \big[ k_2^{(m} k_3^n k_4^{p)} (\cZ_{1234,5,6,7}+{\rm symm}(2,3,4)) + (
2,3,4|2,3,4,5,6,7) \big] \ ,\cr
}$$
become as compact as \rewrtC.

Moreover, there are three topologies of refined GEIs at seven points,
\eqnn\rewrtDalt
$$\eqalignno{
E_{1|23|4,5,6,7} &= \cZ_{23|1,4,5,6,7} + \cZ_{3|12,4,5,6,7} - \cZ_{2|13,4,5,6,7} + k_3^m \cZ^m_{123,4,5,6,7}\cr
&\quad{} - k_2^m \cZ_{132,4,5,6,7}^m + \big[ (s_{34} \cZ_{1234,5,6,7}-s_{24}\cZ_{1324,5,6,7}) + (4\leftrightarrow 5,6,7) \big]  \cr
E_{1|2|34,5,6,7} &= \cZ_{2|1,34,5,6,7}  + \cZ_{2|13,4,5,6,7} - \cZ_{2|14,3,5,6,7} \cr
&\quad{}- s_{23} (\cZ_{1243,5,6,7}+\cZ_{1423,5,6,7})  + s_{24} (\cZ_{1234,5,6,7}+\cZ_{1324,5,6,7})  \cr
&\quad{} + k_2^m ( \cZ^m_{12,34,5,6,7}  - \cZ^m_{213,4,5,6,7} + \cZ^m_{214,3,5,6,7})  &\rewrtDalt  \cr
&\quad{} + \big[ s_{25} (\cZ_{125,34,6,7}-\cZ_{3125,4,6,7}  + \cZ_{4125,3,6,7}) + (5\leftrightarrow 6,7) \big]  \ ,
\cr
E^m_{1|2|3,4,5,6,7} &= \cZ^m_{2|1,3,4,5,6,7}  + \big[ k_3^m \big\{ \cZ_{2|13,4,5,6,7}  -  k_2^p \cZ^p_{213,4,5,6,7} \big\}+ (3\leftrightarrow 4,5,6,7) \big]  \cr
&\ \ \ + k_2^p \cZ^{pm}_{12,3,4,5,6,7} + \big[ s_{23} \big\{ \cZ^m_{123,4,5,6,7} 
- k_4^m \cZ_{4123,5,6,7} - k_5^m \cZ_{5123,4,6,7} \cr
& \ \ \ \ \ \ \ \ \ \ \ \ \ \ \ \ \ \ \ - k_6^m \cZ_{6123,4,5,7}   - k_7^m \cZ_{7123,4,5,6} 
\big\} + (3\leftrightarrow 4,5,6,7) \big]   \, ,
}$$
and integration-by-parts identities lead to the compact representations \rewrtD\ or \rewrtE.

The above GEIs, in turn, will be used as input in the monodromy-variation 
equations to bootstrap the eight-point shuffle-symmetric functions.

\newsubsec\appCeight Eight points

In a similar fashion, it is possible to find all the solutions to 
the scalar shuffle-symmetric functions from the monodromy
variations \assumption\ using various change-of-basis identities such as
\snterms. A long but straightforward analysis leads to,
\eqnn\scalarEight
$$\eqalignno{
\cZ_{12345,6,7,8}& = - \p V_3(1,2,3,4,5)\,,&\scalarEight\cr
\cZ_{123,456,7,8} &=
-\p V_1( 1, 2,3) V_2(1, 4, 5, 6)
-\p V_2(4, 1, 2,3) V_1(1, 5, 6)\cr
&\quad{} +\p V_2(6, 1, 2,3) V_1(1, 4, 5)
 - \p V_3(4, 5, 1, 2,3) + \p V_3(4, 6, 1, 2,3)\cr
&\quad{}+ \p V_3(6, 4, 1, 2,3) - \p V_3(6, 5, 1, 2,3)\,,\cr
\cZ_{1234,56,7,8} &=
- \p V_2(1,2,3,4) V_1(1,5,6)
+ \p V_3(6,1,2,3,4)
- \p V_3(5,1,2,3,4)\,,\cr
\cZ_{123,45,67,8} &=
- \p V_1( 1, 2,3) V_1(1, 4, 5) V_1(1, 6, 7)
+ \p V_2( 5, 1, 2,3) V_1(1, 6, 7) \cr
&- \p V_2( 6, 1, 2,3) V_1(1, 4, 5)
 - \p V_2( 4, 1, 2,3) V_1(1, 6, 7)
 + \p V_2( 7, 1, 2,3) V_1(1, 4, 5)\cr
&+ \p V_3( 4, 7, 1, 2,3)
- \p V_3( 4, 6, 1, 2,3) + \p V_3( 5, 6, 1, 2,3)
- \p V_3( 5, 7, 1, 2,3) \cr
 & + 
 \p V_3( 7, 4, 1, 2,3) - \p V_3( 6, 4, 1, 2,3)
+ \p V_3( 6, 5, 1, 2,3)
- \p V_3( 7, 5, 1, 2,3)\,,\cr
\cZ_{12,34,56,78} & =
       \gg1(1,2) \V1(1,3,4) \V1(1,5,6) \V1(1,7,8)\cr
&       - \V1(1,3,4) \V1(1,5,6) \DV1(7,1,2)
       + \V1(1,3,4) \V1(1,5,6) \DV1(8,1,2)\cr
&       - \V1(1,3,4) \V1(1,7,8) \DV1(5,1,2)
       - \V1(1,5,6) \V1(1,7,8) \DV1(3,1,2)\cr
&       + \V1(1,5,6) \V1(1,7,8) \DV1(4,1,2)
       + \V1(1,3,4) \V1(1,7,8) \DV1(6,1,2)\cr
&       - \V1(1,3,4) \DV2(5,7,1,2)
       + \V1(1,3,4) \DV2(5,8,1,2)
       + \V1(1,3,4) \DV2(6,7,1,2)\cr
&       - \V1(1,3,4) \DV2(6,8,1,2)
       - \V1(1,3,4) \DV2(7,5,1,2)
       + \V1(1,3,4) \DV2(7,6,1,2)\cr
&       + \V1(1,3,4) \DV2(8,5,1,2)
       - \V1(1,3,4) \DV2(8,6,1,2)
       - \V1(1,5,6) \DV2(3,7,1,2)\cr
&       + \V1(1,5,6) \DV2(3,8,1,2)
       + \V1(1,5,6) \DV2(4,7,1,2)
       - \V1(1,5,6) \DV2(4,8,1,2)\cr
&       - \V1(1,5,6) \DV2(7,3,1,2)
       + \V1(1,5,6) \DV2(7,4,1,2)
       + \V1(1,5,6) \DV2(8,3,1,2)\cr
&       - \V1(1,5,6) \DV2(8,4,1,2)
       - \V1(1,7,8) \DV2(3,5,1,2)
       + \V1(1,7,8) \DV2(3,6,1,2)\cr
&       + \V1(1,7,8) \DV2(4,5,1,2)
       - \V1(1,7,8) \DV2(4,6,1,2)
       - \V1(1,7,8) \DV2(5,3,1,2)\cr
&       + \V1(1,7,8) \DV2(5,4,1,2)
       + \V1(1,7,8) \DV2(6,3,1,2)
       - \V1(1,7,8) \DV2(6,4,1,2)\cr
&       - \DV3(3,5,7,1,2)
       + \DV3(3,5,8,1,2)
       + \DV3(3,6,7,1,2)
       - \DV3(3,6,8,1,2)\cr
&       - \DV3(3,7,5,1,2)
       + \DV3(3,7,6,1,2)
       + \DV3(3,8,5,1,2)
       - \DV3(3,8,6,1,2)\cr
&       + \DV3(4,5,7,1,2)
       - \DV3(4,5,8,1,2)
       - \DV3(4,6,7,1,2)
       + \DV3(4,6,8,1,2)\cr
&       + \DV3(4,7,5,1,2)
       - \DV3(4,7,6,1,2)
       - \DV3(4,8,5,1,2)
       + \DV3(4,8,6,1,2)\cr
&       - \DV3(5,3,7,1,2)
       + \DV3(5,3,8,1,2)
       + \DV3(5,4,7,1,2)
       - \DV3(5,4,8,1,2)\cr
&       - \DV3(5,7,3,1,2)
       + \DV3(5,7,4,1,2)
       + \DV3(5,8,3,1,2)
       - \DV3(5,8,4,1,2)\cr
&       + \DV3(6,3,7,1,2)
       - \DV3(6,3,8,1,2)
       - \DV3(6,4,7,1,2)
       + \DV3(6,4,8,1,2)\cr
&       + \DV3(6,7,3,1,2)
       - \DV3(6,7,4,1,2)
       - \DV3(6,8,3,1,2)
       + \DV3(6,8,4,1,2)\cr
&       - \DV3(7,3,5,1,2)
       + \DV3(7,3,6,1,2)
       + \DV3(7,4,5,1,2)
       - \DV3(7,4,6,1,2)\cr
&       - \DV3(7,5,3,1,2)
       + \DV3(7,5,4,1,2)
       + \DV3(7,6,3,1,2)
       - \DV3(7,6,4,1,2)\cr
&       + \DV3(8,3,5,1,2)
       - \DV3(8,3,6,1,2)
       - \DV3(8,4,5,1,2)
       + \DV3(8,4,6,1,2)\cr
&       + \DV3(8,5,3,1,2)
       - \DV3(8,5,4,1,2)
       - \DV3(8,6,3,1,2)
       + \DV3(8,6,4,1,2)\,.
}$$
The sheer size of the solution for $\cZ_{12,34,56,78}$ can be traced back to
the length of the intermediate identity \snterms\ used in its derivation.
Fortunately, the combinatorics of such solutions can be understood in terms of
the multi-word rhomap \mrhomap, and an efficient algorithm to generate them at
arbitrary multiplicity will be provided below.

\subsubsec Vectorial shuffle-symmetric functions

The monodromy variation of $\cZ^m_{1234,5,6,7,8}$ can be written in the
canonical basis of GEIs as
\eqnn\DfourtopExp
$$\eqalignno{
D\cZ^m_{1234,5,6,7,8} &= \Omega_{14}E^m_{1|234,5,6,7,8}
- \Big[ k_5^m\Omega_{45}E_{1|2345,6,7,8} + (5\leftrightarrow
6,7,8)\Big]\cr
&= \Omega_{14}\V2(1,2,3,4)\ell^m
+ k_2^m \Omega_{14}\DV2(2,3,4,1) &\DfourtopExp\cr
&\quad{} - k_3^m \Omega_{14}\big(\DV2(3,2,4,1) + \DV2(3,4,2,1)\big)
+ k_4^m\Omega_{14}\DV2(4,3,2,1)\cr
&\quad{}+ \Big[k_5^m\big(\Omega_{14}\V2(1,2,3,4)\gg1(1,5) -
\Omega_{45}\V3(1,2,3,4,5)\big) + (5\leftrightarrow 6,7,8)\Big]\, .
}$$
Note that there is no linear combination of $\D2V2(i,j,k,l)$-functions
that integrates to $\Omega_{14}\DV2(2,3,4,1)$ as can be checked using
$D \D2V2(i,j,k,l) = 2\Omega_{li}\DV2(i,j,k,l)$. However, the
integration of $\Omega_{14}\V2(1,2,3,4)\ell^m$ produces correction
terms since
\eqn\trickEight{
D(-\DV2(1,2,3,4)\ell^m) = \Omega_{14}\V2(1,2,3,4)\ell^m +
\DV2(1,2,3,4)\sum_{j=2}^8\Omega_{1j}k_j^m.
}
By adding and subtracting the sum on the right-hand side prior to integration
produces corrections to the other terms $\sim k_i^m$. For example,
the new $k_2^m$ terms can be ``integrated'' as
\eqnn\newReqInt
$$\eqalignno{
\int\Big(\Omega_{14}\DV2(2,3,4,1) - \Omega_{12}\DV2(1,2,3,4)\Big) &=
\half\Big(\D2V2(1,4,3,2) + \D2V2(2,1,3,4) &\newReqInt\cr
&\qquad{}+ \D2V2(2,3,1,4) + \D2V2(1,2,3,4)\Big)\,.
}$$
Repeating the same steps as in the previous analyses yields,
\eqnn\ansEight
$$\eqalignno{
\cZ^m_{1234,5,6,7,8} &= -\ell^m\DV2(1,2,3,4) + k_4^m\D2V2(1,2,3,4)
&\ansEight\cr
&+ \half k_2^m\big[\D2V2(1,4,3,2) + \D2V2(2,1,3,4)
+ \D2V2(2,3,1,4) + \D2V2(1,2,3,4)\big]\cr
&- \half k_3^m\big[\D2V2(1,4,2,3) + \D2V2(3,4,2,1)
+ \D2V2(4,1,2,3) - \D2V2(1,2,3,4)\big]\cr
&- \Big[ k_5^m\big(\DV2(1,2,3,4)\gg1(1,5) + \DV3(5,1,2,3,4)\big) +
(5\leftrightarrow 6,7,8)\Big]\,,
}$$
which can be rewritten as,
\eqnn\NewansEight
$$\eqalignno{
\cZ^m_{1234,5,6,7,8} &= -\ell^m\DV2(1,2,3,4)
- \half k_1^m\D2V2(1,2,3,4) + \half k_4^m\D2V2(1,2,3,4)\cr
&+ \half k_2^m\big[\D2V2(1,4,3,2) + \D2V2(2,1,3,4)
+ \D2V2(2,3,1,4)\big]\cr
&- \half k_3^m\big[\D2V2(1,4,2,3) + \D2V2(3,4,2,1)
+ \D2V2(4,1,2,3)\big]\cr
&+ \Big[ k_5^m \V2(1,2,3,4)(\gg2(1,5) - \gg2(4,5))
+ (5\leftrightarrow 6,7,8)\Big]\,&\NewansEight\cr
}$$
after using the weight-four version of \surpr,
\eqn\wgtfive{
\DV2(1,2,3,4)\gg1(1,5) + \DV3(5,1,2,3,4)
= -\V2(1,2,3,4)(\gg2(1,5)-\gg2(4,5)) - \half\D2V2(1,2,3,4)\,.
}
The expression for $\cZ^m_{123,45,6,7,8}$ can be obtained similarly
and a long analysis leads to
\eqnn\messeight
$$\eqalignno{
\cZ^m_{123,45,6,7,8} &=
   - \ell^m\big(\V1(1,4,5)\DV1(1,2,3)
          + \DV2(4,1,2,3)
          - \DV2(5,1,2,3)\big) &\messeight\cr
       &+ \half \big[ k_1^m \big(
          {-} \V1(3,4,5)\D2V1(1,2,3)
          + \D2V2(1,2,3,4)
          - \D2V2(1,2,3,5)
          \big) + (1\leftrightarrow3)\big]\cr
       & + \half k_2^m  \big(
           \V1(3,4,5)\D2V1(2,1,3)
	  - \D2V2(2,1,3,4)
          + \D2V2(2,1,3,5)\cr
&\hskip25pt{}          + \V1(1,4,5)\D2V1(2,3,1)
          - \D2V2(2,3,1,4)
          + \D2V2(2,3,1,5)
          \big)\cr
      & + \big[k_4^m \big(
          \gg1(1,4) \DV2(5,1,2,3)
          - \DV1(1,2,3) \DV1(1,5,4)
          + \DV3(5,4,1,2,3)\cr
&\qquad{}   - \half \big(\V1(3,4,5)\D2V1(1,2,3)
          +\D2V2(1,2,3,5)
          + \D2V2(1,2,4,3)\cr
&\qquad{} + \D2V2(1,4,2,3)
          + \D2V2(3,2,1,4)\big)
          \big) - (4\leftrightarrow5)\big]\cr
      & + \big[k_6^m \big(
            \gg2(4,6)\V2(1,2,3,4)
          - \gg2(1,6)\V1(1,3,2)\V1(1,4,5)\cr
&\qquad{}          - \gg2(3,6)\V1(1,2,3)\V1(3,4,5)
          - \gg2(5,6)\V2(1,2,3,5)
          \big) + (6\leftrightarrow7,8)\big] \, .
}$$
Solving the monodromy variation of $\cZ^m_{12,34,56,7,8}$ along similar
lines yields a long formula which we suppress (it can be downloaded in
\Zfunctions).
This completes the bootstrap procedure for the vectorial shuffle-symmetric
eight-point functions. From the above solutions we can derive the
eight-point vectorial GEIs which in turn allow to bootstrap the vectorial 
${\cal Z}$-functions at nine points.

\subsubsec Tensorial and refined functions

Given their sizes, the tensorial and refined shuffle-symmetric functions will
be omitted. Their explicit expansions are available to download as plain text
files in \Zfunctions, and are ready to be used with {\tt FORM} \FORM.

\subsec A closed formula for scalar shuffle-symmetric functions

The solutions for the scalar shuffle-symmetric functions can be generated via
a conjectural closed formula. This empirical observation is
based on the word-invariant map ${\cal I}(\ldots)$
defined in the appendix \clikeapp\ and is given by
\eqn\cZalgo{
\cZ_{A,B,C,D} =- A\odot \cI(\emptyset|B,C,D)\,,
}
where the $\odot$ operation is defined by
\eqnn\newconc
\eqnn\fourslotmap
$$\eqalignno{
aA\odot(B,C,D,E)&\equiv(BaA | aC,aD,aE)_V\,,&\newconc\cr
(A | B,C,D)_V &\equiv
\p V_{|A|-2}(A)V_{|B|-2}(B)V_{|C|-2}(C)V_{|D|-2}(D)\,,&\fourslotmap
}$$
with the understanding that $V_0(i,j)\equiv 1$ and $\p V_{-1}(i)\equiv -1$.

For example, let us consider $\cZ_{123,45,67,8}=
- 123\odot \cI(\emptyset|45,67,8)$.
A straightforward application of the recursions in appendix \clikeapp\ leads to,
\eqnn\threeslots
$$\eqalignno{
-\cI(\emptyset|45,67,8) &=
-(\emptyset|67,45,8) - (6|45,7,8) -
(64|5,7,8)+(65|4,7,8)&\threeslots\cr
&\quad{}+(7|45,6,8)+(74|5,6,8)-(75|4,6,8)-(4|67,5,8)\cr
&\quad{}-(46|7,5,8)+(47|6,5,8)+(5|67,4,8)+(56|7,4,8)-(57|6,4,8)\,.
}$$
Now, using the definition \newconc\ yields
\eqnn\concthree
$$\eqalignno{
-123\odot\cI(\emptyset|45,67,8) &=
-(123|167,145,18)_V - (6123|145,17,18)_V - (64123|15,17,18)_V \cr
&\quad{}+(65123|14,17,18)_V+(7123|145,16,18)_V+(74123|15,16,18)_V\cr
&\quad{}-(75123|14,16,18)_V-(4123|167,15,18)_V-(46123|17,15,18)_V\cr
&\quad{}+(47123|16,15,18)_V+(5123|167,14,18)_V+(56123|17,14,18)_V\cr
&\quad{}-(57123|16,14,18)_V\,. &\concthree
}$$
Finally, the definition \fourslotmap\ leads to the
correct expression for $\cZ_{123,45,67,8}$ from \scalarEight.

Two comments are in order. First, one may notice that
the combinatorics of the permutations in \concthree\ is closely
related to the change-of-basis identity expressing $-C_{3|12,45,67,8}$
in a basis of $C_{1|A,B,C}$ originally derived in \partI. For instance,
\eqnn\Ccan
$$\eqalignno{
-C_{3|12,45,67} &=
       - C_{1|23,45,67}
       - C_{1|236,7,45}
       - C_{1|2364,5,7}&\Ccan\cr
&\quad{}
       + C_{1|2365,4,7}
       + C_{1|237,6,45}
       + C_{1|2374,5,6}\cr
&\quad{}       - C_{1|2375,4,6}
       - C_{1|234,5,67}
       - C_{1|2346,7,5}\cr
&\quad{}       + C_{1|2347,6,5}
       + C_{1|235,4,67}
       + C_{1|2356,7,4}\cr
&\quad{}       - C_{1|2357,6,4}\,.
}$$
Comparing the expressions \concthree\ and \Ccan\ we see that
$C_{1|23A,B,C}\to (A123,1B,1C,18)$ maps one expression into the other.

Second, the right-hand side of the algorithm $\cZ_{A,B,C,D} =- A\odot
\cI(\emptyset|B,C,D)$ is not manifestly symmetric under exchange of the words
$A\leftrightarrow B,C,D$. Therefore, it must generate identities among the
various functions $V_n(\ldots)$ and their generalizations $\p V_n( \ldots)$.
Consider, for example, the five-point function $\cZ_{1,23,4,5}$ and evaluate
it in the two inequivalent orderings using the algorithm \cZalgo; we see that
$-23\odot\cI(\emptyset|1,4,5)$ and $-1\odot\cI(\emptyset|23,4,5)$ give rise
to
\eqn\diffords{
\cZ_{23,1,4,5}=-\p V_0(2,3)\,,\qquad
\cZ_{1,23,4,5}= V_1(1,2,3) -\DV0(1,2) + \DV0(1,3)\,,
}
which are, of course, equal. A bit less obvious is the equality of both
$\cZ_{1,234,5,6}$ and $\cZ_{234,1,5,6}$ under the algorithm \cZalgo, as
this implies
\eqnn\notman
$$\eqalignno{
V_2(1,2,3,4) &= V_1(1,2,3)\DV0(1,4) - V_1(1,3,4)\DV0(1,2) - \DV1(1,2,4)
+ \DV1(1,3,2)\cr
&\quad{}+ \DV1(1,3,4) - \DV1(1,4,2) - \DV1(2,3,4)\,.
}$$
These observations imply that it is advantageous to write
the expansion of the scalar GEIs in a certain ``canonical'' order such as:
$E_{1|234,5,6}= \cZ_{1,234,5,6} + \cZ_{12,34,5,6} + \cZ_{123,4,5,6}
+\cZ_{412,3,5,6}-\cZ_{14,23,5,6}+\cZ_{143,2,5,6}$ since
it takes the shortest form $E_{1|234,5,6} = V_2(1,2,3,4)$.

\appendix{B}{The Jacobi theta expansion of $g^{(n)}(z,\tau)$}
\applab\appgs

\noindent In this appendix we will list, for convenience, the explicit
expansions of the coefficients of the Kronecker--Eisenstein series in terms of
Jacobi theta functions for the first few cases.

Recall that the functions $g^{(n)}(z,\tau)$ admit the following recursive
expansion \BroedelVLA\
\eqn\grec{
g^{(n)}(z,\tau) = {1\over n}\sum_{j=1}^n
{\cal E}_j(z,\tau)g^{(n-j)}(z,\tau)\,,\quad g^{(1)}(z,\tau) \equiv
E_1(z,\tau)\,,\quad g^{(0)}(z,\tau)\equiv1\,,
}
where
${\cal E}_j(z,\tau) \equiv (-1)^j \big( {\rm G}_j(\tau) - E_j(z,\tau)\big)$
and $E_{n+1}(z,\tau) = (-1)^n {1\over n!} \p^{n+1} \log\t_1(z,\tau)$
with ${\rm G}_j(\tau)$ denoting the Eisenstein series \holeis.
It is a matter of tedious algebra to get:
\eqnn\cool
$$\eqalignno{
g^{(1)}(z,\tau) & = {\theta_1^{(1)}(z,\tau) \over \theta_1(z,\tau)} &\cool\cr
g^{(2)}(z,\tau) & = {1\over 2!}{\theta_1^{(2)}(z,\tau) \over \theta_1(z,\tau)} - {1\over 3!}\mythetaZ3\cr
g^{(3)}(z,\tau) & = {1\over 3!}{\theta_1^{(3)}(z,\tau) \over \theta_1(z,\tau)} - {1\over 3!}\mythetaZ3 g^{(1)}(z,\tau)\cr
g^{(4)}(z,\tau) & =
{1\over 4!} {\theta_1^{(4)}(z,\tau) \over \theta_1(z,\tau)}
- {1\over 3!}\mythetaZ3 g^{(2)}(z,\tau)
- {1\over 5!}\mythetaZ5\cr
g^{(5)}(z,\tau) & =
{1\over 5!} {\theta_1^{(5)}(z,\tau) \over \theta_1(z,\tau)}
- {1\over 3!}\mythetaZ3 g^{(3)}(z,\tau)
- {1\over 5!}\mythetaZ5 g^{(1)}(z,\tau)\,.
}$$
The surprisingly simple pattern above arises from non-trivial cancellations such as
\eqn\simpleE{
{\rm G}_4 - \half {\rm G}_2^2 = -{1\over 30}\mythetaZ5\,,\qquad
8{\rm G}_6 - 6 {\rm G}_2 {\rm G}_4 + {\rm G}_2^3 = -{1\over 105}\mythetaZ7\,,
}
where the expansion
of the Eisenstein series in terms of Jacobi theta functions reads \kiritsis
\eqnn\GstoJac
$$\eqalignno{
{\rm G}_2(\tau) &=-{1\over 3}\mythetaZ3\,,\qquad
{\rm G}_4(\tau) =-{1\over 30}\mythetaZ5 + {1\over 18}\biggl(\mythetaZ3\biggr)^2 &\GstoJac\cr
{\rm G}_6(\tau) &= -{1\over 840}\mythetaZ7 + {1\over 120}\mythetaZ5\mythetaZ3
- {1\over 108}\biggl(\mythetaZ3\biggr)^3\,.\cr
}$$

\subsec Laurent series expansion of the $g^{(n)}(z,\tau)$-functions

The Laurent expansion of $g^{(n)}(z,\tau)$ follows from \grec\ and \weil\ (note ${p\choose0}\equiv1$)
\eqn\weilEn{
E_n(z,\tau) = {1\over z^n} + (-1)^n \sum_{m=1}^\infty {2m-1\choose n-1}
{\rm G}_{2m}(\tau) z^{2m-1}\,.
}
More explicitly,
\eqnn\gnexpansion
$$\eqalignno{
g^{(1)}(z,\tau) &= {1\over z} - {\rm G}_2z - {\rm G}_4z^3 - {\rm G}_6 z^5 +{\cal O}(z^7) &\gnexpansion\cr
g^{(2)}(z,\tau) &= - {\rm G}_2 + \half\big({\rm G}_2^2 -5{\rm G}_4\big)z^2 +\big({\rm G}_2{\rm G}_4 - {7\over 2}{\rm G}_6\big)z^4 +{\cal O}(z^6) \cr
g^{(3)}(z,\tau) &= \half\big({\rm G}_2^2 -5{\rm G}_4\big)z +
\half \big(5 {\rm G}_2{\rm G}_4 - {35\over 3}{\rm G}_6 -{1\over3}{\rm G}_2^3\big)z^3 + {\cal O}(z^5) \cr
g^{(4)}(z,\tau) &= - {\rm G}_4 + \half\big(5 {\rm G}_2{\rm G}_4 - {35\over 3}{\rm G}_6 -{1\over3}{\rm G}_2^3\big)z^2 +{\cal O}(z^4) \, .\cr
}$$

\appendix{C}{String correlators and GEIs}
\applab\convencorr

\noindent
For convenience, in this appendix we quote from
part III of this series of papers a representation of the one-loop
correlators utilizing GEIs for $n=4,5,6,7$, as they are
frequently referred to in this work. 
\eqnn\fourcor
\eqnn\fivecor
\eqnn\sixcor
\eqnn\sevencor
$$\eqalignno{
{\cal K}_4(\ell)&= C_{1|2,3,4} E_{1|2,3,4}  &\fourcor
\cr
{\cal K}_5(\ell)&= C^m_{1|2,3,4,5} E^m_{1|2,3,4,5} + \big[ C_{1|23,4,5} s_{23} E_{1|23,4,5} + (2,3|2,3,4,5)\big]\, ,
&\fivecor\cr
{\cal K}_6(\ell)&= 
 {1\over 2} C^{mn}_{1|2,3,4,5,6} E^{mn}_{1|2,3,4,5,6}
 -  \big[ P_{1|2|3,4,5,6} E_{1|2|3,4,5,6}
+ (2\leftrightarrow 3,\ldots,6) \big] \cr
& \! \! \! +  \big[ s_{23} C^m_{1|23,4,5,6} E^m_{1|23,4,5,6}
+ (2,3|2,3,\ldots,6) \big]  &\sixcor \cr
& \! \! \!   + \Big( \big[ s_{23} s_{45} C_{1|23,45,6} E_{1|23,45,6}
+ {\rm cyc}(3,4,5) \big] + (6\leftrightarrow 5,4,3,2) \Big)  \cr
& \! \! \!  + \Big( \big[ s_{23}s_{34} C_{1|234,5,6}  E_{1|234,5,6}
+ {\rm cyc}(2,3,4) \big]
+ (2,3,4|2,3,\ldots,6) \Big)\cr
\cK_{7}(\ell) &=
{1\over6}C^{mnp}_{1|2,3,4,5,6,7}E^{(s)mnp}_{1|2,3,4,5,6,7}&\sevencor\cr
&+\half C^{mn}_{1|23,4,5,6,7} E^{(s)mn}_{1|23,4,5,6,7} + (2,3|2,3,4,5,6,7)\cr
&+\big[C^{m}_{1|234,5,6,7} E^{(s)m}_{1|234,5,6,7}
+ C^{m}_{1|243,5,6,7} E^{(s)m}_{1|243,5,6,7} \big]+ (2,3,4|2,3,4,5,6,7)\cr
&+ \big[C^{m}_{1|23,45,6,7} E^{(s)m}_{1|23,45,6,7} + {\rm cyc}(2,3,4)\big]+(6,7|2,3,4,5,6,7)\cr
&+\big[C_{1|2345,6,7} E^{(s)}_{1|2345,6,7} + {\rm perm}(3,4,5)\big]+(2,3,4,5|2,3,4,5,6,7)\cr
&+\big[C_{1|234,56,7} E^{(s)}_{1|234,56,7} + C_{1|243,56,7} E^{(s)}_{1|243,56,7} + {\rm cyc}(5,6,7)\big]
+ (2,3,4|2,3,4,5,6,7)\cr
&+\big[ C_{1|23,45,67} E^{(s)}_{1|23,45,67} + {\rm cyc}(4,5,6) \big]+ (3\leftrightarrow 4,5,6,7)\cr
&- P^m_{1|2|3, \ldots,7} E^{(s)m}_{1|2|3, \ldots,7} + (2\leftrightarrow3,4,5,6,7)\cr
&- P_{1|23|4, \ldots,7} E^{(s)}_{1|23|4, \ldots,7}+(2,3|2,3,4,5,6,7)\cr
&- \big[P_{1|2|34,5,6,7} E^{(s)}_{1|2|34,5,6,7}+{\rm cyc}(2,3,4)\big]+(2,3,4|2,3,4,5,6,7)\,.
}$$
By the symmetric role of GEIs and BRST (pseudo-)invariants,
these representations manifest the double-copy structure of
one-loop open-superstring amplitudes \MafraIOJ.
Other properties of one-loop correlators including locality are
manifest in various alternative representations given in part III.

\listrefs

\bye